\documentclass[twocolumn]{aastex61}

\definecolor{carnelian}{rgb}{0.7, 0.11, 0.11}
\definecolor{ao(english)}{rgb}{0.0, 0.5, 0.0}

\received{\today}
\revised{}
\accepted{}
\submitjournal{ApJ}

\shorttitle{Non-Maxwellian analysis of \textit{IRIS} observations}
\shortauthors{Dud\'{i}k et al.}

\begin{document}

\title{Non-Maxwellian analysis of the transition-region line profiles \\ observed by the \textit{Interface Region Imaging Spectrograph}}

\correspondingauthor{Jaroslav Dud\'{i}k}
\email{dudik @asu.cas.cz}

\author[0000-0003-1308-7427]{Jaroslav Dud\'{i}k}
\affil{Astronomical Institute of the Czech Academy of Sciences, Fri\v{c}ova 298, 251 65 Ond\v{r}ejov, Czech Republic}

\author{Vanessa Polito}
\affiliation{Smithsonian Astrophysical Observatory, 60 Garden Street, MS 58, Cambridge, MA 02138, USA}

\author{Elena Dzif\v{c}\'{a}kov\'{a}}
\affiliation{Astronomical Institute of the Czech Academy of Sciences, Fri\v{c}ova 298, 251 65 Ond\v{r}ejov, Czech Republic}

\author{Giulio Del Zanna}
\affiliation{Department of Applied Mathematics and Theoretical Physics, CMS, University of Cambridge, Wilberforce Road, Cambridge CB3 0WA, United Kingdom}
%

\author{Paola Testa}
\affiliation{Smithsonian Astrophysical Observatory, 60 Garden Street, MS 58, Cambridge, MA 02138, USA}

\begin{abstract}
We investigate the nature of the spectral line profiles for transition region ions observed with the \textit{Interface Region Imaging Spectrograph (IRIS)}. In this context, we have analyzed an active-region observation performed by \textit{IRIS} in its 1400\,\AA~spectral window. The transition-region lines are found to exhibit significant wings in their spectral profiles, which can be well-fitted with non-Maxwellian $\kappa$-distribution. The fit with a $\kappa$-distribution can perform better than a double Gaussian fit, especially for the strongest line, \ion{Si}{4} 1402.8\,\AA. Typical values of $\kappa$ found are about 2, occurring in a majority of spatial pixels where the transition region lines are symmetric, i.e., the fit can be performed. Furthermore, all five spectral lines studied (from \ion{Si}{4}, \ion{O}{4} and \ion{S}{4}) appear to have the same FWHM irrespective of whether the line is an allowed or an intercombination transition. A similar value of $\kappa$ is obtained for the electron distribution by fitting of the line intensities relative to \ion{Si}{4} 1402.8\,\AA, if photospheric abundances are assumed. The $\kappa$-distributions however do not remove the presence of non-thermal broadening. Instead, they actually increase the non-thermal width. This is because for $\kappa$-distributions the transition-region ions are formed at lower temperatures. The large observed non-thermal width lowers the opacity of the \ion{Si}{4} line sufficiently enough for this line to become optically thin.
\end{abstract}


%
\section{Introduction}
\label{Sect:1}

The solar transition region (hereafter, TR) is an interface between the cool solar chromosphere and the overlying hot corona. Typical TR temperatures span a few times 10$^4$\,K to almost 10$^6$\,K, resulting in strong ultraviolet (UV) emission lines from heavy ions, such as \ion{C}{4}, \ion{Si}{4}, and \ion{O}{4}. The TR is highly inhomogeneous and also temporally variable. Recent advances in understanding of this enigmatic part of the solar atmosphere enabled by the \textit{Interface Region Imaging Spectrograph} \citep[\textit{IRIS},][]{DePontieu14} include, among others, observational confirmation of the existence of short-lived TR loops \citep{Hansteen14} predicted by \citet{Feldman83,Feldman87} as well as observational signatures of accelerated electrons at footpoints of coronal loops \citep{Testa14}.

Even in the absence of clearly transient phenomena, and aside of the ubiquitous observed redshifts \citep[e.g.,][]{Brekke97,Polito16b} spectroscopic observations in the UV contain a number of features that have not been fully understood. These include \textit{(i)} the fact that some lines can be much stronger than predicted, assuming ionization equilibrium and Maxwellian plasma, \textit{(ii)} the large non-thermal line widths, as well as \textit{(iii)} peculiar line profiles exhibiting enhanced wings. An example of the first category are the intensities of the allowed \ion{Si}{4} lines at 1393.76\,\AA~and 1402.77\,\AA, which are stronger by a factor of $\approx$5 or higher compared to the neighboring intercombination lines of \ion{O}{4} \citep[e.g.,][]{Doyle84,Hayes87,Judge95,Curdt01,Doschek01,DelZanna02,Peter14Sci,Polito16b,Doschek16}. These lines are now observed by \textit{IRIS} within its 1400\,\AA~spectral channel at spectral resolution of about 26 m\AA~and spatial resolution of about 0.33$\arcsec$. The \ion{Si}{4} and \ion{O}{4} lines are formed at similar temperatures, log($T$\,[K])\,=\,4.9 and 5.15, respectively, if collisional ionization equilibrium and a Maxwellian distribution (i.e., no accelerated particles) are assumed \citep{Dudik14a}. Under such conditions, the \ion{O}{4} 1401.16\,\AA~line is expected to be stronger than the \ion{Si}{4} 1402.77\,\AA~one for typical TR differential emission measure distributions. There are only a few such observation known to us. One is a case of the RR Telescopii nebula \citep{Harper99,DelZanna02,Keenan02}, while other stars, such as $\alpha$ Centauri, show similar \ion{Si}{4}\,/\,\ion{O}{4} ratios to the Sun \citep[e.g.,][]{Pagano04,Ayres15}. The other one is an umbral footpoint of a coronal loop reported on by \citet{Chitta16}. Transient ionization, i.e., departures from the equilibrium ionic composition, has been invoked as a possible explanation of the discrepant \ion{Si}{4}\,/\,\ion{O}{4} ratios \citep{Doyle13,Olluri15} as well as their correlation with the \ion{Si}{4} intensities \citep[e.g.,][]{Martinez-Sykora16}. There is an extended literature where transient ionization was shown to significantly affect line intensities in TR hydrodynamical modeling \citep[cf.,][]{Raymond78,Dupree79,Noci89,Raymond90,Hansteen93,Spadaro94,Bradshaw03a,Bradshaw03b,Bradshaw04}. Most of the literature discussed \ion{C}{4} lines, which behave similarly to the \ion{Si}{4} ones. Departures from the Maxwellian distribution have also been invoked as another possible explanation. Enhanced high-energy tails in the electron distribution function, modeled by a $\kappa$-distribution, lead to orders-of-magnitude enhancements in the ionization rate, resulting in the TR ions being formed at much lower temperatures \citep{Dzifcakova13a}. This effect is stronger for \ion{Si}{4} than for \ion{O}{4} \citep{Dudik14a}, leading to strong increase of \ion{Si}{4} intensities. 

The typical non-thermal widths observed in TR lines are around 20\,km\,s$^{-1}$ or larger \citep[e.g.,][]{Kjeldseth-Moe77,Doschek77,Dere87,Dere93,Chae98,Peter99,Peter00,Peter01,Keenan02,Akiyama03,Akiyama05,DePontieu15}. We note that a non-thermal width of 20\,km\,s$^{-1}$ translates to about 0.1\,\AA~for the \ion{Si}{4} 1402.77\,\AA~line. The thermal width is a factor of $\approx$\,3 lower, being 6.86\,km\,s$^{-1}$ for the \ion{Si}{4} line at its Maxwellian peak formation temperature \citep[e.g.][]{DePontieu15}. Thus, the line width is dominated by the non-thermal component. These non-thermal widths derived from observations remain unchanged despite the significant advancement in the spatial resolution achieved so far. 
\citet{DePontieu15} and \citet{Testa16} both find peaks of non-thermal width distributions (for \ion{Si}{4} and \ion{Fe}{12} TR emission, respectively) at about $\approx$15\,km\,s$^{-1}$. While the peaks of the distributions are mostly unchanged with the spatial resolution, the full distributions do change \citep[see e.g., discussion in][]{Testa16}. This invariance with respect to the spatial resolution could be due to sub-resolution structures smaller than 250\,km \citep{DePontieu15}, possibly even 3--30\,km in size \citep{Dere87}. Furthermore, several studies found different widths of allowed and intercombination lines. First such reports started with Skylab observations, followed by reports based on HRTS and SOHO/SUMER observations \citep{Kjeldseth-Moe77,Doschek77,Doschek78,Feldman77,Doschek04}. Different widths suggest that the allowed and intercombination lines could be formed in different conditions, for example, at different electron densities \citep{Doschek84} owing to the different dependence of intensity of allowed and intercombination lines on electron density. This issue has recently been discussed and put forward using \textit{IRIS} observations by \citet{Doschek16}, although the same authors show an example of a quiescent spectrum where the profiles of the two main \ion{Si}{4} and \ion{O}{4} lines are similar. However, \citet{Dere93} pointed out that the intercombination lines can be weak so that it could be difficult to accurately measure their line profiles even with the excellent HRTS instrument. Recently, \citet{Polito16b} reported that \ion{Si}{4} and \ion{O}{4} lines had similar profiles. 

Apart from this, profiles of the TR lines also show departures from a simple Gaussian shape. In particular, strong wings are often observed, even at supersonic velocities \citep{Kjeldseth-Moe77,Doschek78,Dere93,Chae98,Peter99,Peter00,Peter01,Peter06}. This is also the case for $\alpha$ Centauri \citep{Pagano04} and even both $\alpha$ Cen A and B components separately \citep{Ayres15}. These non-Gaussian profiles are usually fitted with two Gaussian components, a narrow and a broader one. The contribution of the broader Gaussian to the total intensity is up to 30\%, with the largest values found in the middle TR \citep{Peter01,Chae98}, i.e., at formation temperatures of \ion{Si}{4} and \ion{O}{4}. These two Gaussian components have been interpreted as the line emission originating in two distinct magnetic structures: The narrow one in closed magnetic loops, while the broader one in open coronal funnels \citep{Peter99,Peter00,Peter01}.

In this paper, we attempt to unify the solution of these three spectroscopic challenges by using the non-Maxwellian $\kappa$-distributions to analyze a particular \textit{IRIS} observation of a bright closed TR loop. The $\kappa$-distributions exhibit significant power-law tails at high velocities or energies. They have been detected in the solar corona \citep{Dudik15} using line intensity ratios. In solar flares, profiles of \ion{Fe}{16} and \ion{Fe}{23} lines observed by \textit{Hinode}/EIS \citep{Culhane07} can be confidently fitted with $\kappa$-distributions \citep{Jeffrey16,Jeffrey17}. The power-law component of the bremsstrahlung emission during flares can also be approximated by $\kappa$-distributions \citep{Oka13,Oka15,Battaglia13,Battaglia15}.

The $\kappa$-distributions are expected in the presence of turbulence if the diffusion coefficient is inversely proportional to velocity \citep{Hasegawa85,Laming07,Bian14}. A turbulence could occur if the plasma is moving, or if it has redshifts \citep{Jeffrey17}, which is an ubiquitous phenomenon in the TR. The presence of $\kappa$-distributions in the TR would lead to enhancements of \ion{Si}{4} intensities with respect to the neighboring \ion{O}{4} ones \citep{Dudik14a}. Power-law distributions or enhanced high-energy particles are also expected in a closed loop, if such a loop is heated by reconnecting current-sheets along its length \citep{Gontikakis13,Gordovskyy13,Gordovskyy14}. Wave-particle interaction involving whistler waves also lead to enhancement of high-energy particles in a closed loop geometry \citep{Vocks08,Vocks16}. The fundamental reason for the existence of power-law tails is that the cross-section for Coulomb collisions and the collision frequency decrease with velocity as $v^{-4}$ and $v^{-3}$, respectively. Particles with progressively higher energies are thus progressively more collisionless. Furthermore, density and temperature gradients in plasma can lead to appearance of the high-energy tails of the distribution \citep{Roussel-Dupre80a,Shoub83,Ljepojevic88}. Non-Maxwellians are expected to occur if the electron mean-free path is larger than about $10^{-2}$ of the local pressure scale-length at any point along a given magnetic field line \citep{Scudder13}. Such conditions should be common in stellar coronae; however, they could occur also in the TR.

Having been thus motivated, in this work we first investigate whether the $\kappa$-distributions could fit the TR line profiles, and how such a fit compares to the classical double Gaussian case. The fit to the TR lines observed by \textit{IRIS} should be feasible, since the $\kappa$-distributions were recently used successfully by \citet{Jeffrey16,Jeffrey17} to fit the coronal lines observed by the \textit{Hinode}/EIS instrument \citep{Culhane07}, even though EIS has fewer points per profile than \textit{IRIS}, as well as a much larger instrumental width, limiting its use for such purposes. Instead, \textit{IRIS} not only has a very small instrumental width, but also samples the line profiles very well.


This paper is organized as follows. The non-Maxwellian $\kappa$-distributions are described in Sect. \ref{Sect:2}. There, we also derive the theoretical $\kappa$-profile of a spectral line, and we discuss the relation of individual fit parameters, such as the characteristic width and $\kappa$ to the observed FWHM of the line and the ion temperature. The \textit{IRIS} observations analyzed are described in Sect. \ref{Sect:3}. Gaussian and $\kappa$-fits are performed in Sects. \ref{Sect:4} and \ref{Sect:5}, while the fitting of the line intensities relative to \ion{Si}{4} is performed in Sect. \ref{Sect:6}. The results are discussed in Sect. \ref{Sect:7} and a summary is given in Sect. \ref{Sect:8}.

%
\begin{figure*}[!ht]
	\centering
	\includegraphics[width=5.8cm]{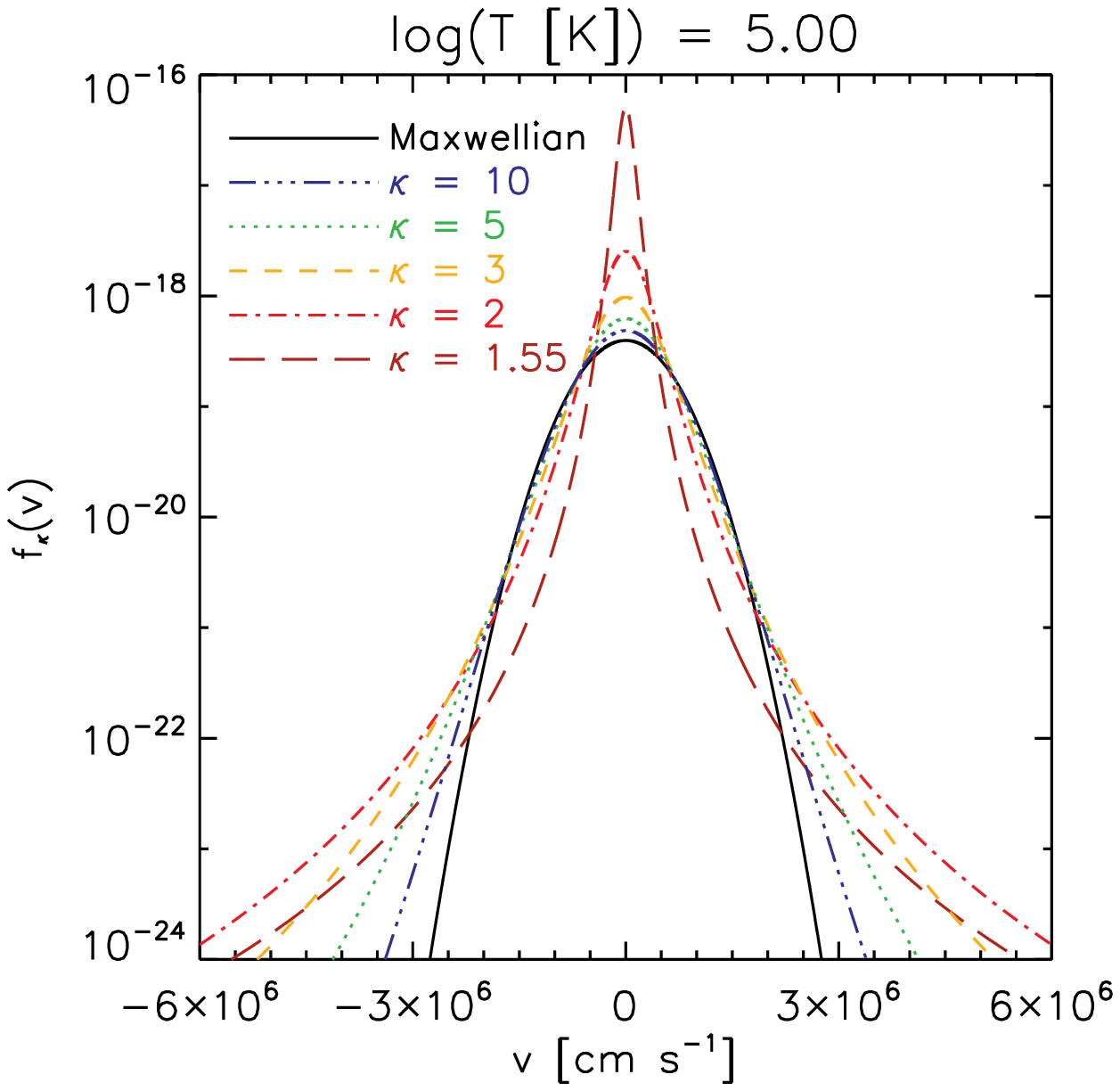}
	\includegraphics[width=5.8cm]{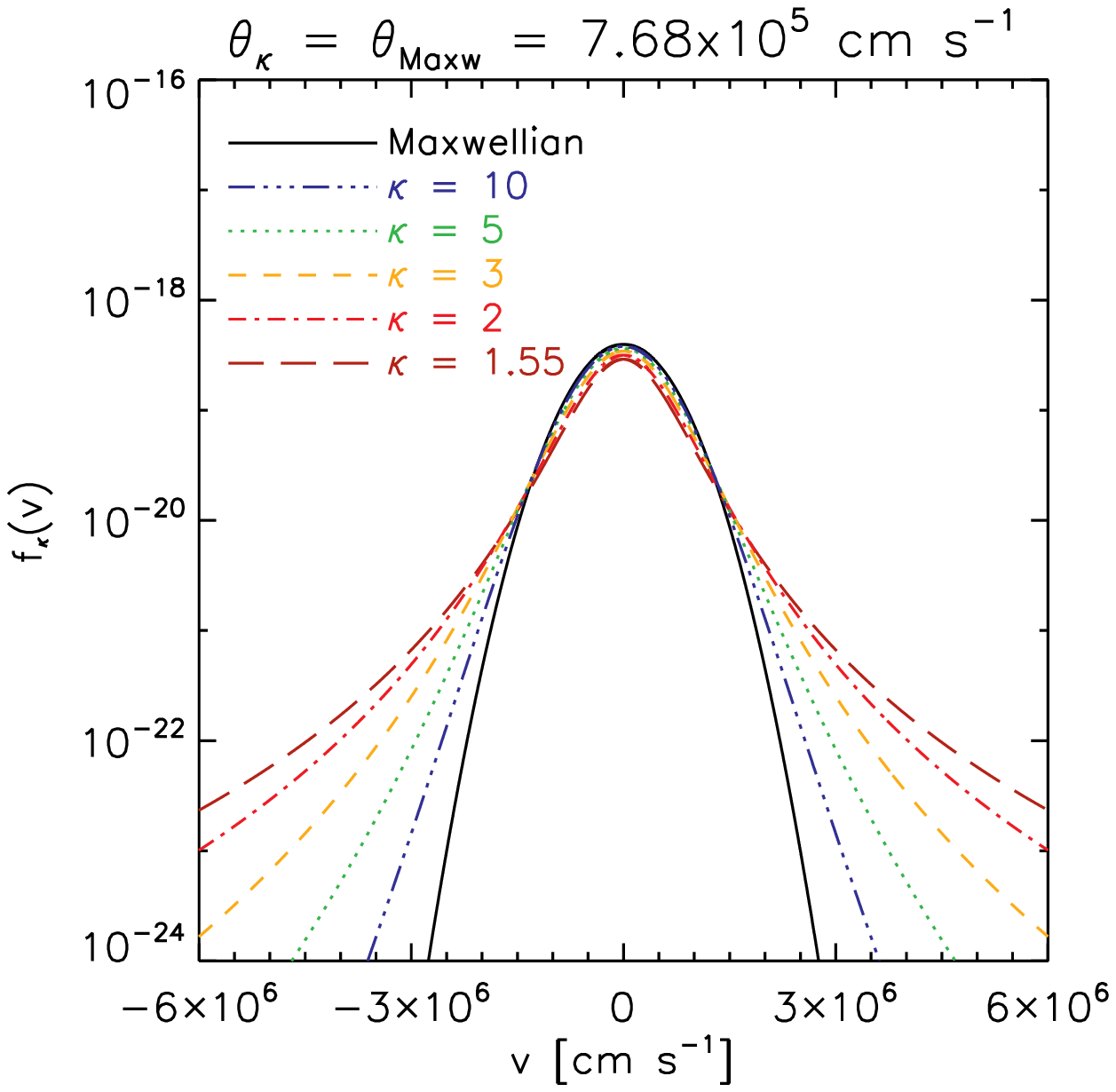}
	\includegraphics[width=5.8cm]{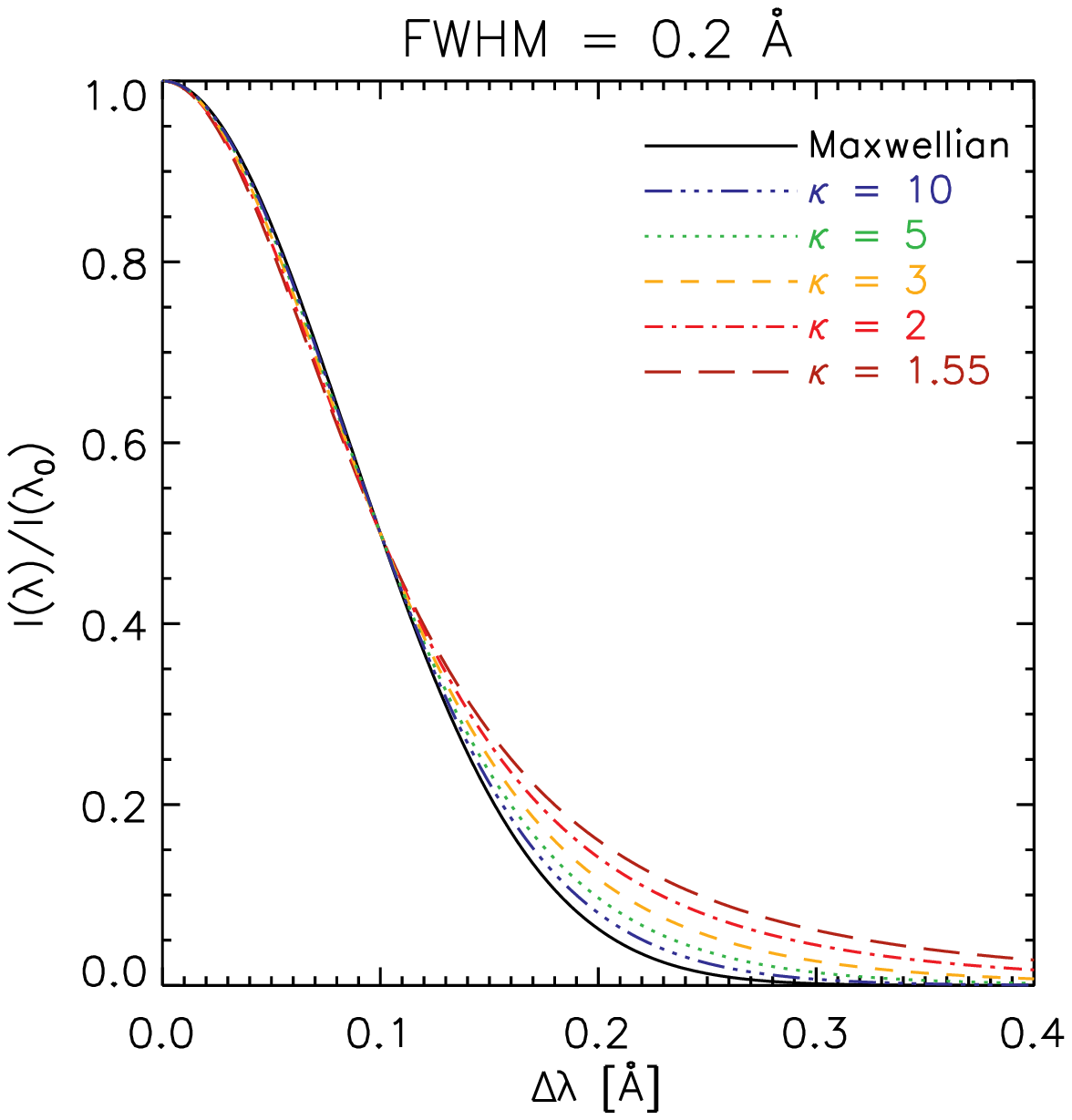}

\caption{The $\kappa$-distribution of particle velocities and the resulting line profile. \textit{Left} and \textit{middle}: The $\kappa$-distributions with varying value of $\kappa$ plotted for constant $T$ and $\theta_\kappa$, respectively. \textit{Right}: Normalized line profile with constant FWHM$_\kappa$ as a function of $\kappa$. Only a half of the profile is shown.
\label{Fig:kappa_v}}
\end{figure*}
%
%

%
\section{The non-Maxwellian $\kappa$-distributions and the line profiles}
\label{Sect:2}

%
\subsection{The $\kappa$-distributions}
\label{Sect:2.1}

The $\kappa$-distribution is a distribution of particle velocities $v$ or energies $E$\,=\,$mv^2/2$. It is characterized by a power-law high-energy tail \citep{Olbert68,Vasyliunas68a,Vasyliunas68b,Owocki83,Livadiotis09,Livadiotis13}. In its energy form, the isotropic $\kappa$-distribution is given by
\begin{equation}
	f_\kappa(E) \mathrm{d}E = A_{\kappa} \frac{2}{\sqrt{\pi} (k_\mathrm{B}T)^{3/2}} \frac{E^{1/2}\mathrm{d}E}{\left(1+ \frac{E}{(\kappa - 3/2) k_\mathrm{B}T} \right)^{\kappa+1}}\,,
	\label{Eq:Kappa_E}
\end{equation}
where the $A_{\kappa}$\,=\,$\Gamma(\kappa+1)$\,/\,$\left[(\kappa-3/2)^{3/2} \Gamma(\kappa-1/2)\right]$ is the normalization constant and $k_\mathrm{B}$\,=\,1.38 $\times 10^{-16}$ erg\,K$^{-1}$ is the Boltzmann constant. This form of the $\kappa$-distribution is assumed for the electrons when calculating the optically thin synthetic non-Maxwellian spectra \citep[e.g.,][]{Dudik14a,Dzifcakova15} where the ionization, recombination, and excitation transitions are dominated by electron-ion collisions.

In the energy form, the $\kappa$-distribution has two independent parameters, $\kappa$ and $T$. Maxwellian is recovered for $\kappa$\,$\to$\,$\infty$, while the opposite situation of $\kappa$\,$\to$\,3/2 corresponds to strongest deviation from the Maxwellian distribution. The $T$ is the thermodynamic temperature related to the mean energy $\left< E \right> = {3k_\mathrm{B}T}/{2}$, which is independent of $\kappa$. More detailed discussion on the representation of $T$ as temperature in the framework of generalized Tsallis statistical mechanics \citep{Tsallis88,Tsallis09} can be found in \citet{Livadiotis09}. 

The corresponding $\kappa$-distribution in the velocity form is given by \citep[e.g.,][]{Livadiotis15}
\begin{equation}
	  f_\kappa(v) \mathrm{d}v = \frac{C_{\kappa}}{(\pi(\kappa-3/2)\theta^2)^{3/2}} \frac{\mathrm{d}v}{\left(1+ \frac{v^2}{(\kappa-3/2) \theta^2}\right)^{\kappa+1}}\,,
	\label{Eq:Kappa_v}
\end{equation}
where $v$\,=\,$|\textbf{v}|$\,=\,$|\textbf{v}_{\parallel} + \textbf{v}_\perp|$, $C_{\kappa}$\,=\,$\Gamma(\kappa+1)$\,/\,$\Gamma(\kappa-1/2)$ is the normalization constant, $\theta$\,=\,$\sqrt{2k_\mathrm{B}T/m}$ is the thermal velocity, and $m$ is the particle mass. In this form, the quantity $T$ is the thermodynamic temperature, and is independent of $\kappa$ \citep{Livadiotis15}. 

We note that the $\kappa$-distributions are sometimes written in the form \citep[see also][]{Olbert68,Lazar16,Jeffrey16}
\begin{equation}
	  f_\kappa(v) \mathrm{d}v = \frac{C_{\kappa}}{(\pi\kappa\theta_\kappa^2)^{3/2}} \frac{\mathrm{d}v}{\left(1+ \frac{v^2}{\kappa \theta_\kappa^2}\right)^{\kappa+1}}\,,
	\label{Eq:Kappa_v_old}
\end{equation}
which is equivalent to Eq. (\ref{Eq:Kappa_v}) if the quantity $\kappa \theta_\kappa$\,=\,$(\kappa-3/2)\theta$, or, equivalently, if $\theta_\kappa$ is related to $T$ by the expression \citep[c.f.,][]{Lazar15}
\begin{equation}
	T = \frac{m}{k_\mathrm{B}} \int v^2 f_\kappa(v) \mathrm{d}^3 \vec{v} = \frac{m}{2k_\mathrm{B}} \frac{2\kappa}{2\kappa-3} \theta_\kappa^2\,.
	\label{Eq:v_th}
\end{equation}
From a mathematical standpoint, either the $T$ or $\theta_\kappa$ can be chosen as independent variable; if $\theta_\kappa$ is chosen as independent, then $T$ depends on both $\kappa$ and $\theta_\kappa$, and vice versa. Examples of the $\kappa$-distributions for different $\kappa$ and a constant $T$ or $\theta_\kappa$ are shown in Fig. \ref{Fig:kappa_v} left and right, respectively.

The definition of the $\kappa$-distributions in Eq. (\ref{Eq:Kappa_v}) corresponds to the $\kappa$-distributions of the second kind \citep{Livadiotis09} rather than of the first kind, which has been used to fit the Hinode/EIS lines by \citet{Jeffrey17}. The $\kappa$-distributions of the second kind differ from those of the first kind by having the factor $-(\kappa+1)$ in the exponent rather than $-\kappa^*$. 
The two kinds of $\kappa$-distributions are equivalent under transformation $\kappa^*$\,=\,$\kappa+1$ if the velocity scales are related as $\kappa^{*1/2} \theta_{\kappa^*}$\,=\,$\kappa^{1/2} \theta_{\kappa}$ \citep{Livadiotis09}. This means that the lower asymptotic limit of $\kappa^*$ for the $\kappa$-distributions of the first kind is 5/2, not 3/2 as in our case.

Finally we note that the relative number of particles in the high-energy tail and the energy carried by them have been calculated by \citet{Oka13}. For example, in a $\kappa$\,=\,4 distribution, the high-energy tail contains $\approx$20\% of particles, and these carry $\approx$50\% of energy. For $\kappa$\,=\,2, about 35\% of particles are in the high-energy tail, which contains more than 80\% of the energy.

%
\subsection{Non-Maxwellian line profiles}
\label{Sect:2.2}

It is well-known that the line profiles can reflect the ion velocity distribution because of the Doppler effect
\begin{equation}
	\Delta\lambda / \lambda_0 = v_\parallel/c\,,
	\label{Eq:Doppler}
\end{equation}
where $\Delta\lambda = \lambda - \lambda_0$ is the wavelength difference from the rest-wavelength $\lambda_0$, $v_\parallel$ is the velocity component parallel to the line of sight, and $c$ is the speed of light. Combining Eqs. (\ref{Eq:Kappa_v})--(\ref{Eq:Doppler}), assuming isotropy and integrating over \textbf{d}$^2\textbf{v}_\perp$\,=\,2$\pi v_\perp \mathrm{d}v_\perp$ \citep{Jeffrey17}, and normalizing, we obtain the line profile as \citep[cf.,][]{Dzifcakova89}
\begin{equation}
	\frac{I_\kappa(\lambda)}{I(\lambda_0)} = \left(1 + \frac{mc^2 (\lambda - \lambda_0)^2}{2k_\mathrm{B}T(\kappa - 3/2)\lambda_0^2} \right)^{-\kappa}\,,
	\label{Eq:Kappa_profile}
\end{equation}
where $I_\kappa(\lambda)$ is intensity at wavelength $\lambda$. The full-width at half-maximum (FWHM) of this profile is given by
\begin{equation}
	\mathrm{FWHM}^2_\kappa =  8 \lambda_0^2 k_\mathrm{B}T (\kappa-3/2) (2^{1/\kappa} -1) / mc^2\,.
	\label{Eq:FWHM_kappa}
\end{equation}
The corresponding Gaussian ($\kappa$\,$\to$\,$+\infty$) profile is
\begin{equation}
	\frac{I_\mathrm{G}(\lambda)}{I(\lambda_0)} = \mathrm{exp}\left(- \frac{mc^2}{2k_\mathrm{B}T} \frac{(\lambda - \lambda_0)^2}{\lambda_0^2} \right)\,,\\
	\label{Eq:Gauss_profile}
\end{equation}
and its $\mathrm{FWHM}^2_\mathrm{G} = 8 \mathrm{ln}(2) \lambda_0^2 k_\mathrm{B}T / mc^2$.

Following the above, if a line with an observed FWHM is fitted with a $\kappa$-distribution, the temperature $T_\kappa$ derived using Eq. (\ref{Eq:FWHM_kappa}) is related to the temperature $T_\mathrm{Maxw}$ obtained from a Gaussian fit of the same line as \citep[cf.,][]{Dzifcakova89}
\begin{equation}
	T_\mathrm{Maxw}/T_\kappa = (\kappa -3/2) (2^{1/\kappa} -1) / \mathrm{ln}(2)\,.
	\label{T_kappa_ratio}
\end{equation}
That is, the $T_\mathrm{Maxw}$ derived from the same observed FWHM is a lower limit to the ion kinetic temperature $T_\kappa$ if the line profile is given by a $\kappa$-distribution.

The normalized line profile of an emission line arising from plasmas characterized by ion $\kappa$-distribution is shown in Fig. \ref{Fig:kappa_v} \textit{right}. In this image, the line profile is assumed to have the same FWHM, which we chose to be 0.2\,\AA~independently of $\kappa$. It can readily be seen that the changes in the line profile with $\kappa$ are modest, less than $\approx$9.8\% $I(\lambda_0)$ for $\kappa$\,=\,1.55 compared to the Gaussian (Maxwellian) profile. The largest changes occur in the line wings at $\Delta \lambda \approx$ FWHM. The changes of the profile within the line core are small, below 3\% of $I(\lambda_0)$. Considering the uncertainties related to the observations and the presence of an instrumental profile, this behavior makes the ion $\kappa$-distributions difficult to be detected except in strong lines with wings well above the continuum, located in an uncrowded region of the spectrum. The transition-region lines observed with the \textit{IRIS} instrument \citep{Dudik14a} offer such opportunity (Sects. \ref{Sect:4.2} and \ref{Sect:5}).

%
\subsection{Formulae for line profile fitting}
\label{Sect:2.3}

In principle, additional broadening mechanisms can increase the FWHM of the line. To account for this, we consider the characteristic width of the line to be a free parameter. Consequently, we fit the observed spectrum (Sect. \ref{Sect:3} with Gaussian and $\kappa$-profiles using the formulae
\begin{equation}
	I_\mathrm{G}(\lambda) = I_0 \mathrm{exp}\left(-\frac{(\lambda-\lambda_0)^2}{2 w_\mathrm{G}^2} \right)\,
	\label{Eq:gauss_fit}
 \end{equation}
and
\begin{equation}
	I_\kappa(\lambda) = I_0 \left(1 + \frac{(\lambda-\lambda_0)^2}{2(\kappa-3/2) w_\kappa^2} \right)^{-\kappa}\,,
	\label{Eq:kappa_fit}
\end{equation}
where $I_0 = I(\lambda_0)$ is the peak intensity and $w_\kappa$ is the characteristic width. We note that the above formula is different from the one used by \citet{Jeffrey16}. 
A factor of $\kappa-3/2$ is present in the denominator instead of just $\kappa$. This factor is kept for consistency with the theoretical line profile (Eq. \ref{Eq:Kappa_profile}) arising only from the distribution of ion velocities (Sect. \ref{Sect:2.2}). Finally, we keep the factor 2 in the denominator for consistency with the Gaussian profile as implemented in the SolarSoft routine \textit{comp\_gauss.pro}. This factor can be dropped by a unique transformation of $2w_\kappa^2$\,$\to$\,$z_\kappa^{2}$, so it essentially only modifies the resulting characteristic width.

We note that the total intensity $I_\mathrm{tot}$ of the Gaussian profile is 
\begin{equation}
	I_\mathrm{tot} 	= \int\limits_{-\infty}^{+\infty} I_\mathrm{G}(\lambda) \mathrm{d}\lambda
			= w_\mathrm{G} I_0 (2\pi)^{1/2}\,,
	\label{Eq:total_intensity_Gauss}
\end{equation}
while for the $\kappa$-profile we obtain
\begin{equation}
	I_\mathrm{tot} 	
			= w_\kappa I_0  (2\pi)^{1/2} (\kappa-3/2)^{1/2} \frac{\Gamma(\kappa +1/2)}{\Gamma(\kappa +1)}\,.
	\label{Eq:total_intensity_kappa}
\end{equation}

If the line width is given only by the isotropic ion motion due to a $\kappa$-distribution of ion velocities, $w_\kappa$ is given by (Eqs. \ref{Eq:Kappa_v}--\ref{Eq:FWHM_kappa})
\begin{equation}
	w_\kappa^2 = \lambda_0^2 \frac{k_\mathrm{B}T}{mc^2} = \frac{1}{2} \frac{\lambda_0^2}{c^2} \theta^2 = \frac{1}{8} \frac{\mathrm{FWHM}_\kappa^2}{(\kappa-3/2) (2^{1/\kappa} -1)} \,.
	\label{Eq:w_kappa}
\end{equation}
The last equation between $w_\kappa$ and FWHM$_\kappa$ is also valid in a general case, i.e., also for the FWHM derived from observations. We further note that this equation means that the relation between characteristic width $w_\kappa$ of the line profile and the corresponding ion temperature $T$ does not depend on $\kappa$. In the fitting however, $w_\kappa$ is a free parameter alongside $I_0$, $\lambda_0$, and $\kappa$. I.e., in principle, it is possible to obtain a different value of $w_\kappa$ depending on whether the line is fitted with a Gaussian or a $\kappa$-profile.

%
\begin{figure}[!t]
	\centering
	\includegraphics[width=0.5\textwidth]{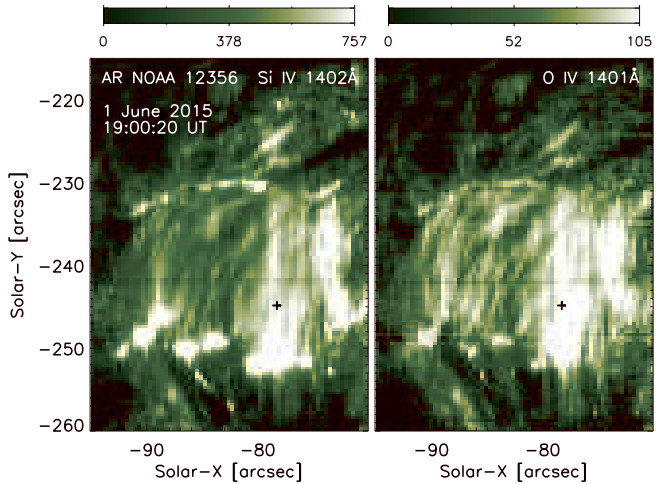}	
\caption{IRIS spectrograph images formed in the \ion{Si}{4} 1402.77\,\AA~(left) and \ion{O}{4} 1401.16\,\AA~(right) spectral lines, showing the active region under study. The intensity units are normalized for the exposure time and are expressed in DN s$^{-1}$. The cross symbols overlaid on the \ion{Si}{4} and \ion{O}{4} images indicate the position of the pixel where we take the TR spectrum analyzed in Sect. \ref{Sect:4}. The time halfway through the \textit{IRIS} raster is indicated on the left panel.
\label{Fig:IRIS_sg}}
\end{figure}
%
%
%
\section{\textit{IRIS} observations}
\label{Sect:3}

Since its launch in 2013, the \textit{IRIS} instrument has provided high spatial (0.33$\arcsec$--0.4$\arcsec$) and temporal ($\approx$\,2\,s) resolution images and spectra of the Sun in the far ultra-violet (FUV) at 1332--1407\,\AA~and near-UV 2783--2835\,\AA~spectral ranges, allowing to investigate the highly dynamical nature of the low solar atmosphere. In this work, we analyze the spectra of the strongest TR lines at around 1400\,\AA~observed by \textit{IRIS} in the active region (AR) NOAA 12356 on 2015 June 1. The observed transitions include the \ion{Si}{4} at 1402.77\,\AA~(log($T_\mathrm{max}$\,[K])\,$\approx$\,4.9), \ion{O}{4} at 1399.77\,\AA, 1401.16\,\AA, and 1404.82\,\AA~(log($T_\mathrm{max}$\,[K]~$\approx$~5.15) and the \ion{S}{4} at 1404.85\,\AA~and 1406.01\,\AA~(log($T_\mathrm{max}$\,[K])~$\approx$~5.0), where $T_\mathrm{max}$ refers to Maxwellian peak formation temperatures of a given ion. It should be noted that the \ion{O}{4} and \ion{S}{4} lines around 1404.8\,\AA~are blended together. A review of the experimental data for these ions indicated that the rest wavelengths should be separated by about 0.04\,\AA~\citep[][Appendix A therein]{Polito16b}. The spectra of these lines for the same \textit{IRIS} observation was analyzed by \cite{Polito16b}, who investigated the use of \ion{O}{4} and \ion{S}{4} lines as density and temperature diagnostics of the plasma from which they are emitted. We also note that the \ion{O}{4} 1397.198\,\AA~and \ion{S}{4} 1398.040\,\AA~lines are not included within the spectral range of the present \textit{IRIS} observation.

The \textit{IRIS} study presented in this work consists of a dense, 96-step single raster which scanned the AR under study over a field of view of 33$\arcsec$ $\times$ 119$\arcsec$ and with an exposure time of $\approx$\,60\,s. The present observation included binning by 2 pixels in both the Solar $Y$ direction as well as in wavelength $\lambda$. The level 2 \textit{IRIS} data were downloaded from the \textit{IRIS} website\footnote{http://iris.lmsal.com/search/} and are obtained from level 0 data after flat-field correction, geometry calibration and dark current subtraction\footnote{http://iris.lmsal.com/documents.html}. In addition, we performed a cosmic ray removal and calibration of the wavelength array as described in \cite{Polito15} and \cite{Polito16a}. The level 2 data are expressed in data number (DN) and, in order to convert them to physical units, one can perform the radiometric calibration detailed in the \textit{IRIS} software note 24. However, the \textit{IRIS} spectrograph has flat response curves in the long-FUV (FUVL) window where the observed lines are located, and thus the calibration factors would not change the line profiles nor affect their relative intensities. Therefore, the spectra of the TR lines analyzed in this work are expressed in data number (DN), as they are measured by the spectrograph. The error associated with the data counts in each detector pixel is obtained by summing in quadrature the photon counting error (which is given by the square root of the photon counts) and the readout noise. We assume a gain (photons DN$^{-1}$) of 4 for the FUV channel and a readout noise of $\approx$ 3.1 DN \citep[see][]{DePontieu14}. 

Figure \ref{Fig:IRIS_sg} shows the intensity images in solar coordinates $X$\,=\,[$-$95.1,$-$70.1]
and $Y$\,=\,[$-$260,$-$215] for the two strongest lines, \ion{Si}{4} 1402.77\,\AA~and \ion{O}{4} 1401.16\,\AA~observed during the dense \textit{IRIS} raster analyzed in this study. The intensity of the \ion{Si}{4} and \ion{O}{4} at each pixel is obtained by summing the total counts over the line profile and normalizing it with respect to the exposure time. Therefore, the units of the images are expressed in DN\,s$^{-1}$. The \ion{Si}{4} and \ion{O}{4} images show the presence of different features, including bundles of cool AR loops and compact brightenings, analyzed and described in detail in \cite{Polito16b}. 

%
%
%
\begin{table*}[!t]
\caption{Statistical moments of the five \textit{IRIS} TR lines, calculated between $\lambda_0$\,$\pm$\,0.4\,\AA~($\lambda_0$\,$\pm$\,0.6\,\AA~for \ion{Si}{4}) for the spatial pixel marked in Fig. \ref{Fig:IRIS_sg}. The lines are labeled using the nominal rest-wavelength \citep[see][]{Polito16b}, indicated for comparison. The Gaussian FWHM$_\mathrm{G}$\,=\,$w\sqrt{8\mathrm{ln}2}$ is indicated rather than $w$ only. The larger value of this FWHM$_\mathrm{G}$ for \ion{Si}{4} is an artifact of the larger wavelength range. See text for details.}
\label{Table:1}
  $$ 
  \begin{tabular}{lcccc}
	\hline
	\hline
	\noalign{\smallskip}
	Line 						& $\lambda_0$ [\AA~] 	&$w\sqrt{8\mathrm{ln}2}$	& $S$			& $K$ \\
	\noalign{\smallskip}
	\hline
	\noalign{\smallskip}
	\ion{O}{4}  1399.78 \AA~			&  1399.83 $\pm$ 0.03	& 0.266	$\pm$ 0.000		&  +0.00 $\pm$ 0.08	& 3.5 $\pm$ 0.2		\\
	\ion{O}{4}  1401.16 \AA~			&  1401.22 $\pm$ 0.03	& 0.256	$\pm$ 0.000		&  +0.04 $\pm$ 0.04	& 3.8 $\pm$ 0.1		\\
	\ion{Si}{4} 1402.77 \AA~			&  1402.82 $\pm$ 0.03	& 0.287	$\pm$ 0.000 \footnote{artefact of the $\lambda_0$\,$\pm$\,0.6\,\AA}
														& --0.07 $\pm$ 0.04	& 5.8 $\pm$ 0.2		\\
	\ion{O}{4}  1404.82 \AA~(bl \ion{S}{4})	&  1404.85 $\pm$ 0.03	& 0.265	$\pm$ 0.000		& --0.08 $\pm$ 0.11 	& 4.0 $\pm$ 0.3		\\
	\ion{S}{4}  1406.06 \AA~			&  1406.10 $\pm$ 0.03	& 0.256 $\pm$ 0.001		&  +0.02 $\pm$ 0.18	& 4.3 $\pm$ 0.6		\\
	\noalign{\smallskip}
	\noalign{\smallskip}
	\hline
  \end{tabular}
  $$ 
\end{table*}
%
%
%
%
%
\begin{table*}[!t]
\caption{Fit parameters with their respective uncertainties obtained for the fit with a $\kappa$-profile (Fig. \ref{Fig:Fits}, \textit{third column}). Derived FWHM$_\kappa$ and $T$ quantities are listed as well.}
\label{Table:2}
  $$ 
  \centering
  \begin{tabular}{lcccccc}
	\hline
	\hline
	\noalign{\smallskip}
	Line 						& $\lambda_0$ [\AA]	& $I_0$ [DN]	& $w_\kappa$ [\AA]	& $\kappa$		& FWHM$_\kappa$ [\AA]	& $T_\mathrm{i}$ [MK] \\
	\noalign{\smallskip}
	\hline
	\noalign{\smallskip}
	\ion{O}{4}  1399.78 \AA~			& 1399.831 $\pm$ 0.001	&  385 $\pm$ 6	&  0.143 $\pm$ 0.010	& 2.16 $\pm$ 0.17	& 0.20 $\pm$ 0.08	& 1.81 $\pm$ 0.26	\\
	\ion{O}{4}  1401.16 \AA~			& 1401.218 $\pm$ 0.000	& 1399 $\pm$ 10	&  0.127 $\pm$ 0.003	& 2.35 $\pm$ 0.08	& 0.20 $\pm$ 0.05	& 1.43 $\pm$ 0.06	\\
	\ion{Si}{4} 1402.77 \AA~			& 1402.820 $\pm$ 0.000	& 4598 $\pm$ 18	&  0.136 $\pm$ 0.001	& 2.16 $\pm$ 0.03	& 0.19 $\pm$ 0.02	& 2.86 $\pm$ 0.06	\\
	\ion{O}{4}  1404.82 \AA~(bl \ion{S}{4})	& 1404.855 $\pm$ 0.001	&  383 $\pm$  6	&  0.163 $\pm$ 0.018	& 1.90 $\pm$ 0.13	& 0.19 $\pm$ 0.13	& 2.35 $\pm$ 0.52	\\
	\ion{S}{4}  1406.06 \AA~			& 1406.103 $\pm$ 0.001	&  282 $\pm$  5	&  0.144 $\pm$ 0.021	& 1.91 $\pm$ 0.18	& 0.17 $\pm$ 0.17	& 3.64 $\pm$ 1.08	\\
	\noalign{\smallskip}
	\noalign{\smallskip}
	\hline
  \end{tabular}
  $$ 
\end{table*}
 
%
%
\section{Line profile analysis in a selected pixel}
\label{Sect:4}

Since the non-Maxwellian analysis is quite involved, we first analyze the spectrum in a single selected pixel along one of the bundle of bright TR loops. This is instructive, as it enables us to show and discuss the differences among fitting all five lines with a single and double Gaussians, and a $\kappa$-distribution. The chosen example pixel is indicated by a cross symbol overlaid on the images in Fig. \ref{Fig:IRIS_sg} and its coordinates within the raster are $[49,101]$ in pixel units. The line fitting of the five TR lines observed by \textit{IRIS} is presented here, while an analysis of line profiles elsewhere together with the spatial distribution of the derived $\kappa$-values is postponed to Sect. \ref{Sect:5}. We note that analysis of a single-pixel spectrum, as opposed to averaging over a spatial box, allows us to take advantage of the high-resolution of \textit{IRIS} and thus to avoid as much as possible the contributions from different plasmas that could be present in a larger spatial region.


%
\subsection{Line profiles: Symmetry and tails}
\label{Sect:4.1}

Following \citet{Jeffrey16}, we first performed the analysis of line profile symmetry and peakedness by calculating the first four moments of the line profile. Aside the peak wavelength $\lambda_0$, the variance $w^2$, skewness $S$, and kurtosis $K$ are defined by \citep[cf.,][]{Jeffrey16}
\begin{eqnarray}
	w^2 &=& 		\frac	{\int_\lambda I(\lambda)(\lambda-\lambda_0)^2 \mathrm{d}\lambda}
					{\int_\lambda I(\lambda) \mathrm{d}\lambda}\,,\\
	\label{Eq:variance}
	S &=& \frac{1}{w^3}	\frac	{\int_\lambda I(\lambda)(\lambda-\lambda_0)^3 \mathrm{d}\lambda}
					{\int_\lambda I(\lambda) \mathrm{d}\lambda}\,,\\
	\label{Eq:skewness}
	K &=& \frac{1}{w^4}	\frac	{\int_\lambda I(\lambda)(\lambda-\lambda_0)^4 \mathrm{d}\lambda}
					{\int_\lambda I(\lambda) \mathrm{d}\lambda}\,,
	\label{Eq:kurtosis}
\end{eqnarray}
where $S$ and $K$ are distribution-normalized. Because of the factor $\lambda-\lambda_0$ raised to the 3rd or 4th power, the values of $S$ and $K$ are dominated by points located further away from $\lambda_0$\,$\pm$\,$w$. Hence, $S$ and $K$ represent measures of the symmetry of the line profile and presence of strong tails, respectively. We note that a Gaussian distribution has $S$\,=\,0 and $K$\,=\,3.

To calculate these values, we subtracted the pseudo-continuum (15.3\,DN; see Sect. \ref{Sect:4.2}) and used a wavelength range of $\lambda_0 \pm$\,0.4\,\AA~for \ion{O}{4} and \ion{S}{4} lines, while for \ion{Si}{4}, we used the wavelength range of $\lambda_0 \pm$0.6\AA. These wavelength ranges are larger than those used by \citet{Jeffrey16} for EIS lines ($\lambda_0 \pm$\,0.2\,\AA). We require a larger interval because of the pronounced wings of the observed TR lines. The $\lambda_0 \pm$\,0.4 (0.6)\,\AA~interval is wide enough to contain the line profile and several pseudo-continuum points at both extremities while avoiding the weak line-like features such as \ion{Fe}{2} occasionally present above the pseudo-continuum.

The values of $w^2$, $S$, and $K$ together with the first moment $\lambda_0$ and their respective uncertainties for the five \textit{IRIS} TR lines are given in Table \ref{Table:1}. This Table indicates that all five lines are red-shifted by a similar amount (43--55\,m\AA) with respect to the rest wavelengths \citep[see discussion in Appendix A of][]{Polito16b}. This small dispersion in the observed redshift is not surprising given that the \textit{IRIS} pixel in the present observation is 25\,m\AA. Table \ref{Table:1} also indicates that all five lines have nearly the same width; furthermore, they are all close to being symmetric, and show the presence of strong tails. 

\begin{figure*}
	\centering
	\includegraphics[width=4.98cm,clip,bb= 0 52 487 340]{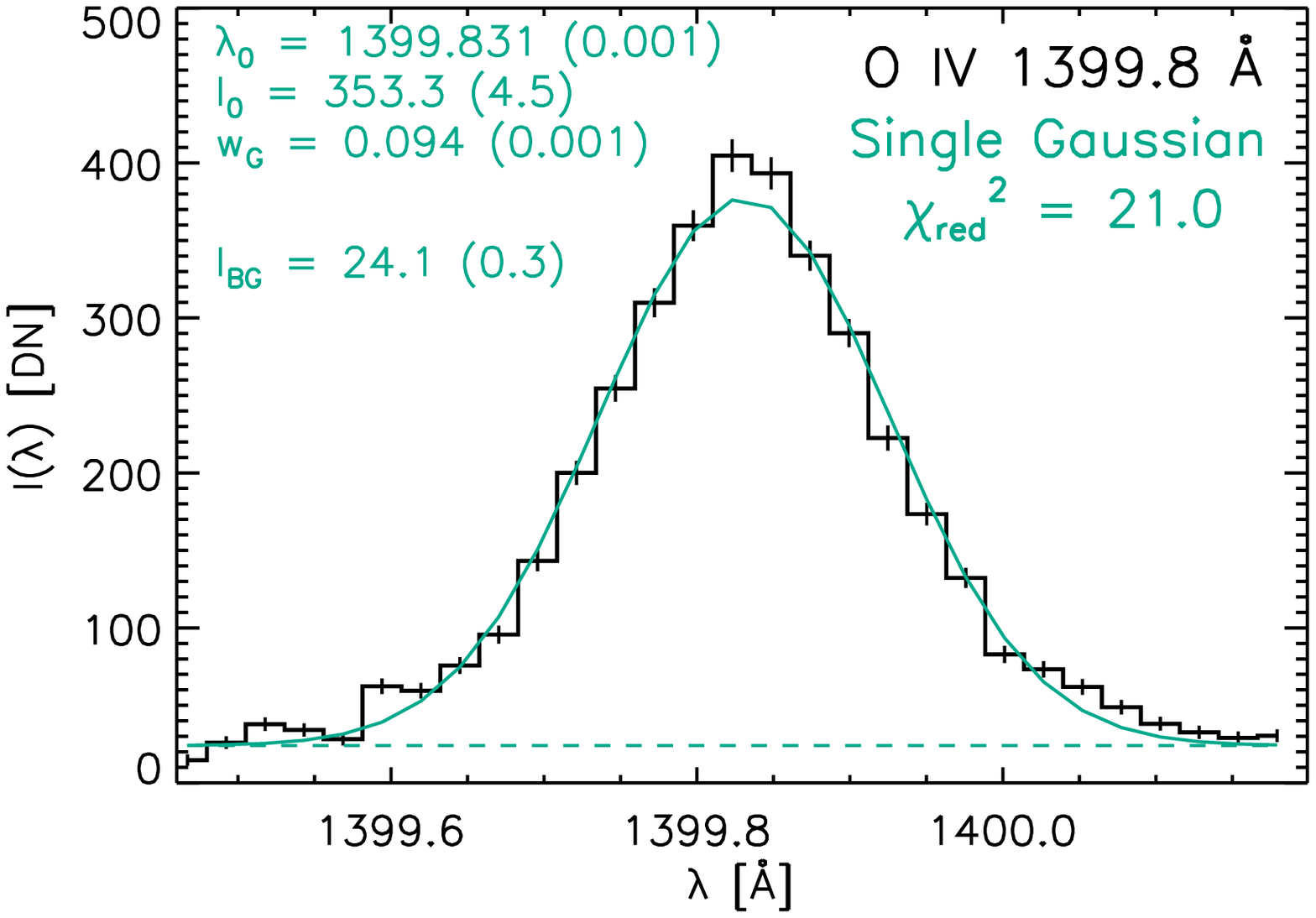}
	\includegraphics[width=4.21cm,clip,bb=75 52 487 340]{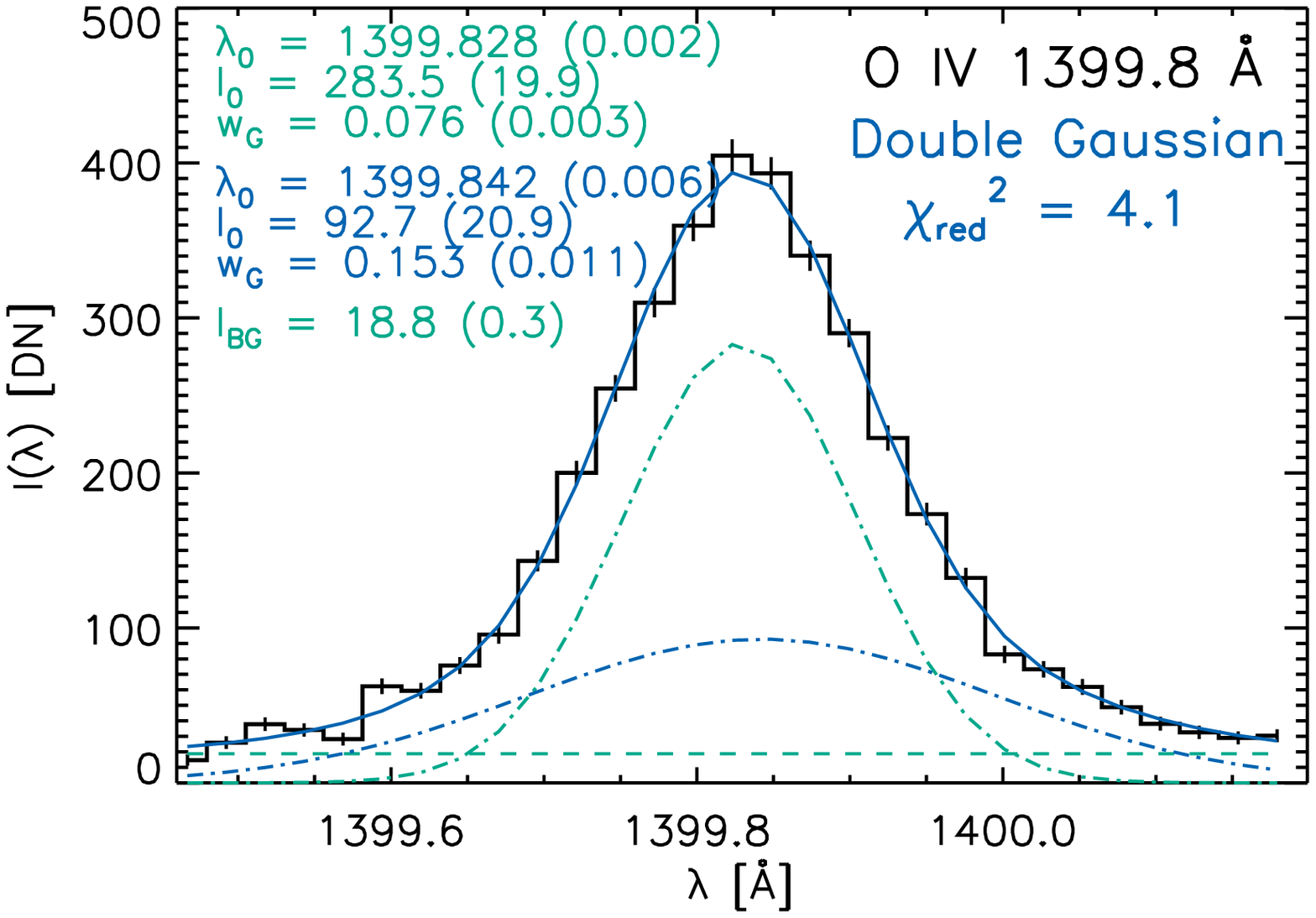}
	\includegraphics[width=4.21cm,clip,bb=75 52 487 340]{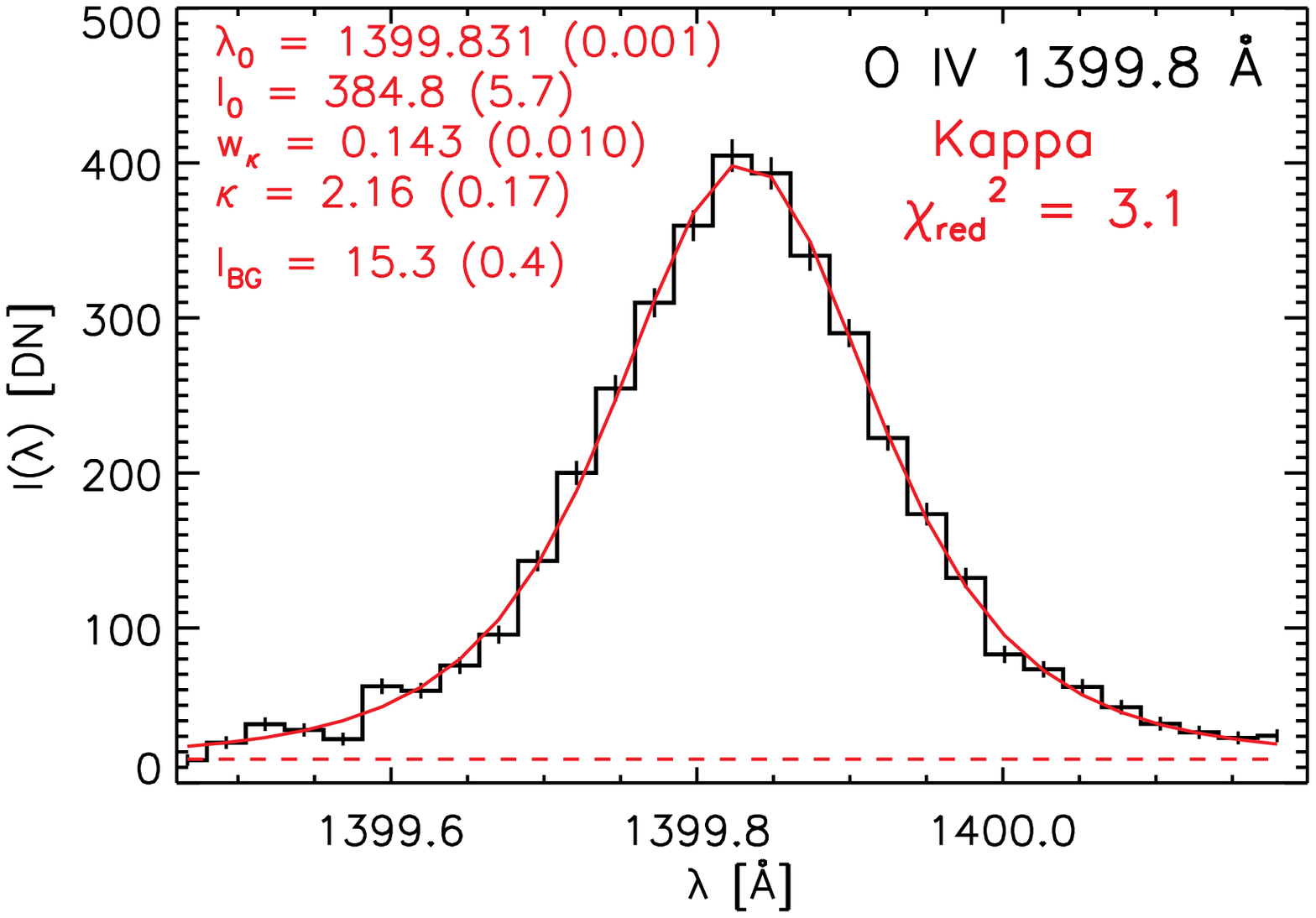}
	\includegraphics[width=4.21cm,clip,bb=75 52 487 340]{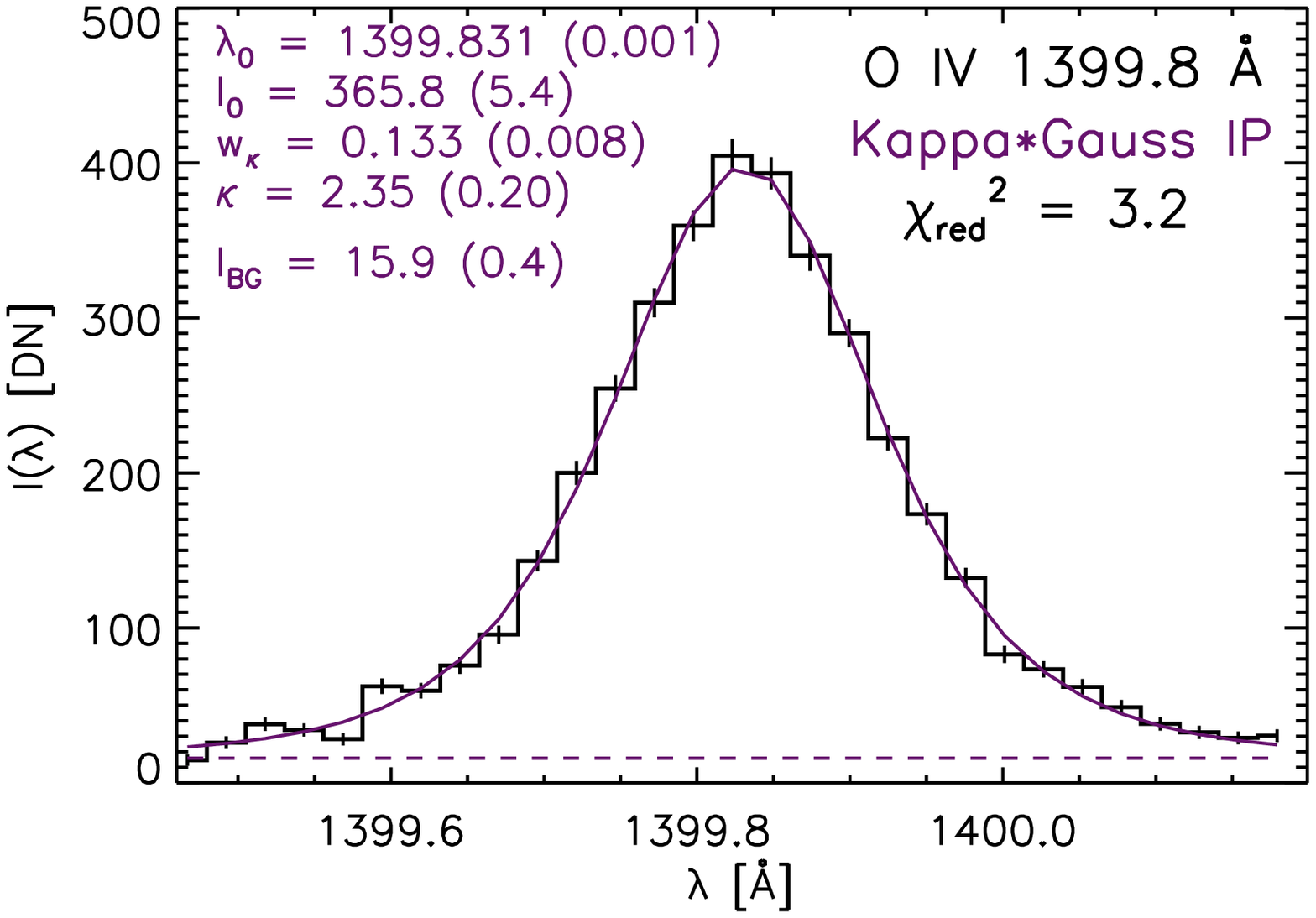}
	\includegraphics[width=4.98cm,clip,bb= 0  0 487 145]{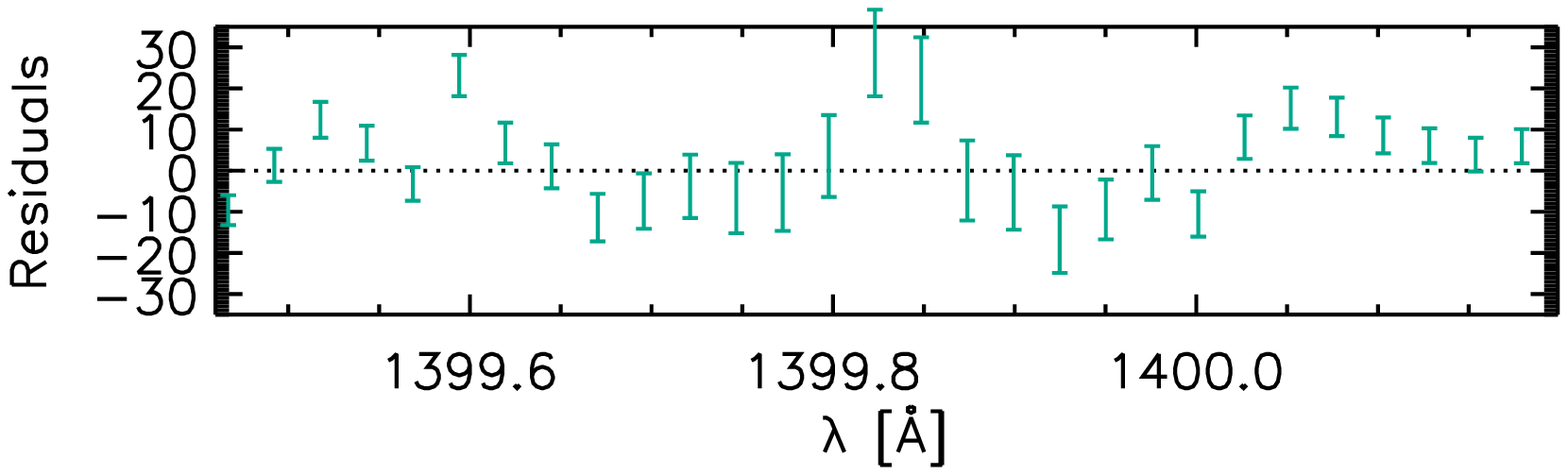}
	\includegraphics[width=4.21cm,clip,bb=75  0 487 145]{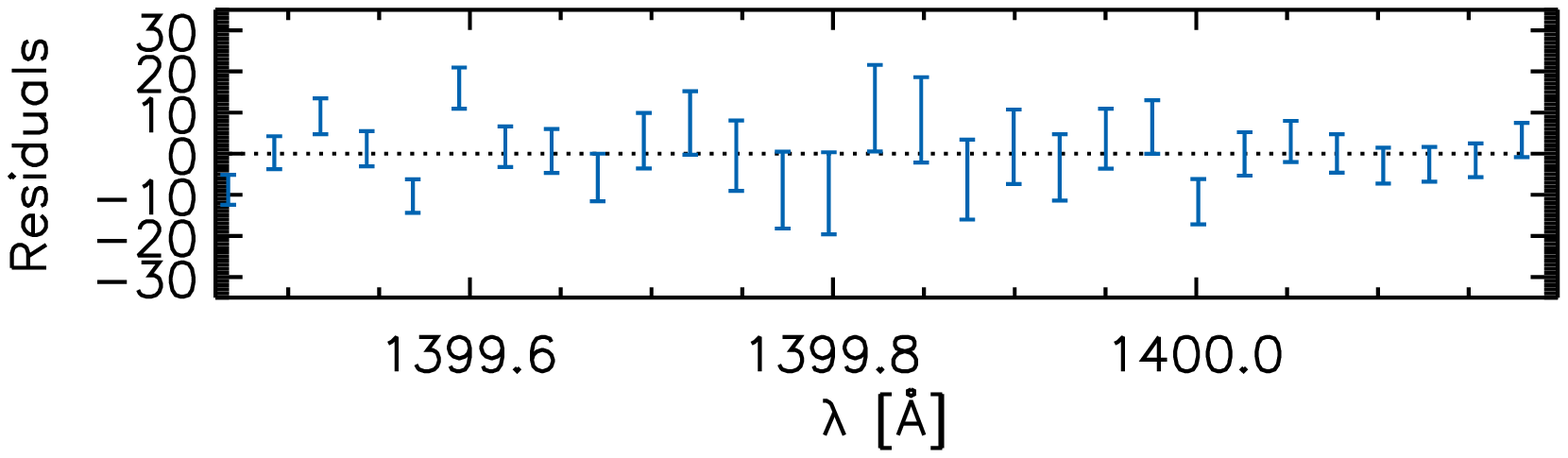}
	\includegraphics[width=4.21cm,clip,bb=75  0 487 145]{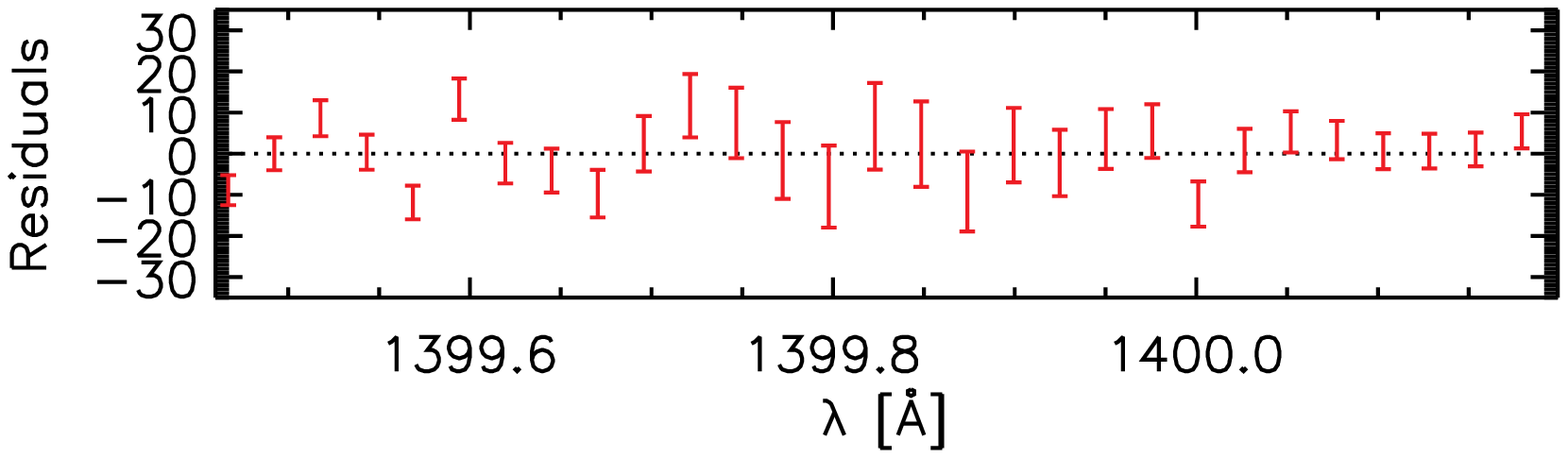}
	\includegraphics[width=4.21cm,clip,bb=75  0 487 145]{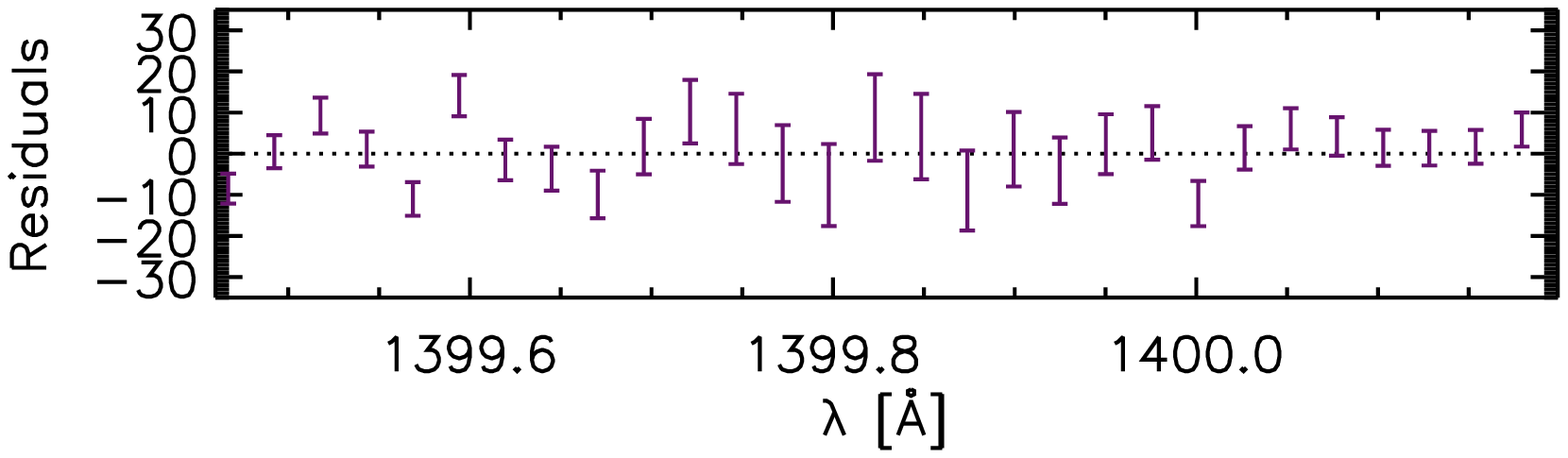}
	\includegraphics[width=4.98cm,clip,bb= 0 52 487 340]{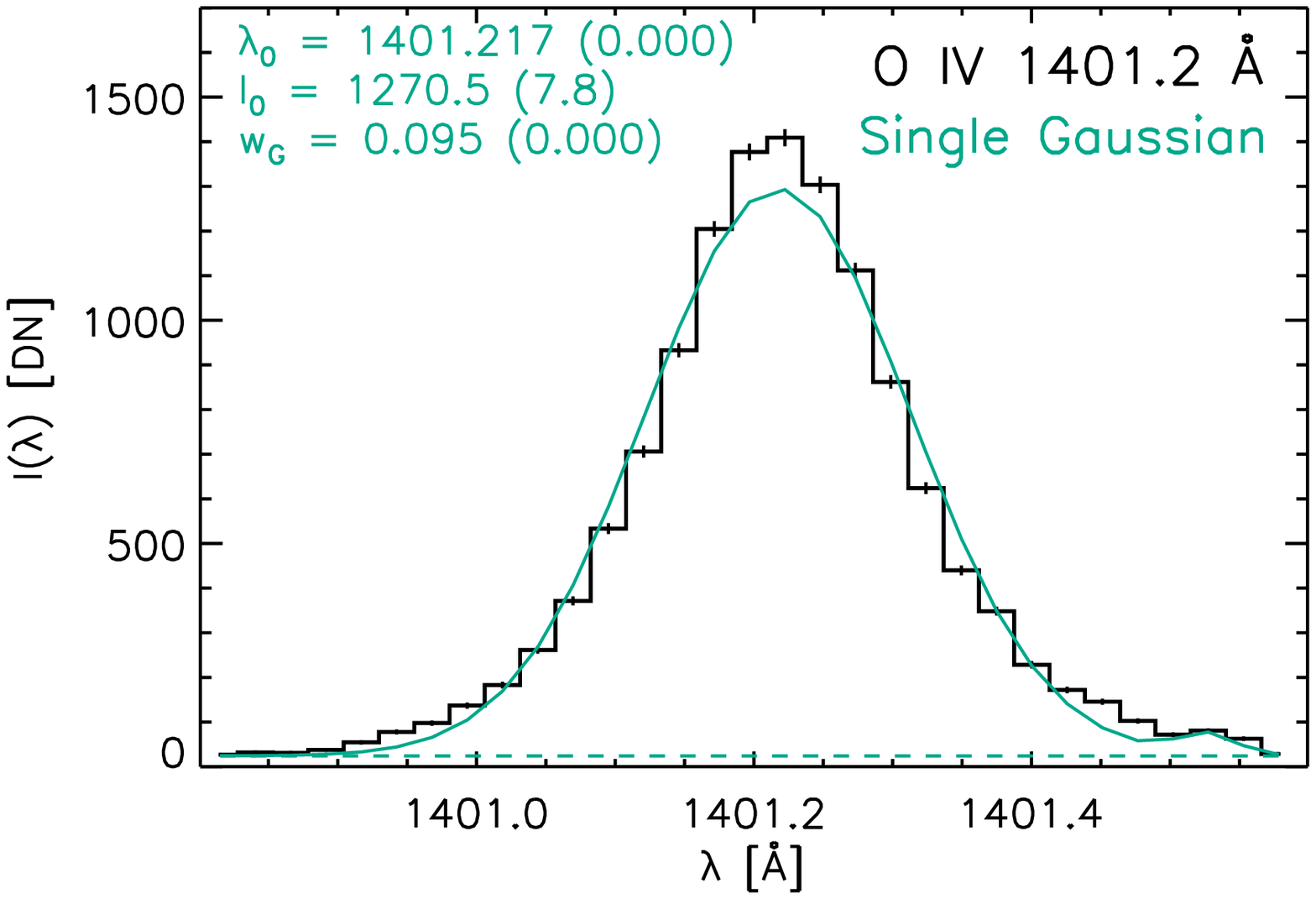}
	\includegraphics[width=4.21cm,clip,bb=75 52 487 340]{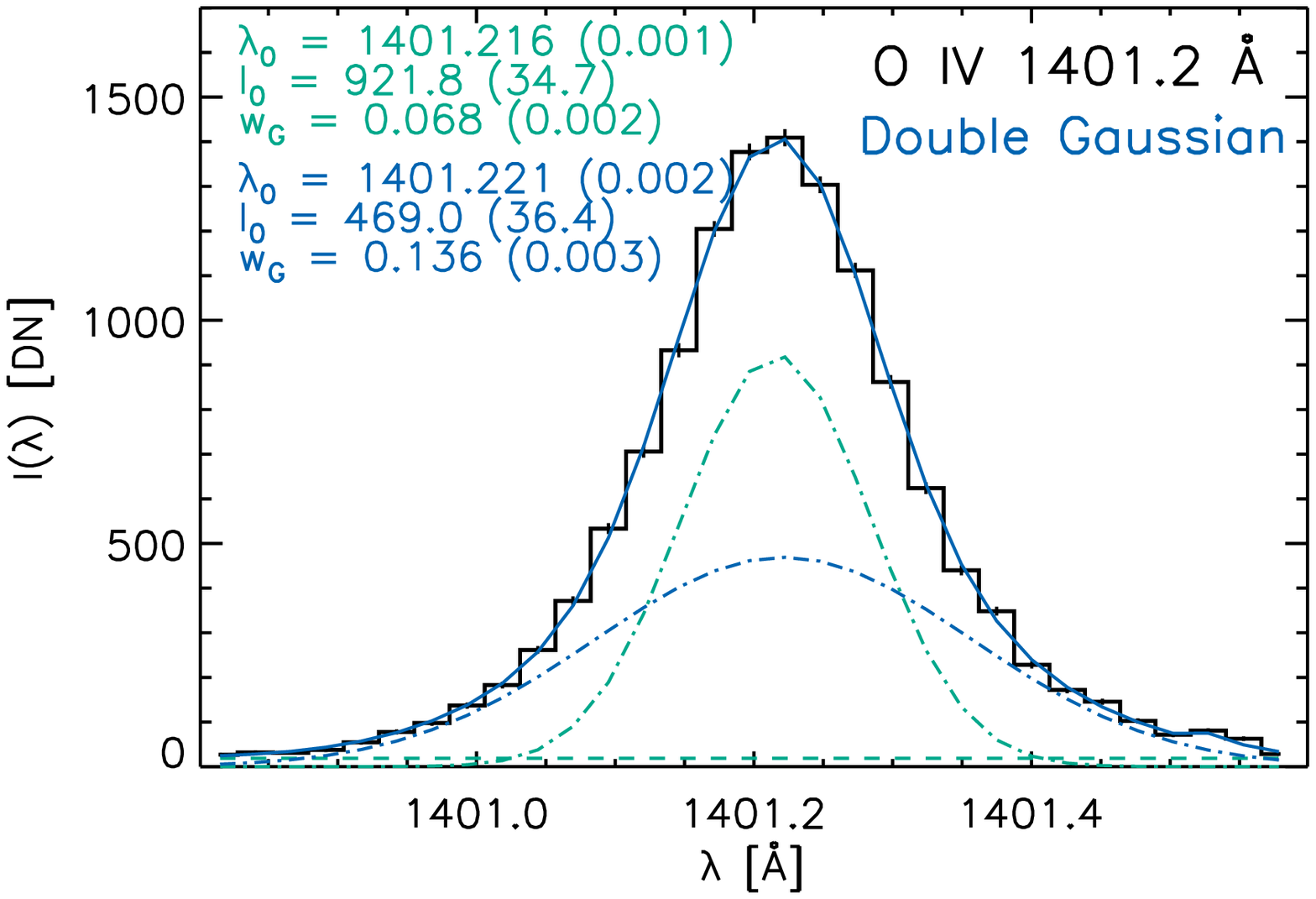}
	\includegraphics[width=4.21cm,clip,bb=75 52 487 340]{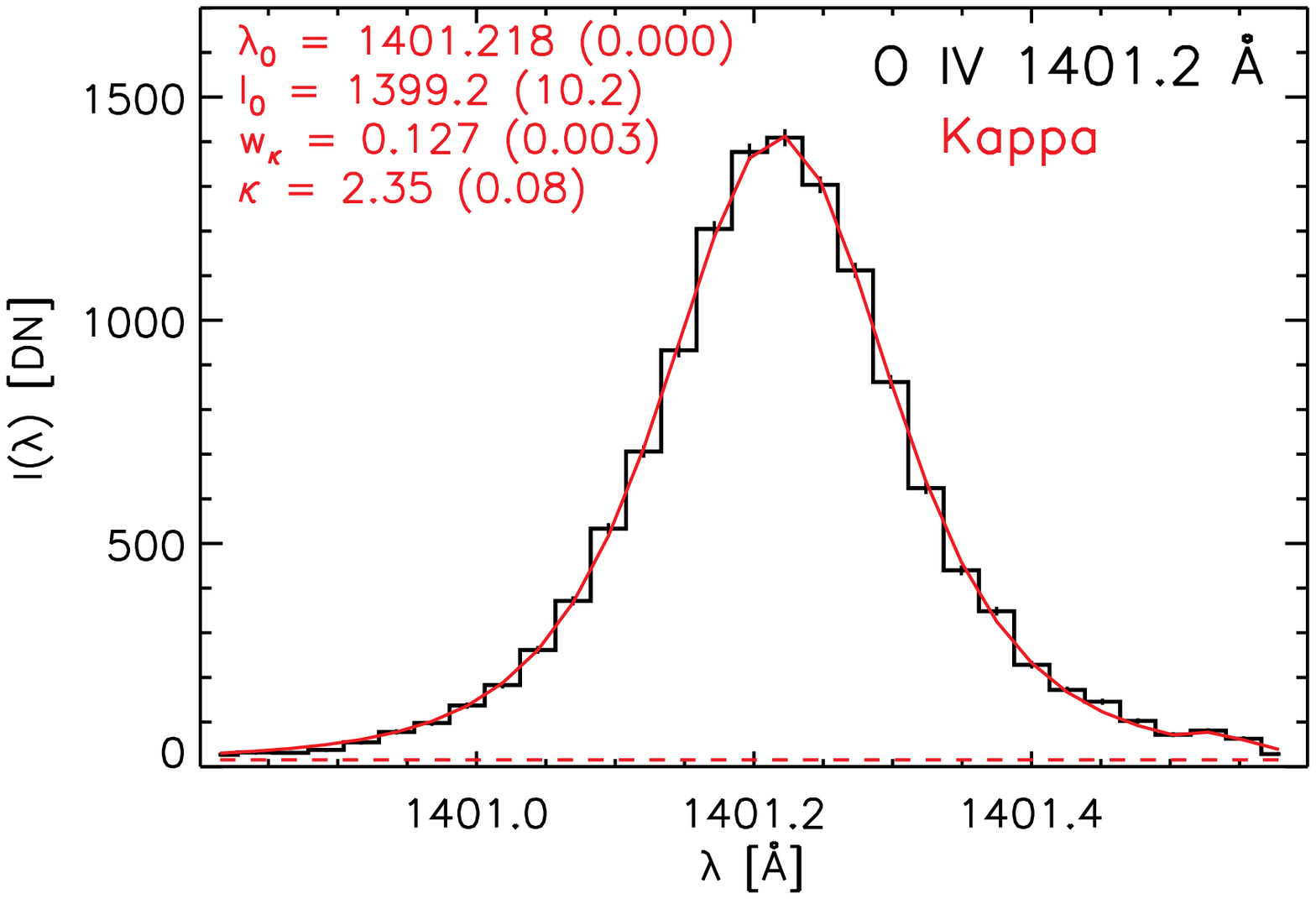}
	\includegraphics[width=4.21cm,clip,bb=75 52 487 340]{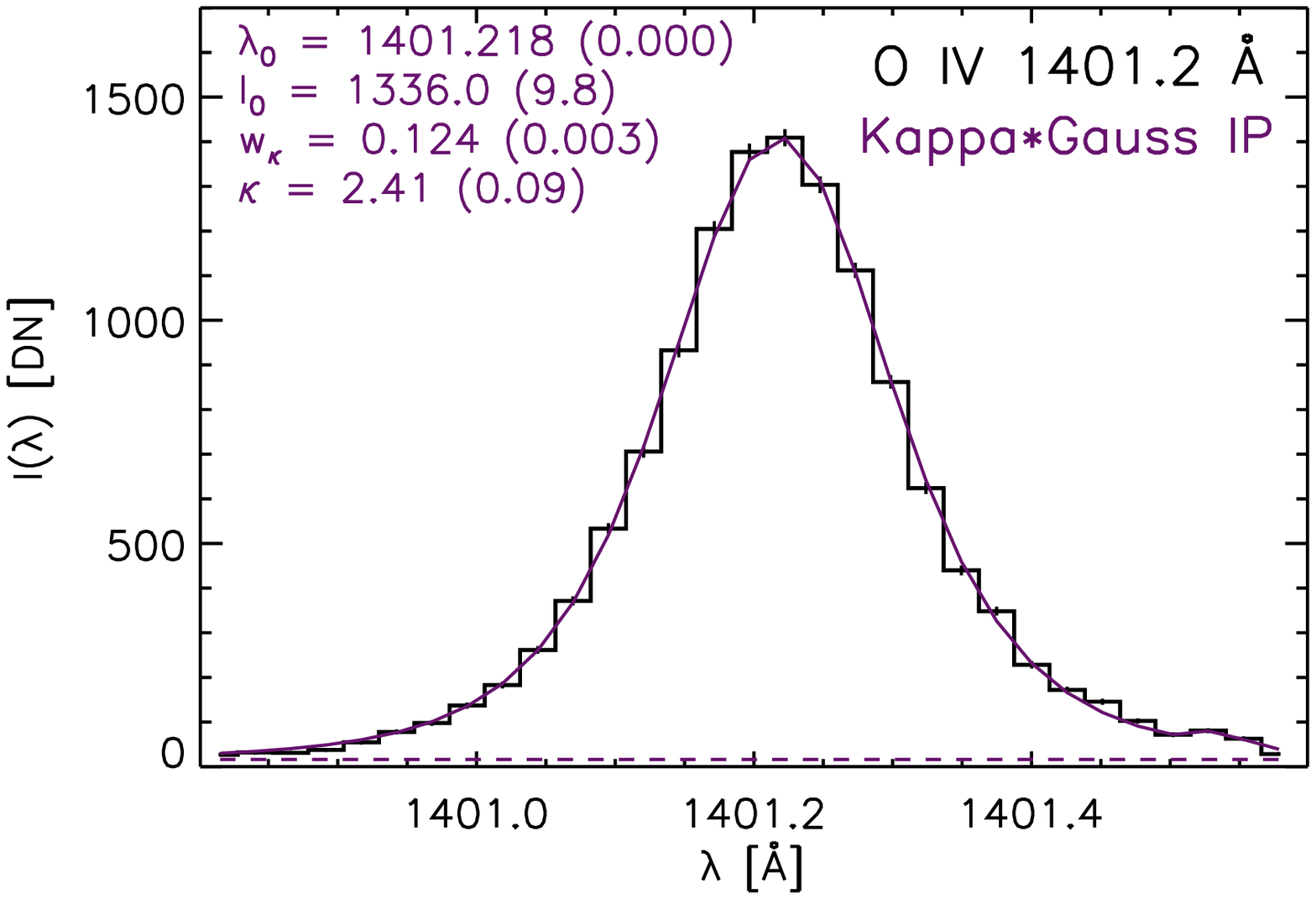}
	\includegraphics[width=4.98cm,clip,bb= 0  0 487 145]{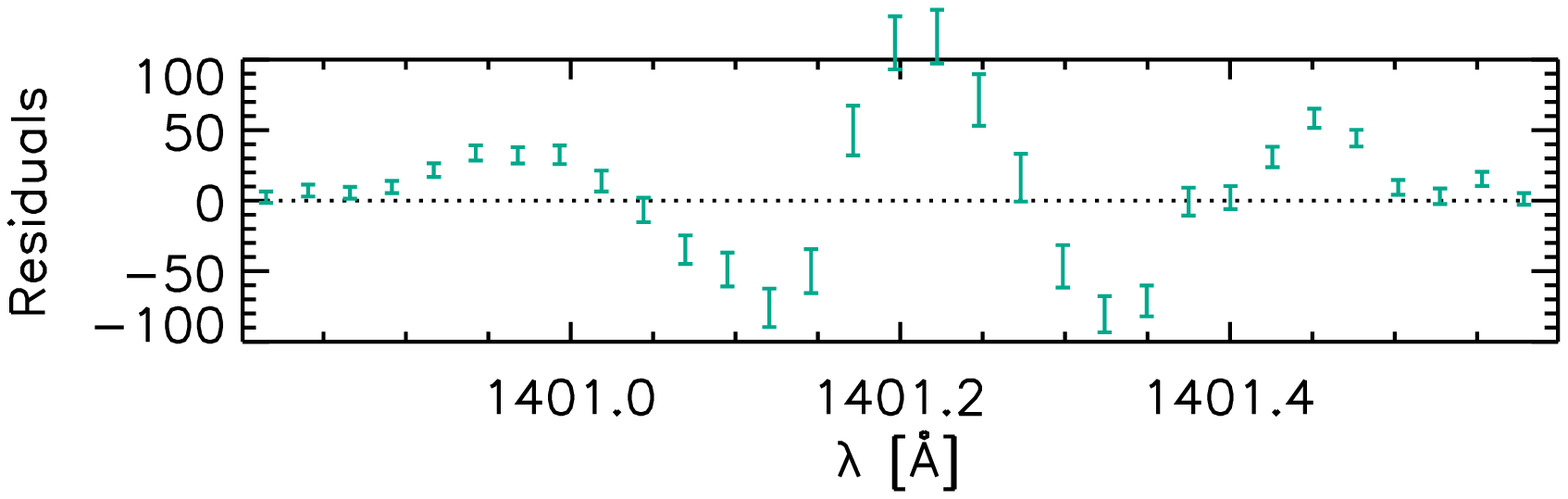}
	\includegraphics[width=4.21cm,clip,bb=75  0 487 145]{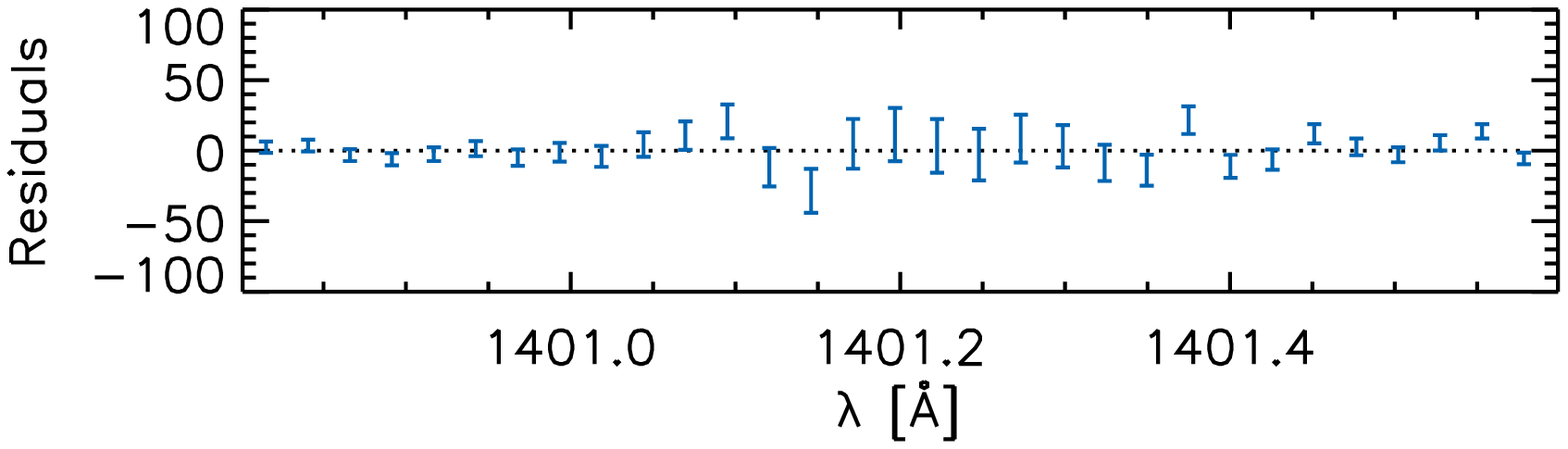}
	\includegraphics[width=4.21cm,clip,bb=75  0 487 145]{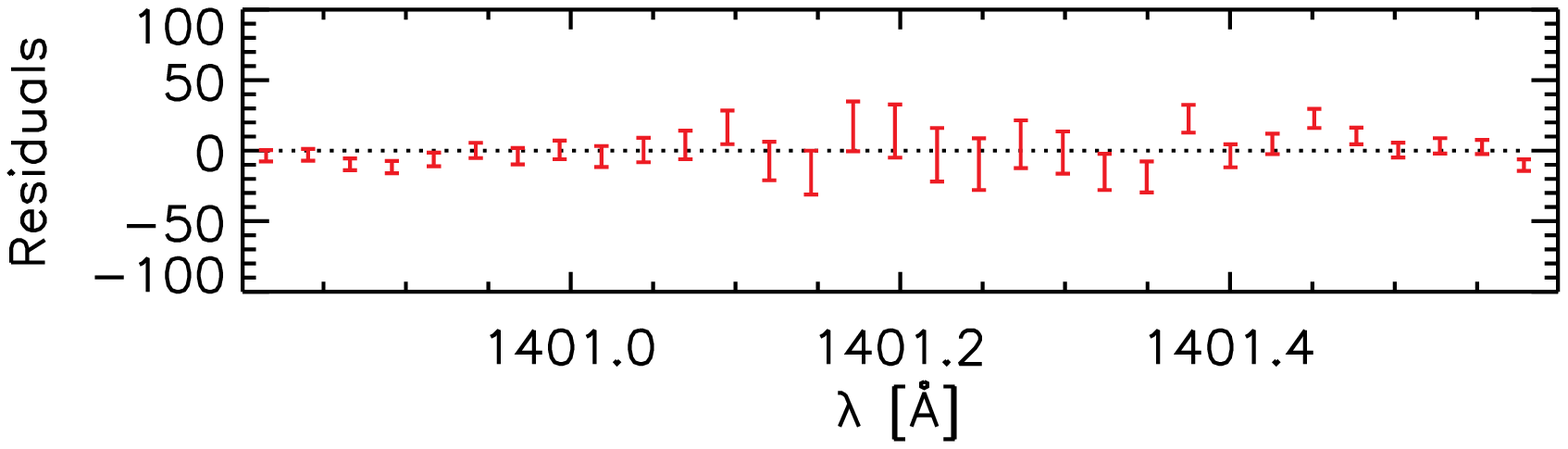}
	\includegraphics[width=4.21cm,clip,bb=75  0 487 145]{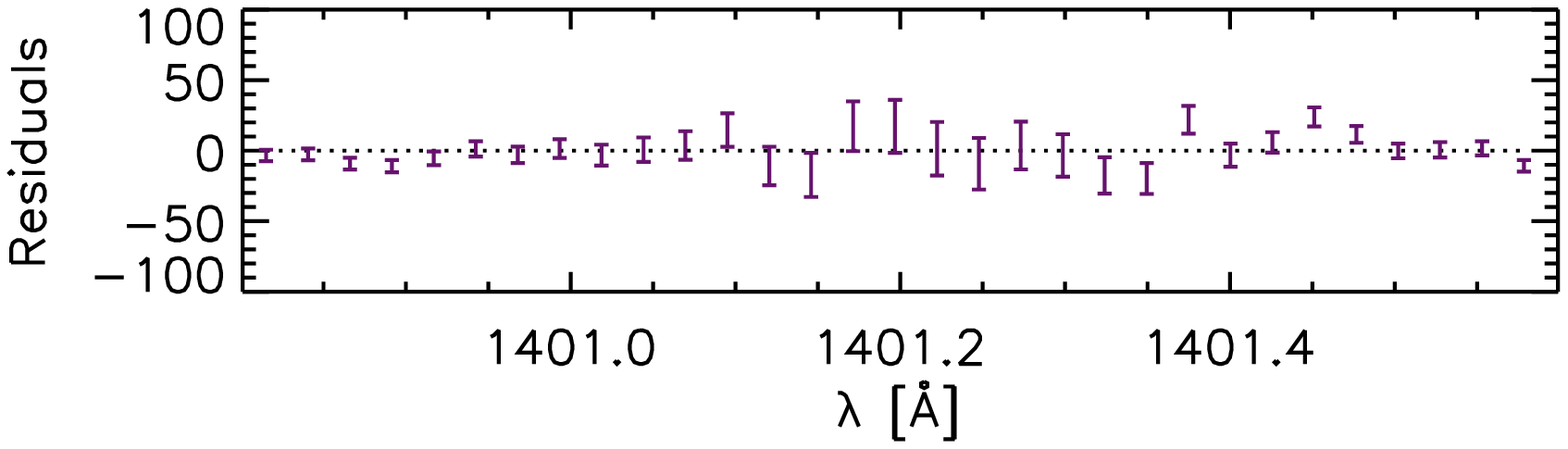}
	\includegraphics[width=4.98cm,clip,bb= 0 52 487 340]{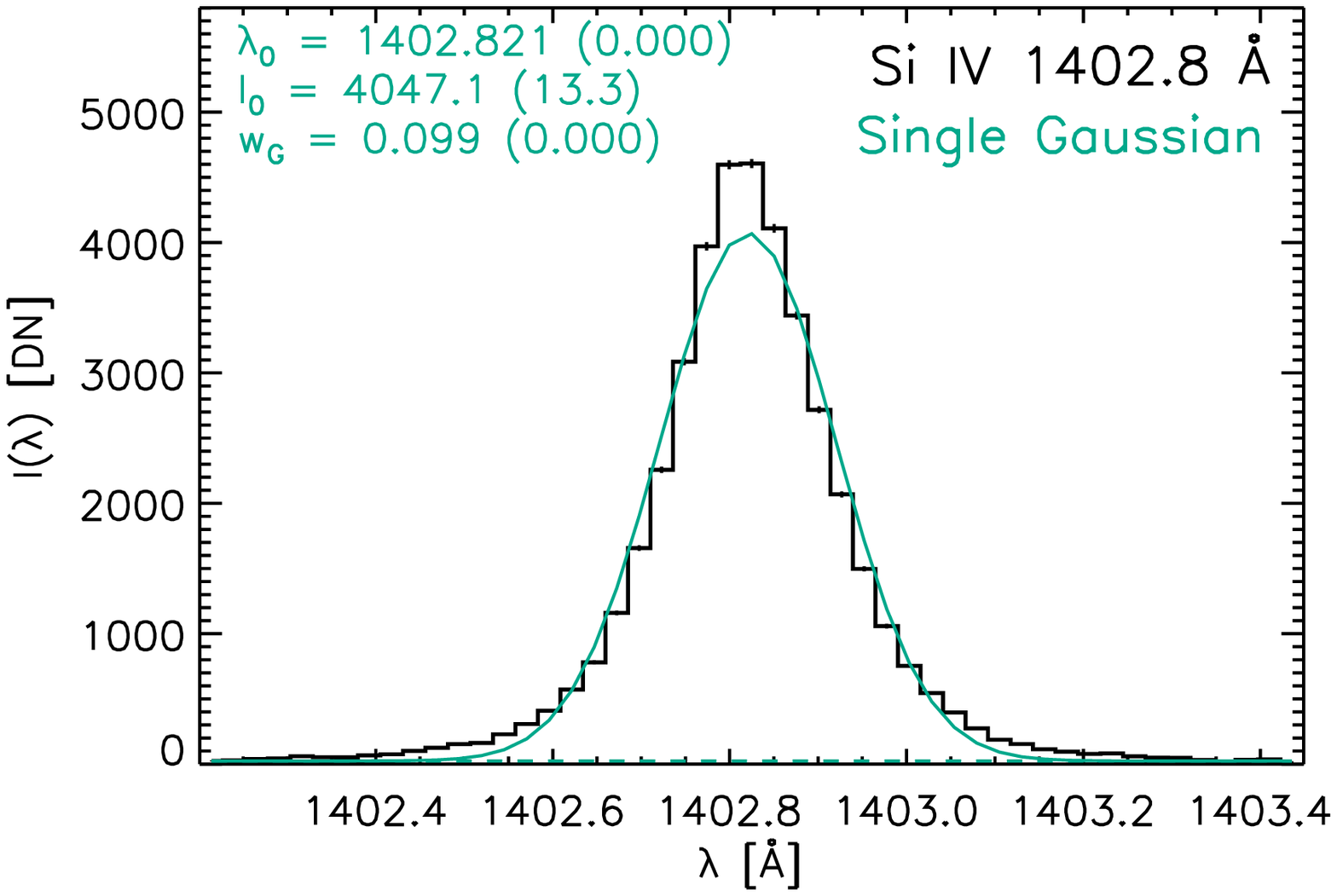}
	\includegraphics[width=4.21cm,clip,bb=75 52 487 340]{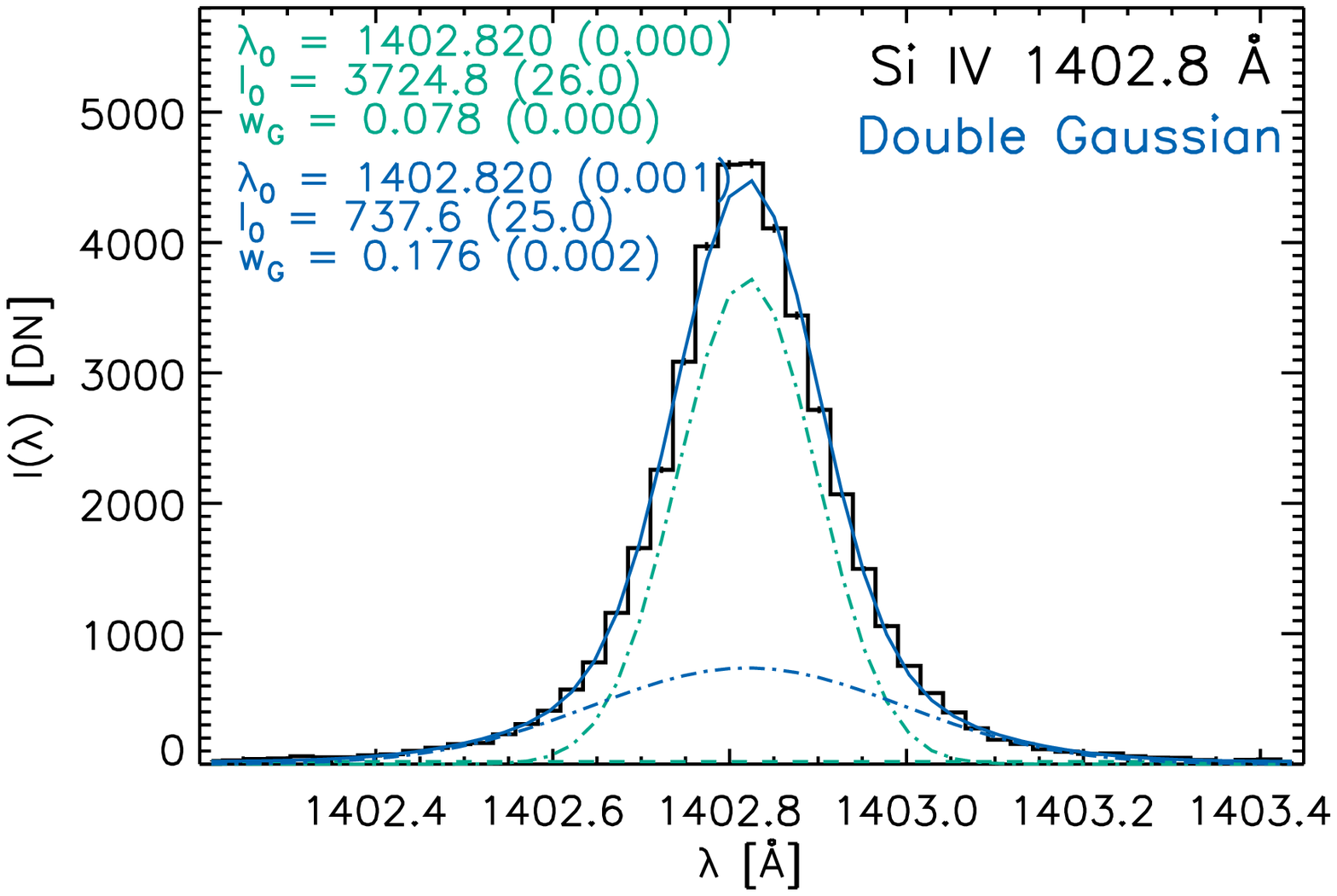}
	\includegraphics[width=4.21cm,clip,bb=75 52 487 340]{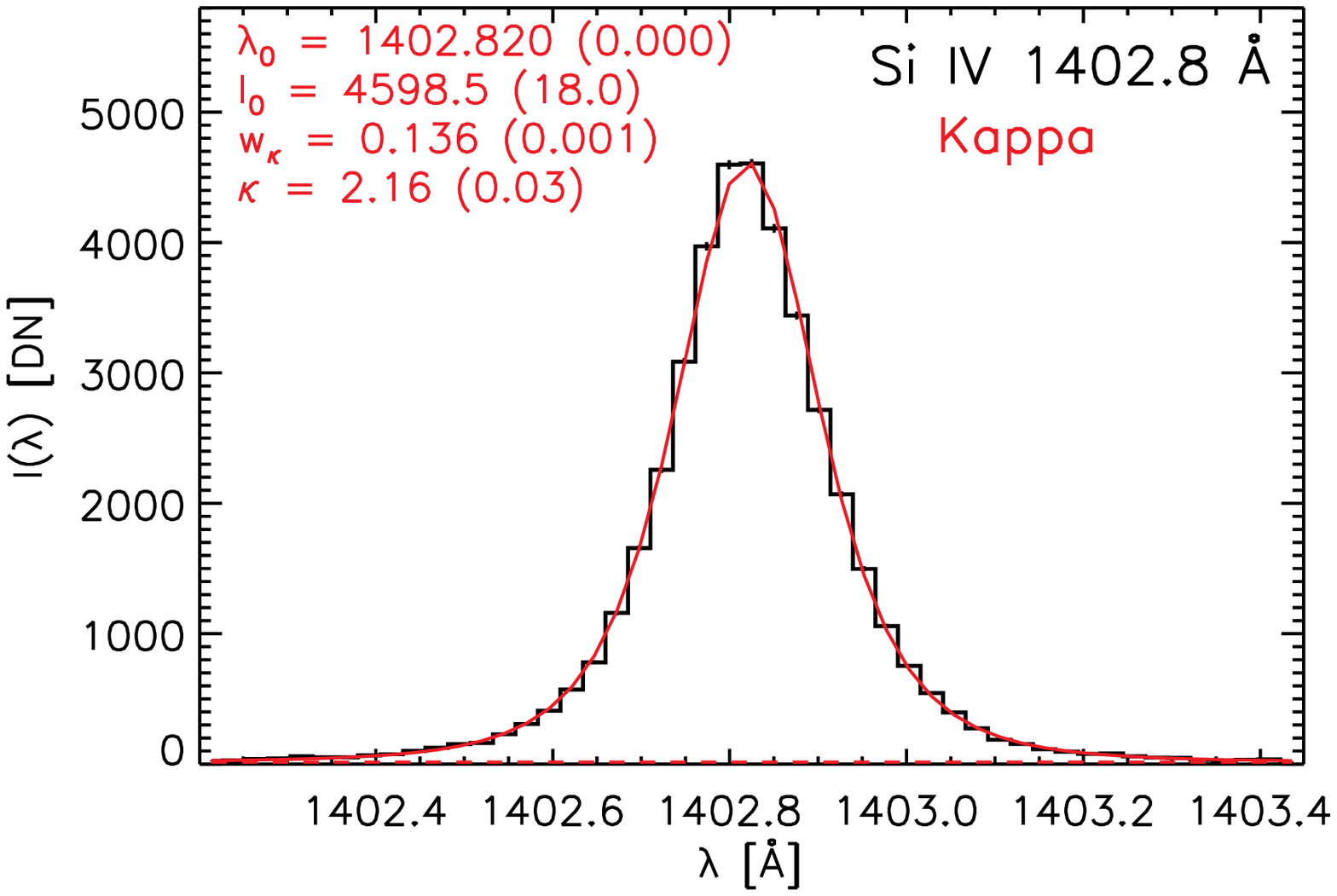}
	\includegraphics[width=4.21cm,clip,bb=75 52 487 340]{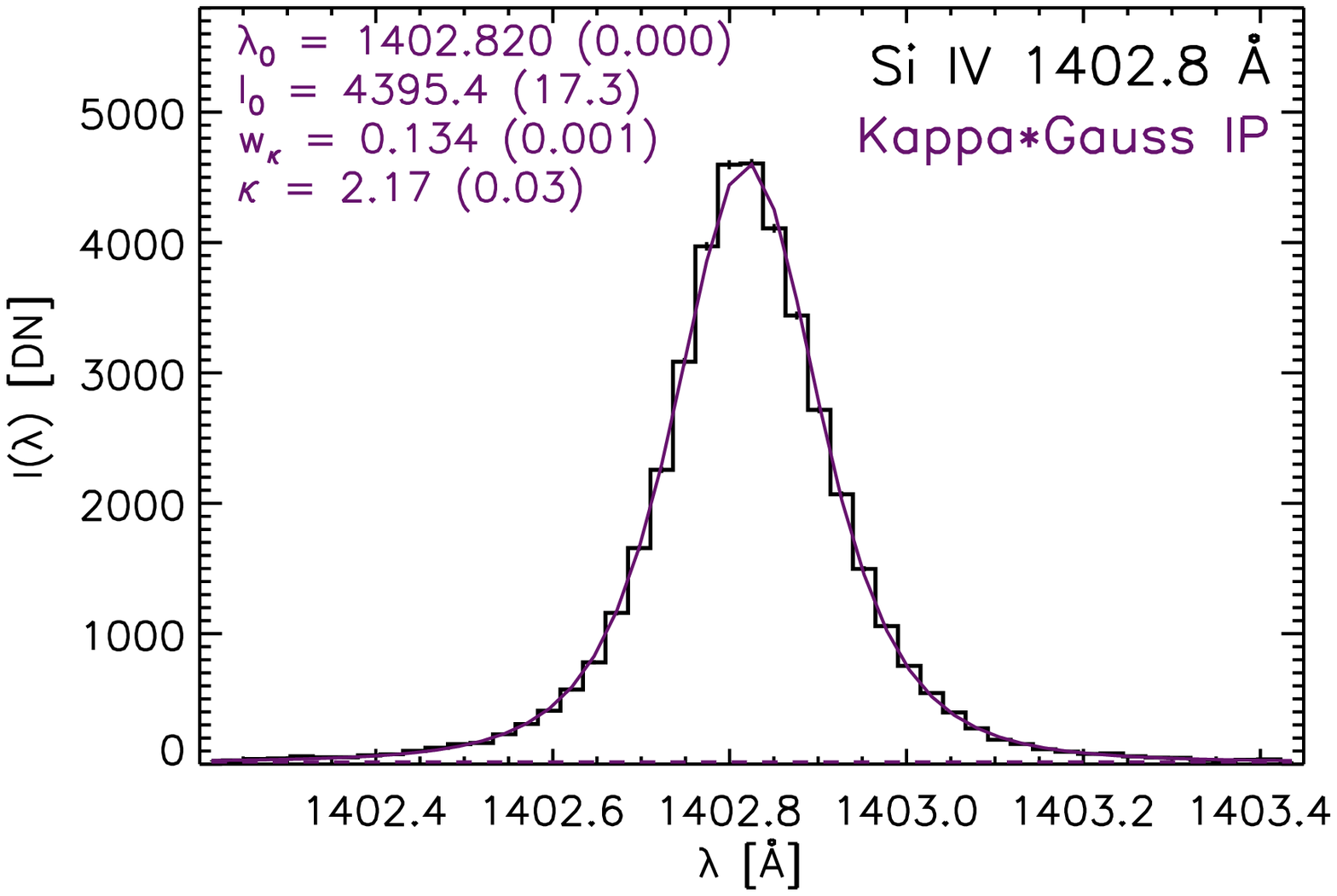}
	\includegraphics[width=4.98cm,clip,bb= 0  0 487 145]{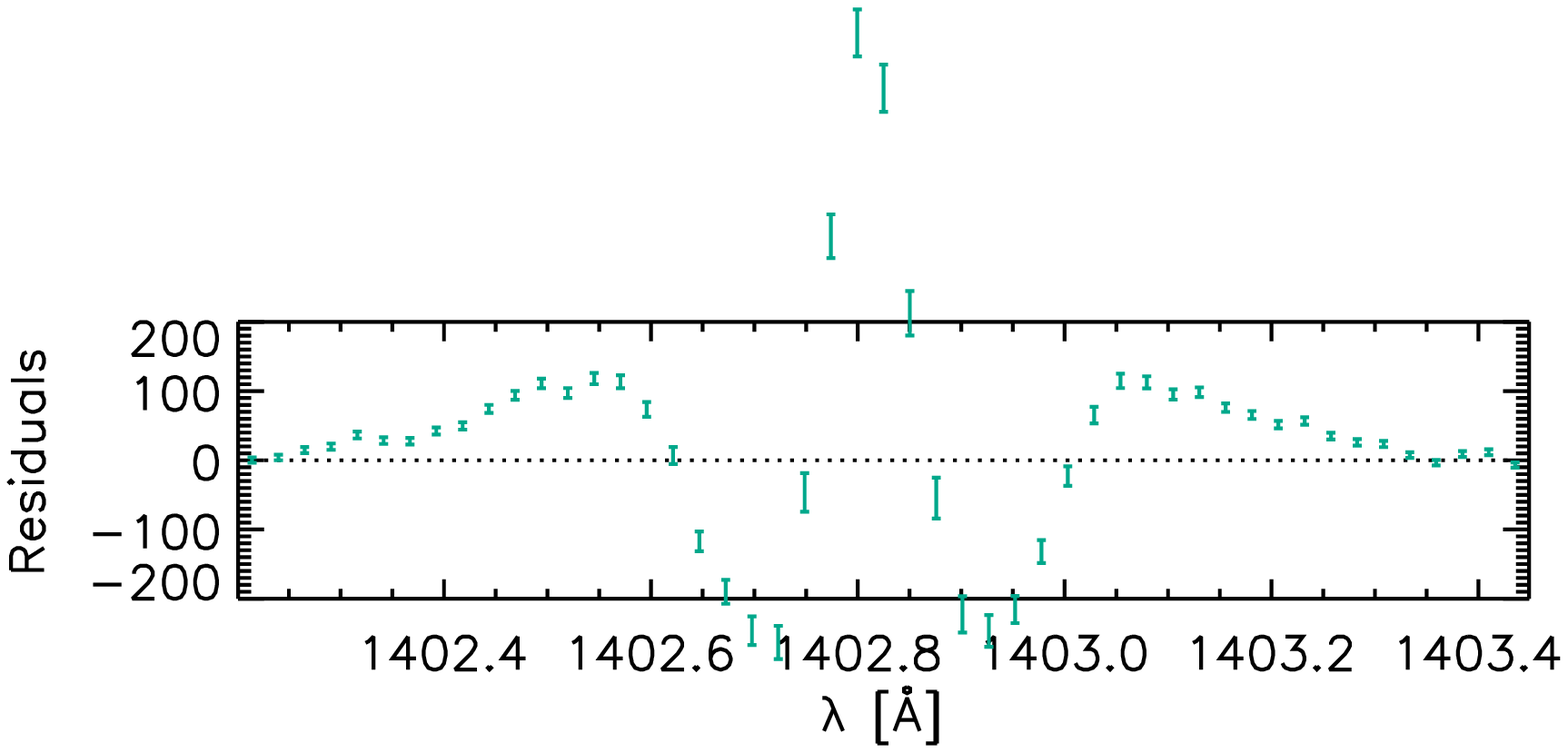}
	\includegraphics[width=4.21cm,clip,bb=75  0 487 145]{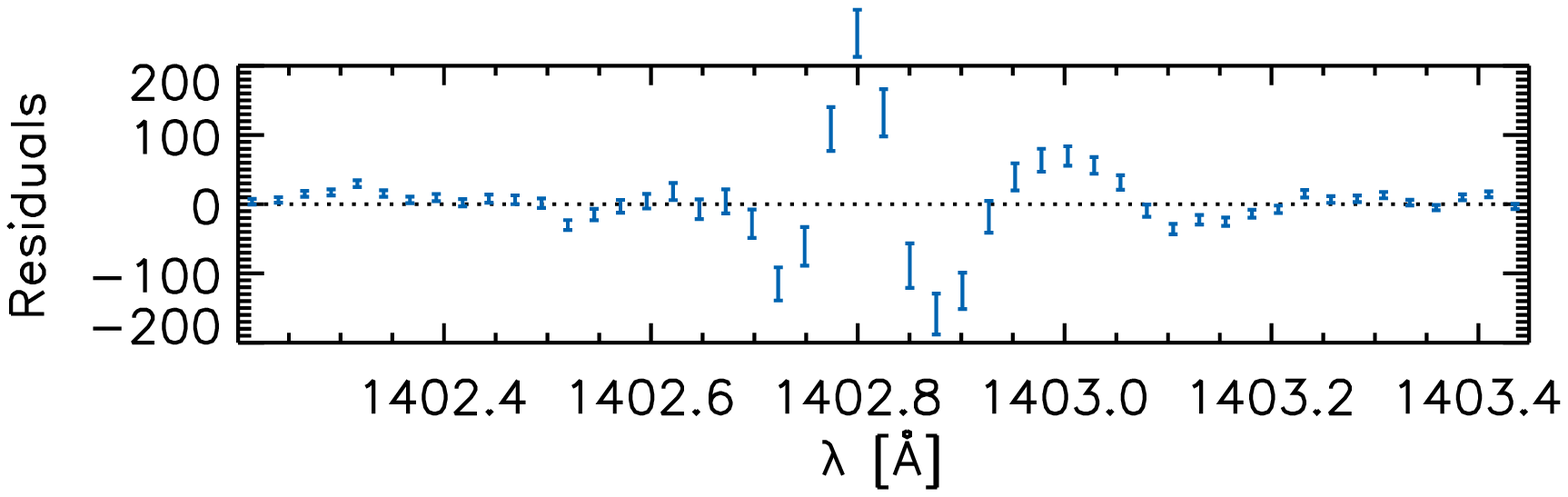}
	\includegraphics[width=4.21cm,clip,bb=75  0 487 145]{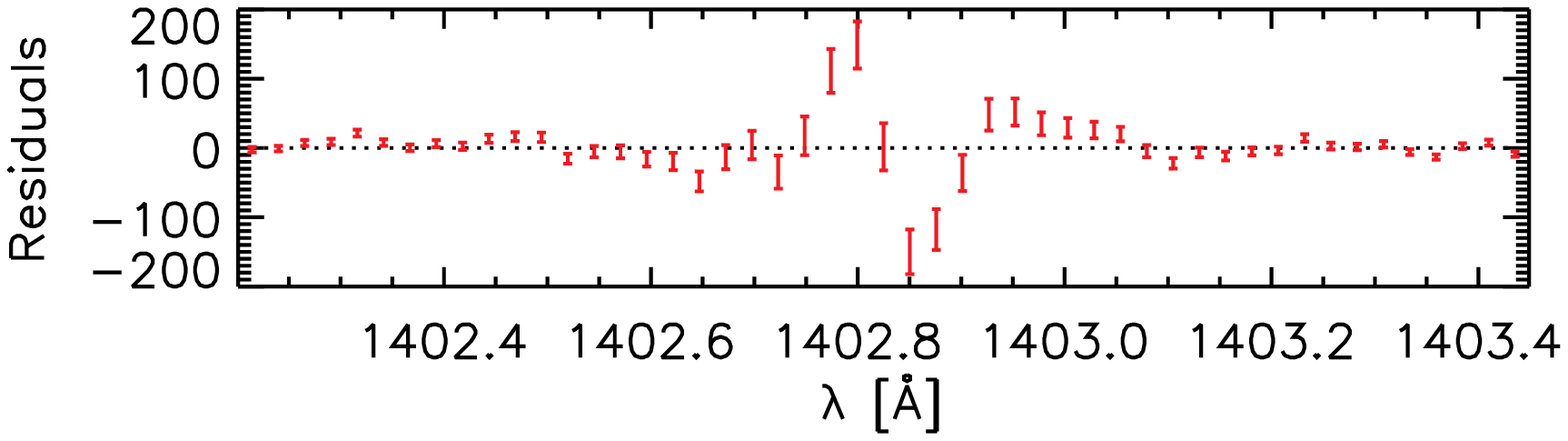}
	\includegraphics[width=4.21cm,clip,bb=75  0 487 145]{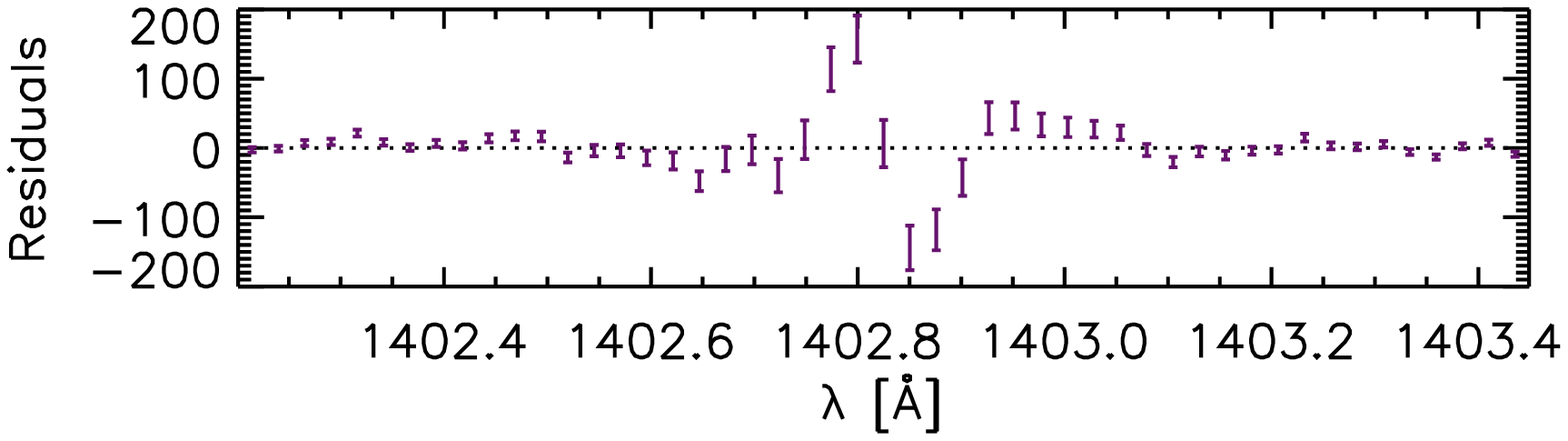}
	\includegraphics[width=4.98cm,clip,bb= 0 52 487 340]{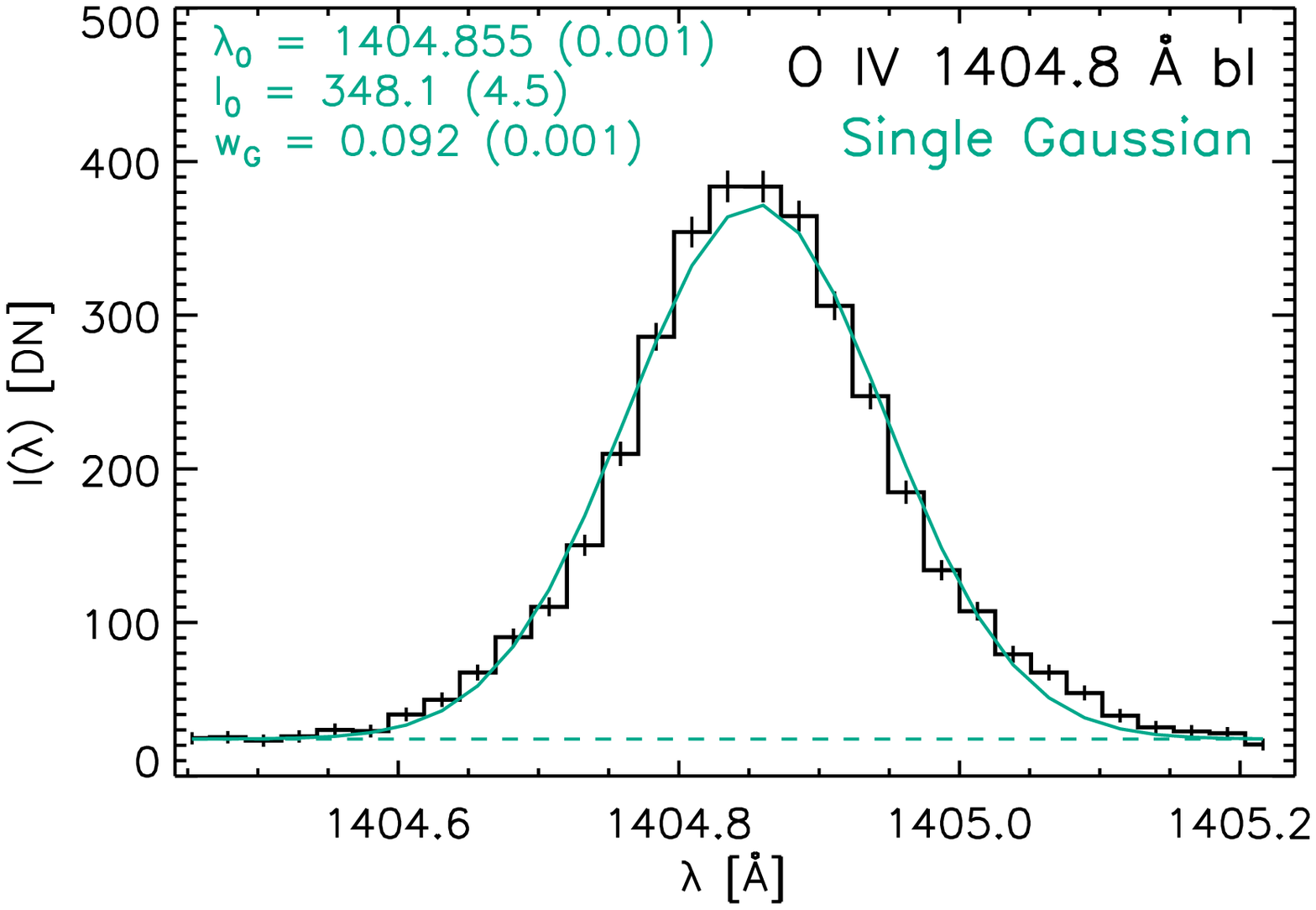}
	\includegraphics[width=4.21cm,clip,bb=75 52 487 340]{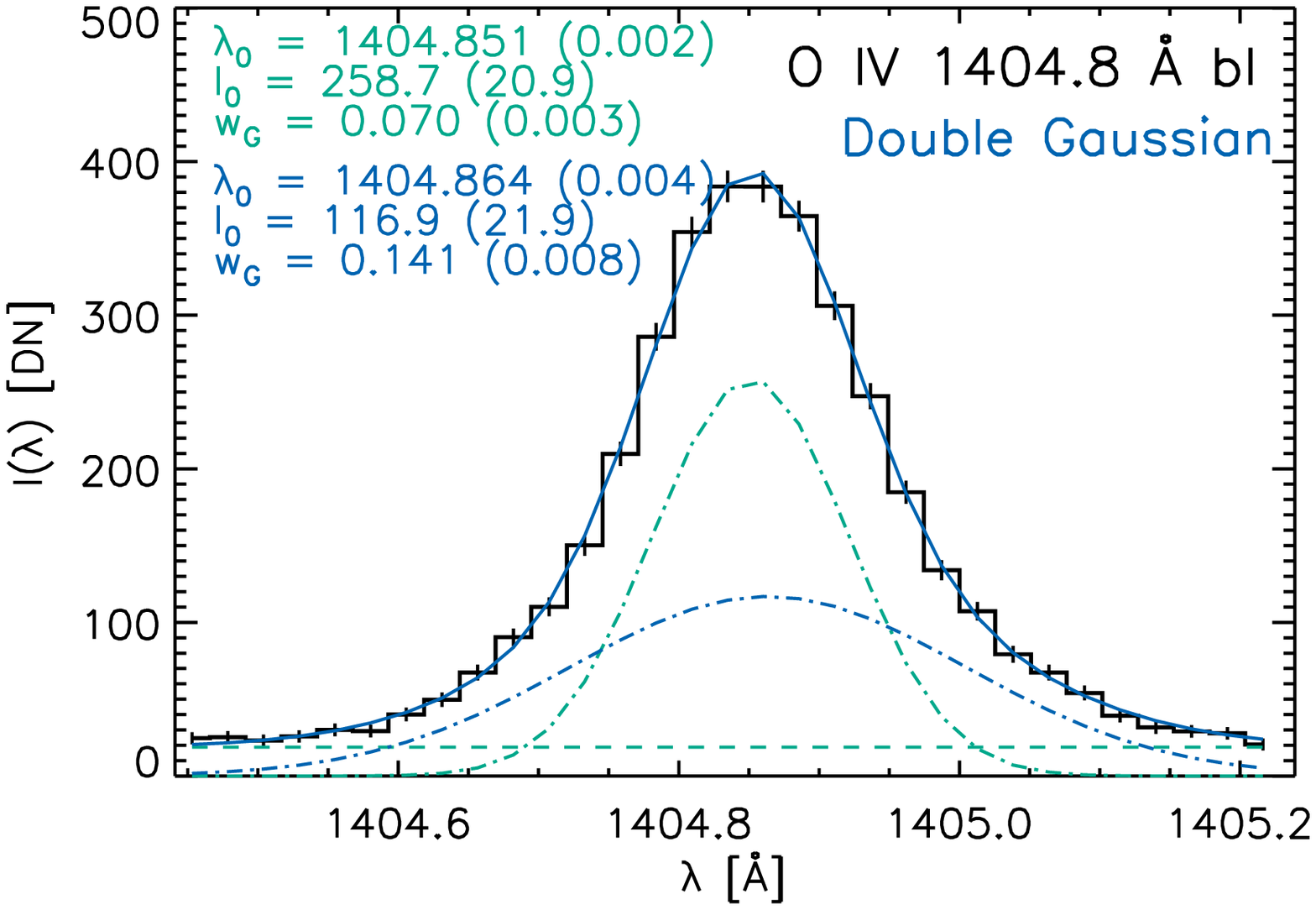}
	\includegraphics[width=4.21cm,clip,bb=75 52 487 340]{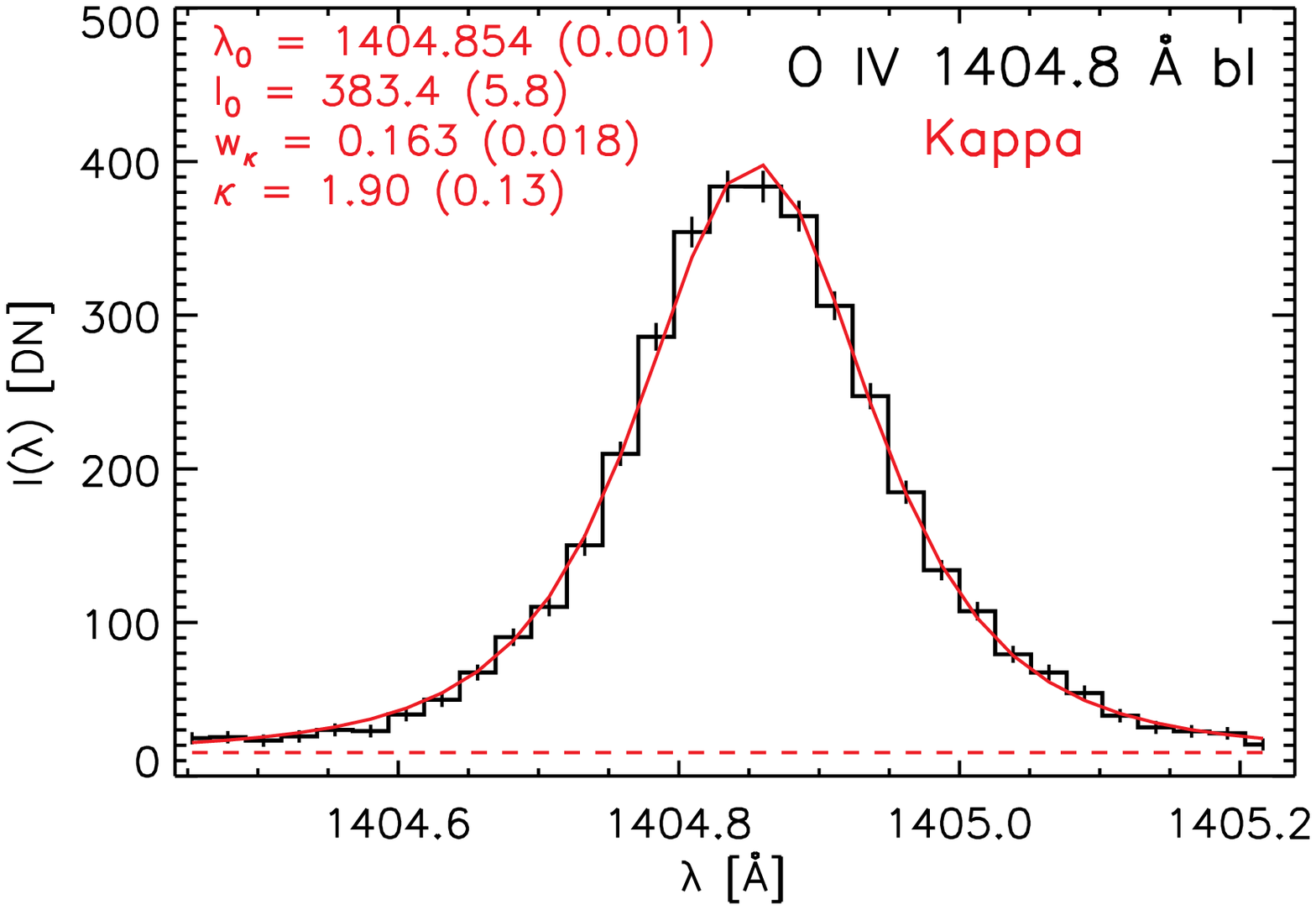}
	\includegraphics[width=4.21cm,clip,bb=75 52 487 340]{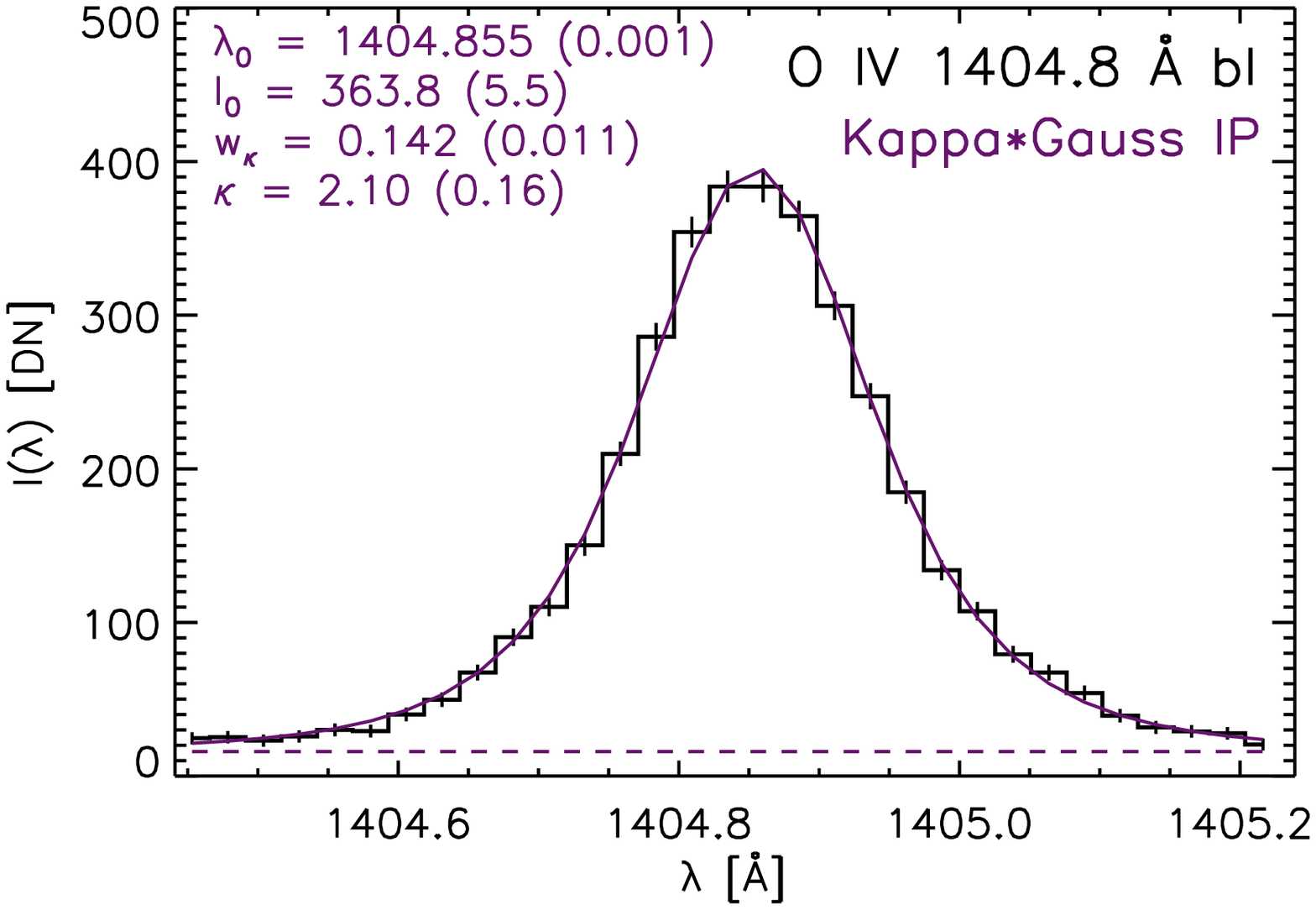}
	\includegraphics[width=4.98cm,clip,bb= 0  0 487 145]{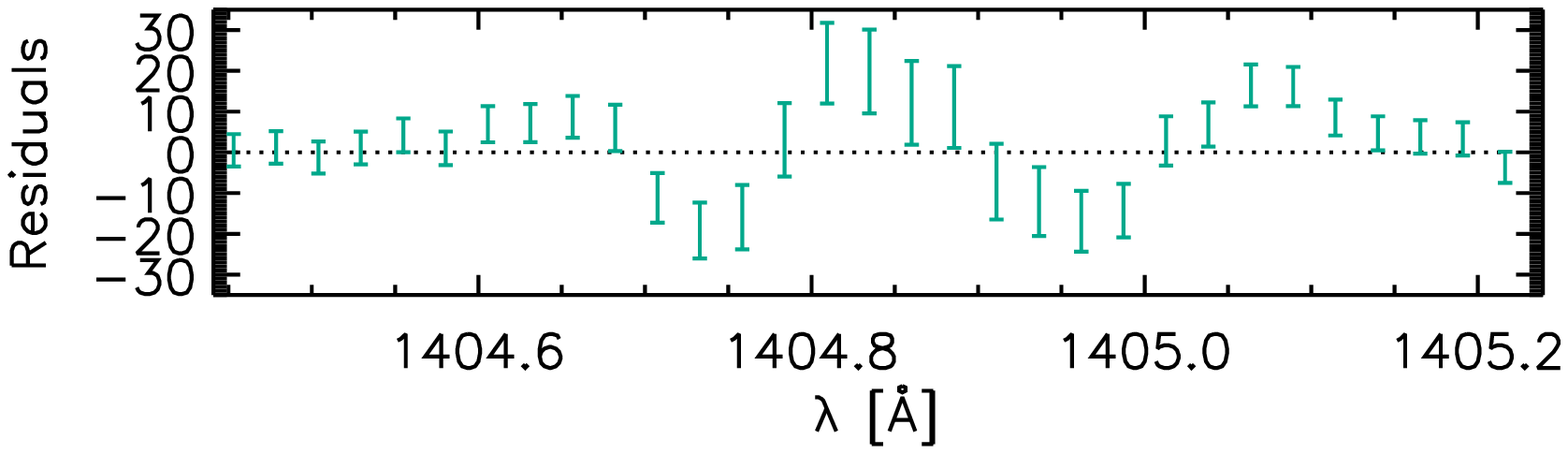}
	\includegraphics[width=4.21cm,clip,bb=75  0 487 145]{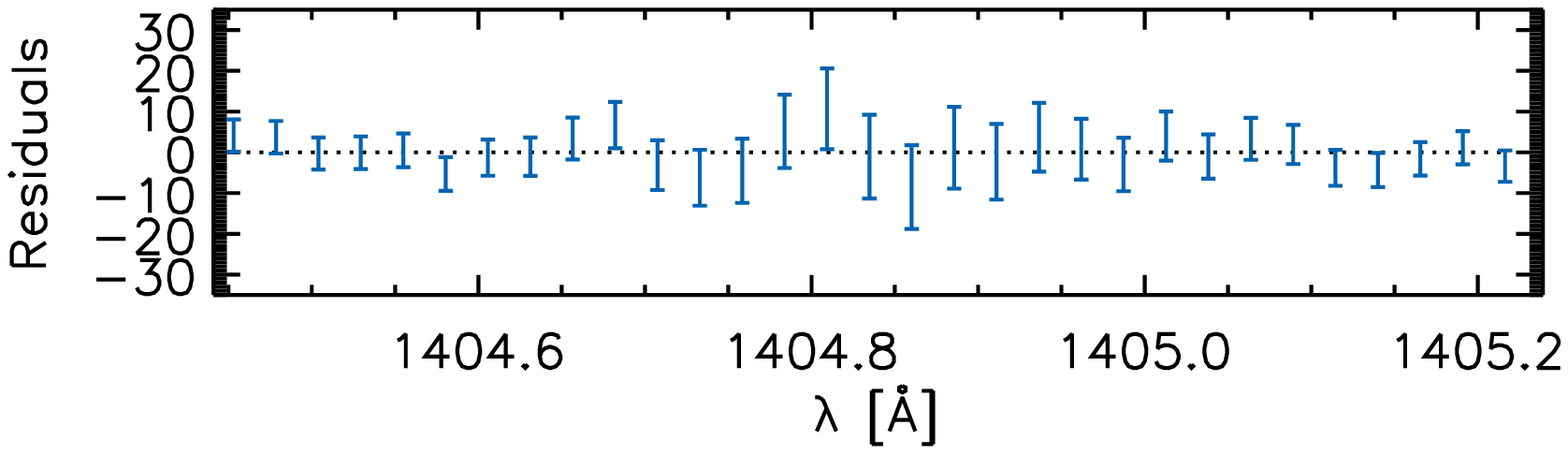}
	\includegraphics[width=4.21cm,clip,bb=75  0 487 145]{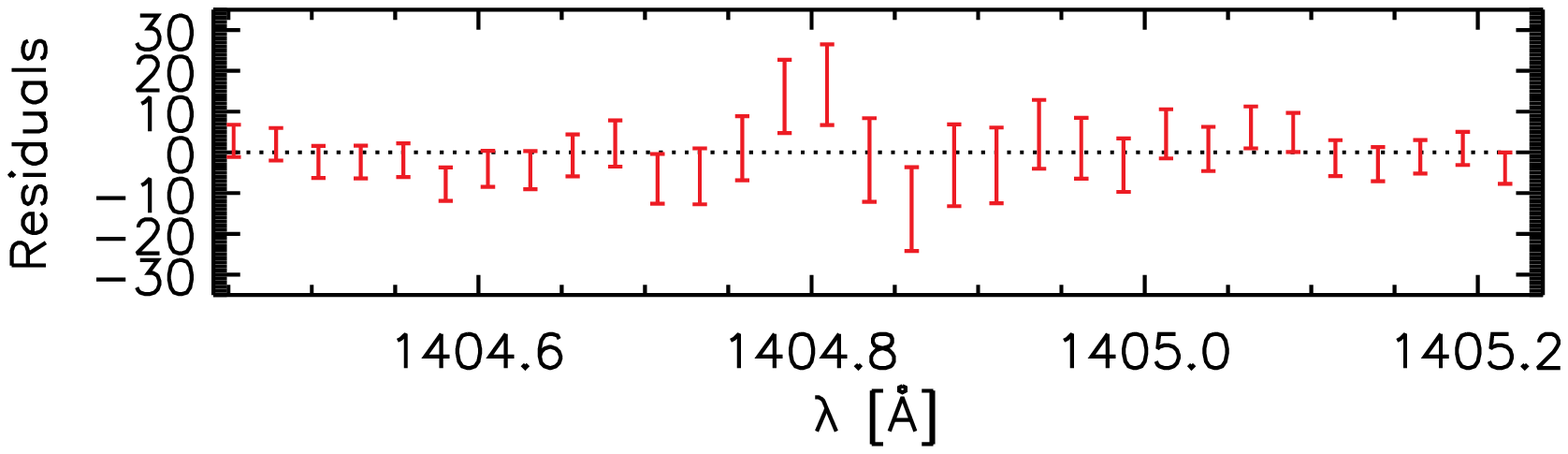}
	\includegraphics[width=4.21cm,clip,bb=75  0 487 145]{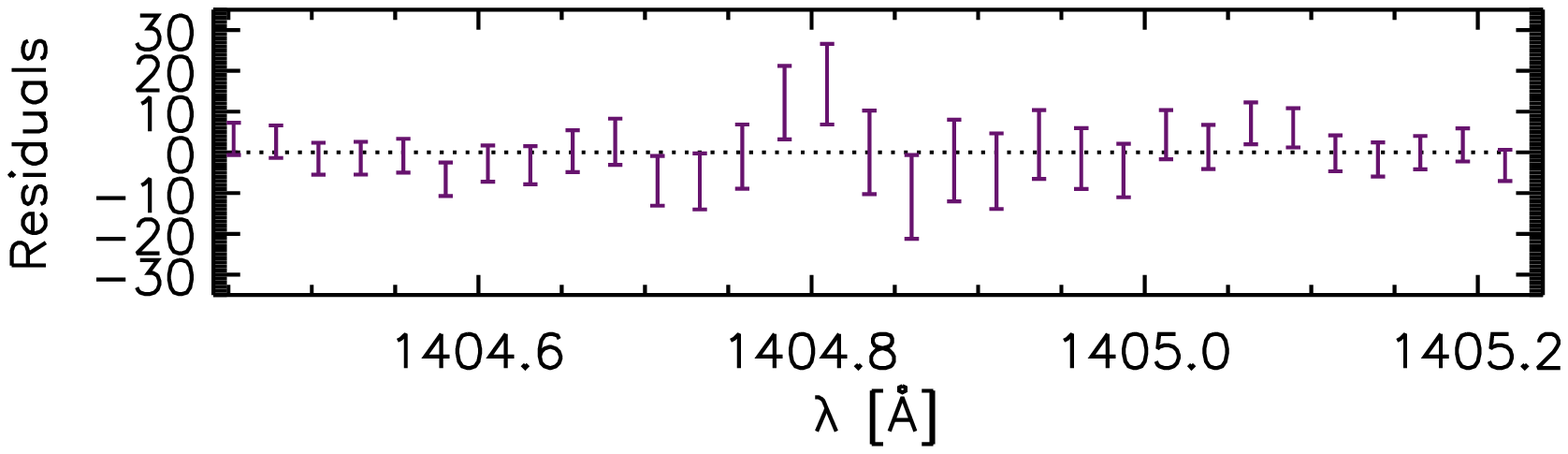}
	\includegraphics[width=4.98cm,clip,bb= 0 52 487 340]{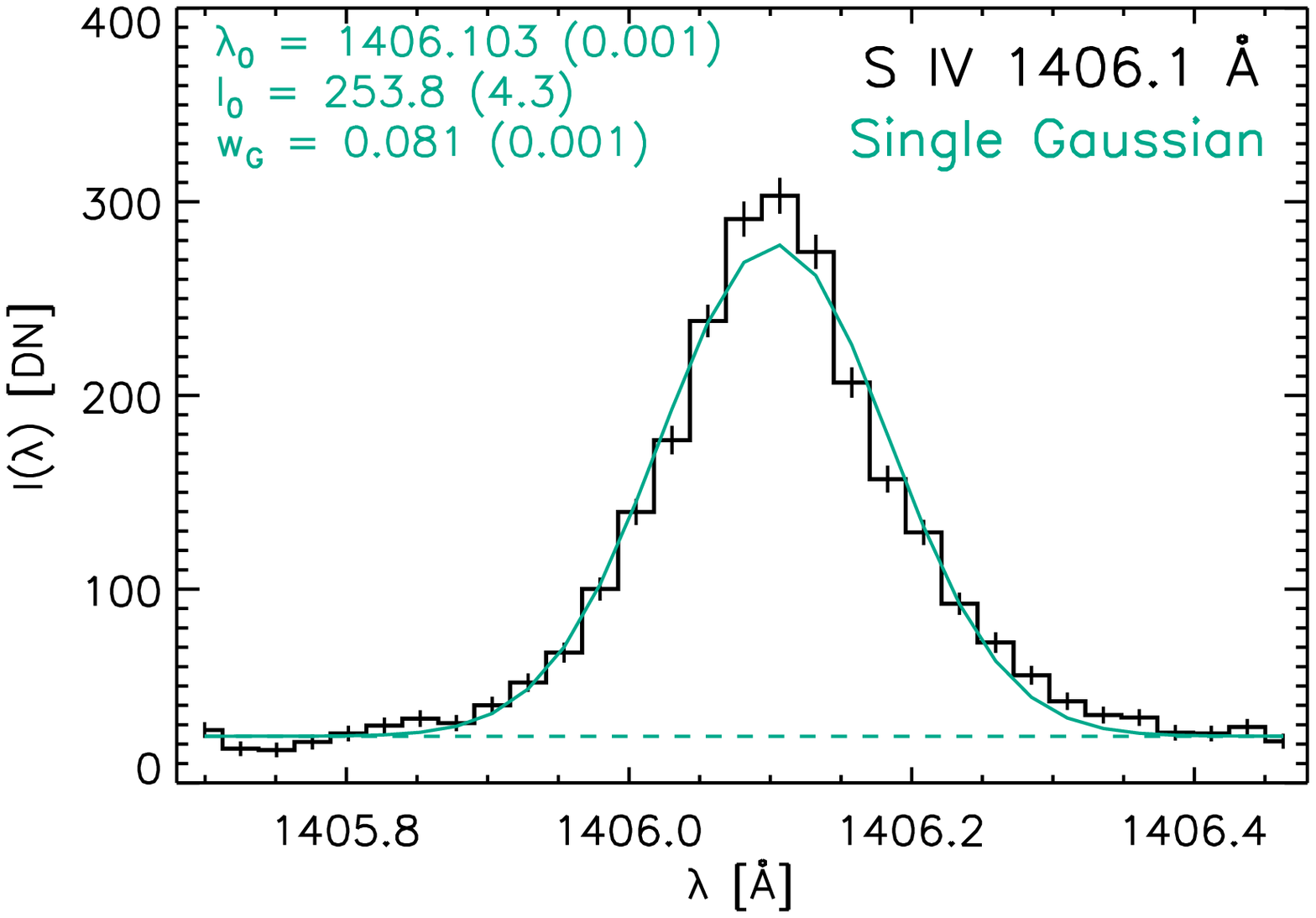}
	\includegraphics[width=4.21cm,clip,bb=75 52 487 340]{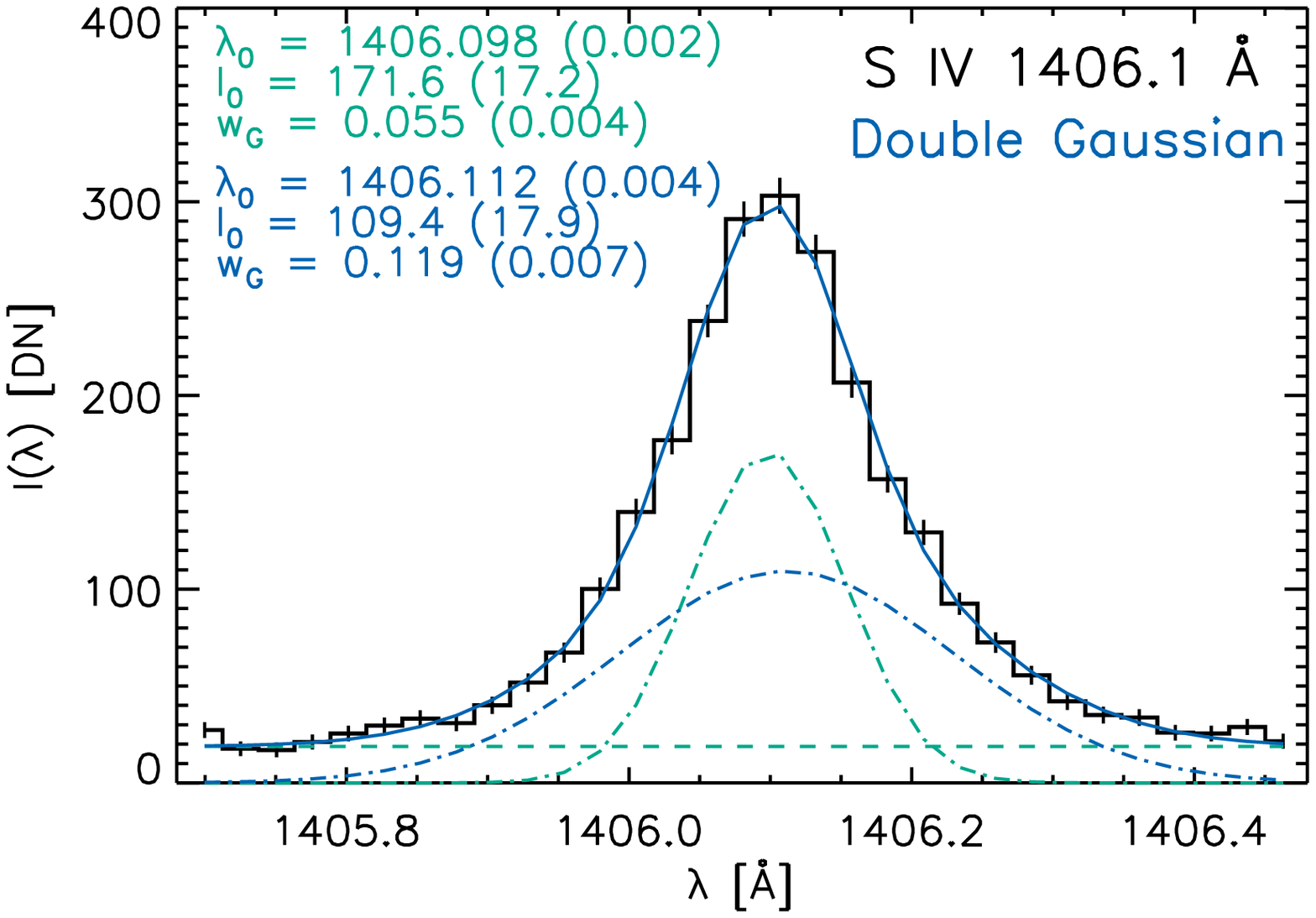}
	\includegraphics[width=4.21cm,clip,bb=75 52 487 340]{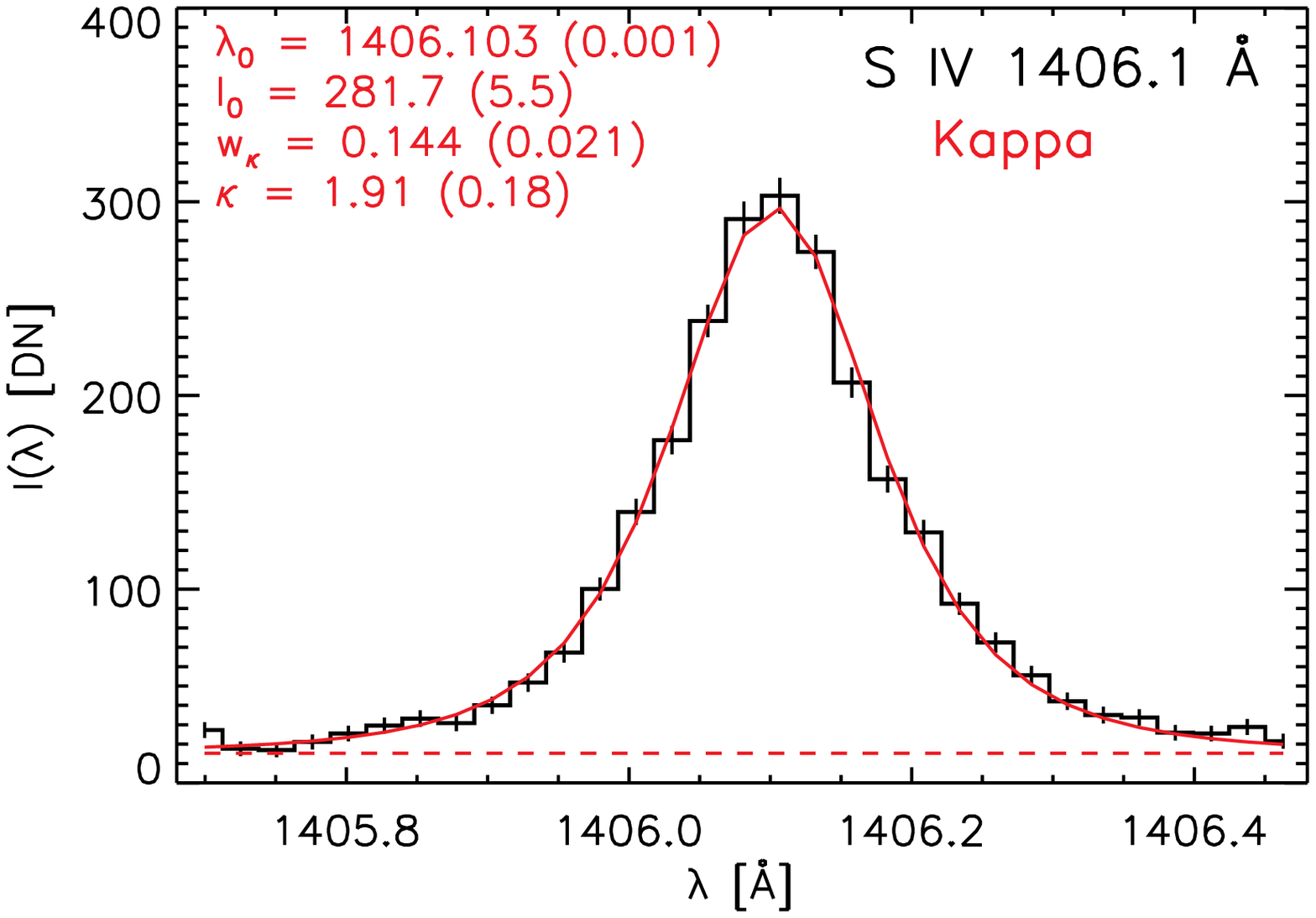}
	\includegraphics[width=4.21cm,clip,bb=75 52 487 340]{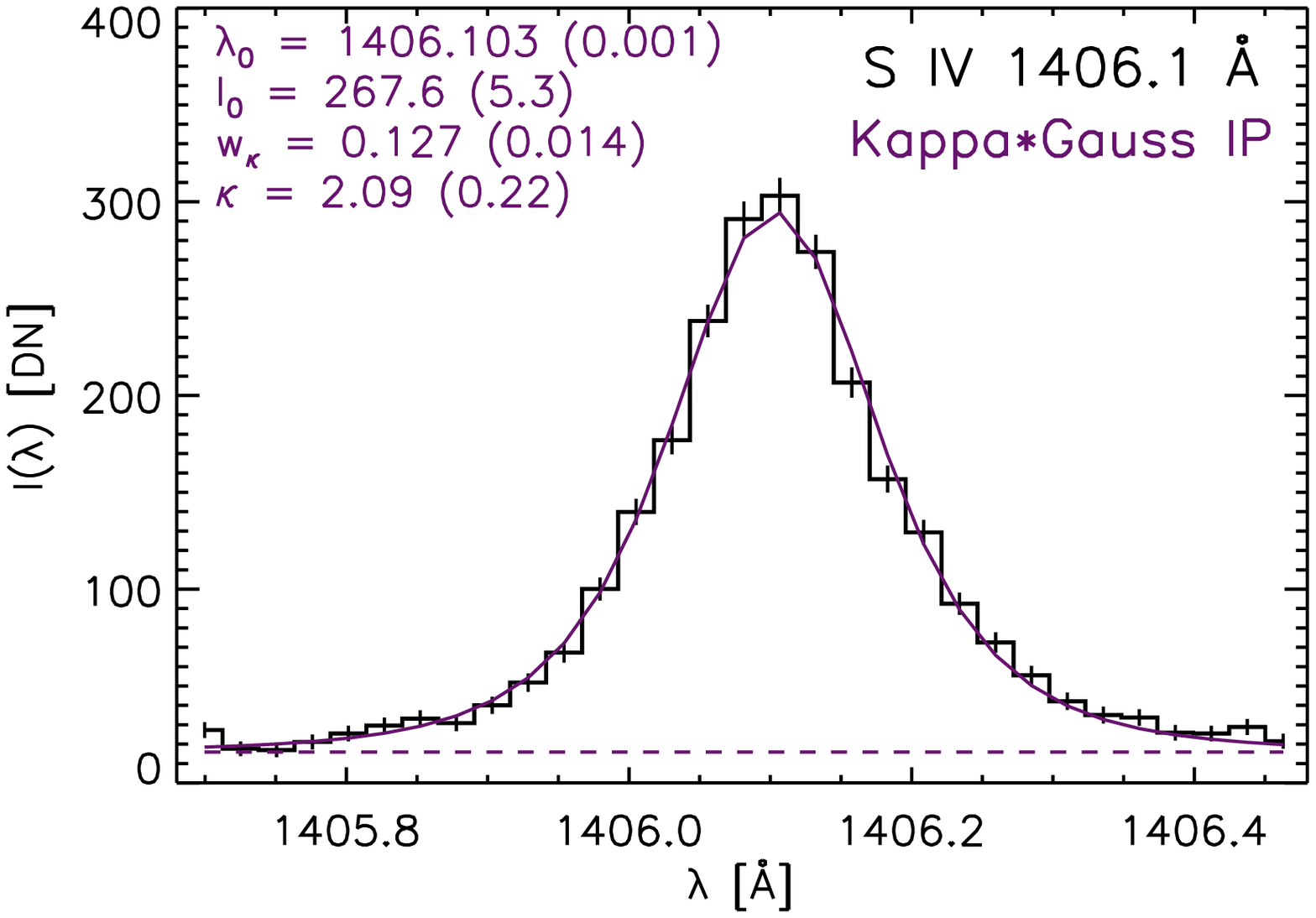}
	\includegraphics[width=4.98cm,clip,bb= 0  0 487 145]{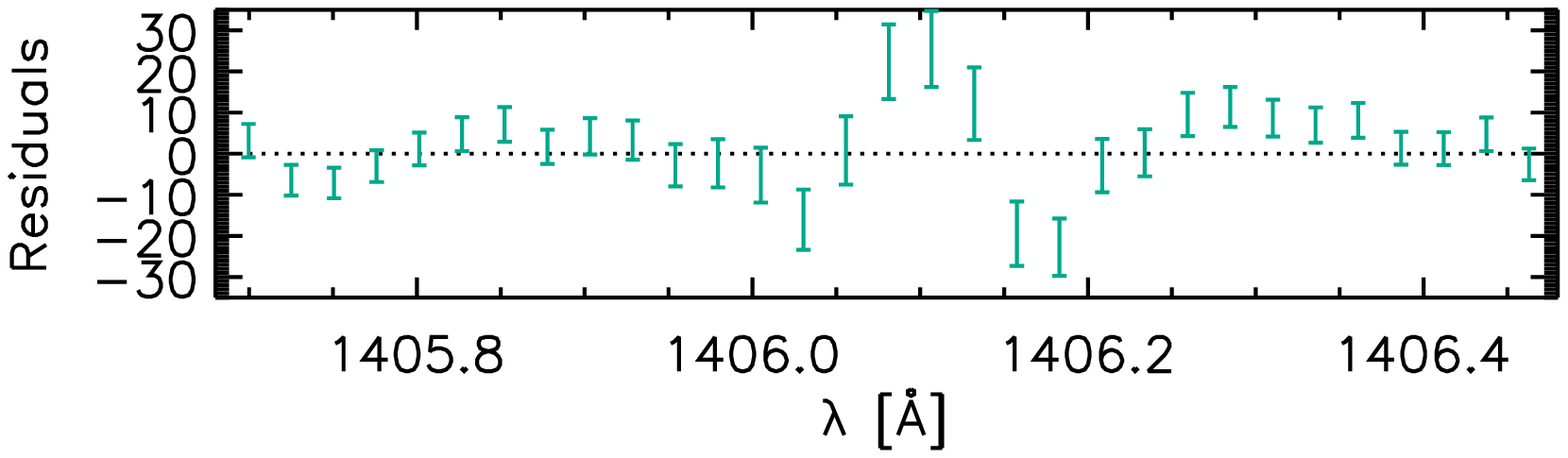}
	\includegraphics[width=4.21cm,clip,bb=75  0 487 145]{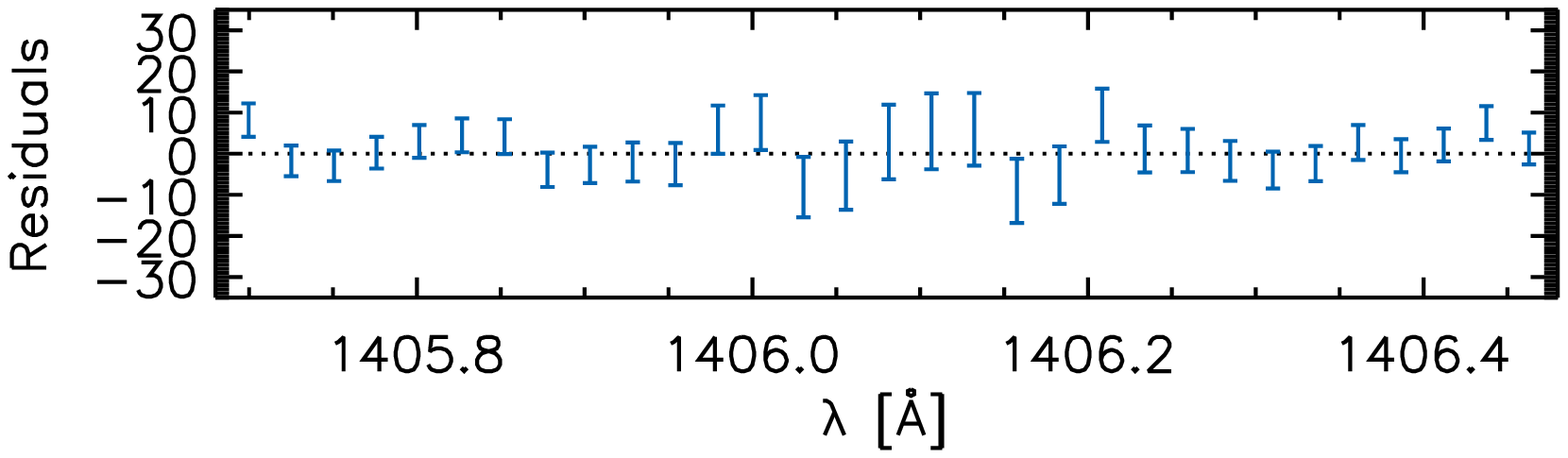}
	\includegraphics[width=4.21cm,clip,bb=75  0 487 145]{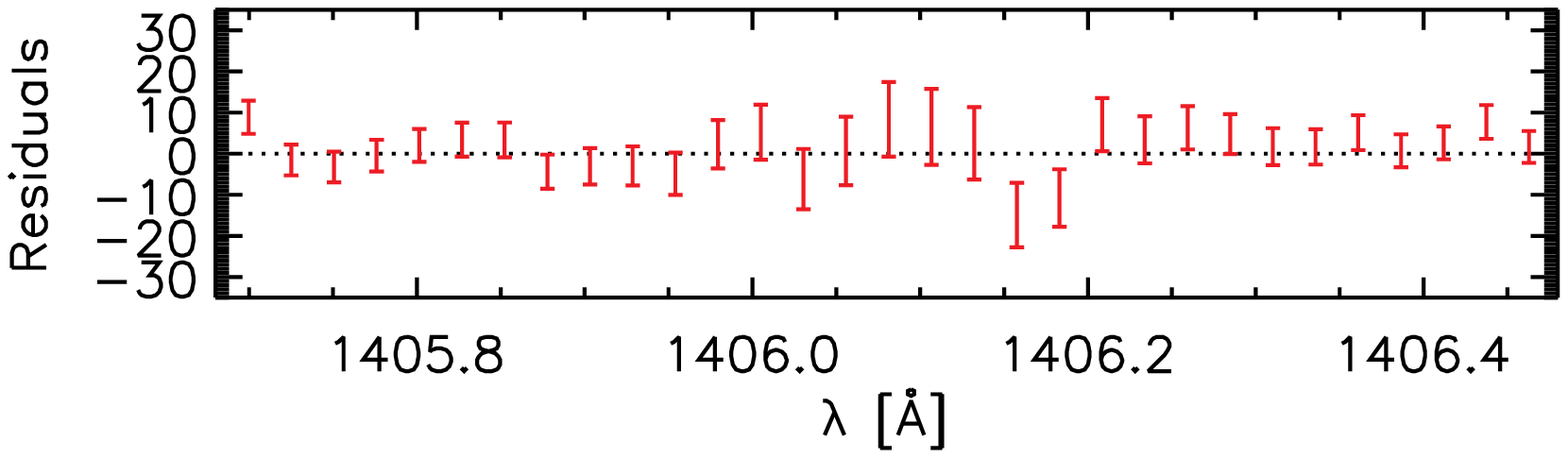}
	\includegraphics[width=4.21cm,clip,bb=75  0 487 145]{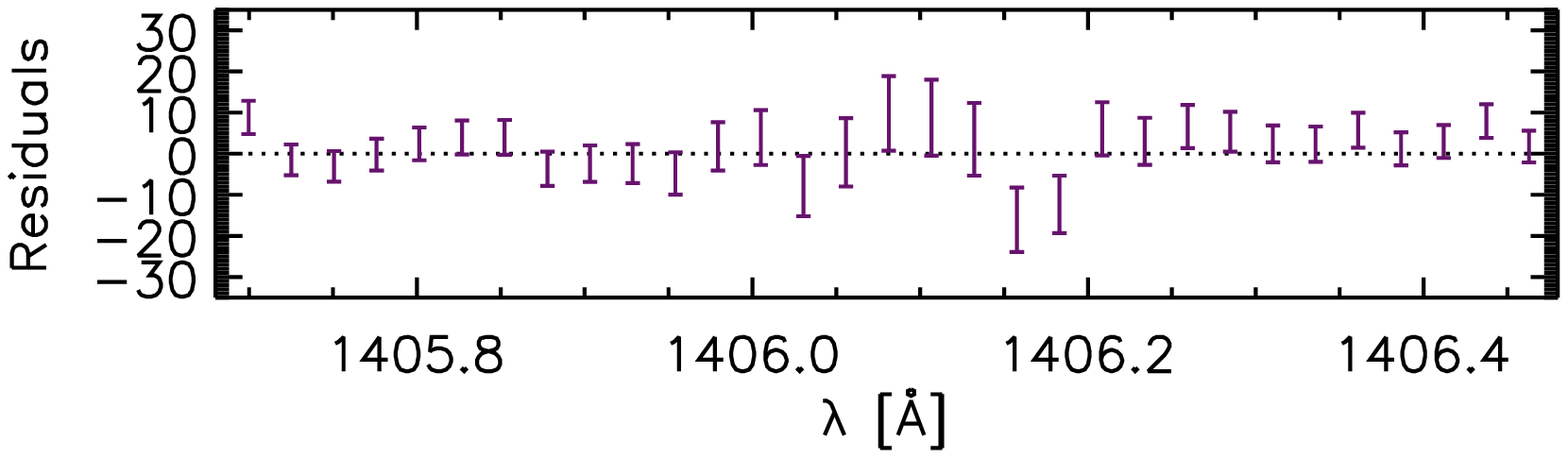}
\caption{\textit{IRIS} TR line fitting for the spectrum of the selected pixel marked in Fig. \ref{Fig:IRIS_sg}. Each of the five rows represent one of the five observed TR lines, while each columns represent a particular type of fit. Fit parameters are listed in each panel, with uncertainties in parentheses. Residuals $I_\mathrm{obs}(\lambda_i) - I_\mathrm{fit}(\lambda_i)$ are also shown. See text for details.
\label{Fig:Fits}}
\end{figure*}

The value of $|S|$ is below the value of 0.1 (within the uncertainties) for all lines, although the value may not be meaningful for the weaker lines (especially \ion{S}{4}) due to its uncertainty. We also note that the strongest \ion{O}{4} line at 1401\,\AA~is blended in the far red wing with a weak but recognizable \ion{S}{1} 1401.5\,\AA~transition (Fig. \ref{Fig:Fits}, second row). However, in this case this \ion{S}{1} transition is too weak to affect the value of $|S|$ significantly. The \ion{S}{4} 1406\,\AA~line does not have visible blends, see \citet[][]{Polito16b}. The \ion{Fe}{2} 1405.6\,\AA~line is however distinguishable outside the \ion{S}{4} 1406.1\,\AA~line.

The kurtosis $K$ is for all five TR lines higher than 3.5. This indicates presence of significant wings, in accordance with the visual inspection of the spectrum. For the strongest line in the spectrum, \ion{Si}{4}, $K$ is 5.8, which is even larger than the value of $K$ for the other lines. This indicates stronger wings of the \ion{Si}{4} line compared to the \ion{O}{4} and \ion{S}{4} ones. Restricting the wavelength range to $\lambda_0 \pm$ 0.4\,\AA~would decrease the value of $K$, this being due to the line still having significant wings (of the order of 50\,DN, i.e., about 3 times the local pseudo-continuum) at the edges of this smaller interval. We note that the value of $w^2$ for the \ion{Si}{4} line is higher than for other lines; however, this is an effect of the larger wavelength range. Restricting it to $\pm$\,0.4\,\AA~would produce $w$\,$\approx$\,0.26, in accordance with other lines.

%
%
%
%
%
\subsection{Line profile fitting with Gaussian and $\kappa$ profiles}
\label{Sect:4.2}

The line profile fitting is performed for the full \textit{IRIS} FUV window around 1400\,\AA~and the selected single-pixel spectrum (Sect. \ref{Sect:3} and Fig. \ref{Fig:IRIS_sg}). The fitting assumes assumes a constant ``background'' pseudo-continuum with intensity $I_\mathrm{BG}$ and the presence of five strong TR spectral lines of \ion{Si}{4}, \ion{O}{4}, and \ion{S}{4}. These are fitted at the same time using the fitting formulae for Gaussian and $\kappa$ line profiles for each line (Eqs. \ref{Eq:Gauss_profile} and \ref{Eq:Kappa_profile}). Values of the fitting parameters and their uncertainties are found using the Levenberg-Marquardt least-squares method \citep[see, e.g][]{Press92} as implemented in the \textit{mcurvefit.pro} IDL routine available under SolarSoft.

The goodness-of-fit is evaluated using the reduced $\chi^2_\mathrm{red}$, given by
\begin{equation}
	\chi^2_\mathrm{red} = \frac{1}{\nu} \sum\limits_{i=0}^{N-1} \frac{(I_\mathrm{obs}(\lambda_i) - I_\mathrm{fit}(\lambda_{i}))^2}{\sigma^2\left(I_\mathrm{obs}(\lambda_i)\right)}\,,
	\label{Eq:chi2_reduced}
\end{equation}
where $N$\,=\,326 is the total number of spectral bins in the present \textit{IRIS} raster, $\nu = N - N_\mathrm{fit} -1$ is the number of degrees of freedom in the fit, $N_\mathrm{fit}$ is the number of free parameters in the fit, $I_\mathrm{obs}(\lambda_i)$\,$\pm$\,$\sigma(I_\mathrm{obs}(\lambda_i))$ are the observed intensities in the spectral bin $i$ and their respective uncertainties (see Sect. \ref{Sect:3}). The Levenberg-Marquardt fitting procedure minimizes the $\chi^2_\mathrm{red}$. Since the fitting is performed at the same time for the pseudo-continuum and all lines, there is only a single value of $\chi^2_\mathrm{red}$ for the entire \textit{IRIS} FUV window around 1400\,\AA. This value is listed for each type of line profile fit in the top panels of Fig. \ref{Fig:Fits}.

We note that the blended \ion{O}{4} line around 1404.82\,\AA~is also fitted at the sametime and in the same manner as other unblended lines. This is done since the blending \ion{O}{4} and \ion{S}{4} transitions are located close in wavelength \citep[see Appendix A of][]{Polito16b} as well as for completeness. Not including this strong line could lead to difficulties with the determination of $I_\mathrm{BG}$ and in turn affect the fit parameters for other lines.

In addition, we included the weak \ion{S}{1} 1401.5\,\AA~line in the far red wing of the \ion{O}{4} 1401.2\,\AA~line, as well as the \ion{Fe}{2} 1405.6\,\AA~line that can be recognized above the pseudo-continuum. In practice, we find that the inclusion of these two lines only has a small impact on the $\chi^2_\mathrm{red}$ value. Similarly, adding other weak lines (some of unknown origin) does not improve the $\chi^2_\mathrm{red}$, and we decided not to include these weak line-like components to fit the small ``lumps and bumps'' above the pseudo-continuum. The influence of these line-like components on the overall $\chi^2_\mathrm{red}$ is further discussed in Appendix \ref{Sect:B}.

We note that the $\chi^2_\mathrm{red}$ has been criticized as a goodness-of-fit measure for non-linear functions \citep{Andrae10}, since the degrees of freedom may not be properly defined using the formula mentioned above. However, we found that the lower the $\chi^2_\mathrm{red}$ obtained, the better the overall fit to the \textit{IRIS} FUV spectrum. This is confirmed by the behavior of the residuals $I_\mathrm{obs}(\lambda_i) - I_\mathrm{fit}(\lambda_{i})$, which are supplied for each fit in this Section. 

\subsubsection{Single and double Gaussian fits}
\label{Sect:4.2.1}

We first tried to fit the observed line profiles using single and double Gaussian components. We note that this is a standard fitting procedure for spectral lines. The results of the fitting are shown in Fig. \ref{Fig:Fits}, which is organized as follows: Each of the five individual observed lines is shown in a given row in black color, with the corresponding error bars in each spectral bin. Colors denote individual fitting components, with the single and double Gaussian fits being shown in the first two columns, while the fitting with a $\kappa$-distribution is shown in the third and fourth columns. Individual parameters obtained from the fitting are listed in each panel, with errors given by the fit shown in parentheses. Residuals $I_\mathrm{obs} - I_\mathrm{fit}$ are also shown at the bottom of each panel. 

We see that the single-Gaussian fitting does a poor job of approximating the peaks of all lines and also the far wings. The situation is the worst for the strongest line, i.e., \ion{Si}{4}, where the peak is underestimated by 200--400 DN. The far wings are also underestimated. The overall $\chi^2_\mathrm{red}$ is 21, indicating poor fit. We however note that single-Gaussian fitting for these non-Gaussian profiles does not show large uncertainties in the width as reported by \citet[][Sect. 2.3.4 therein]{Akiyama05}. We also note that the $w_\mathrm{G}$ obtained from the fitting is nearly the same for all five TR lines, including the allowed \ion{Si}{4} line. This is contrary to most results reported in literature (but see \citeauthor{Polito16b} \citeyear{Polito16b} and Fig. 8 of \citeauthor{Doschek16} \citeyear{Doschek16}).

Double Gaussian fits do a much better job at approximating the observed line profiles. Peaks and far wings of all lines are approximated well, except for the peak of \ion{Si}{4}, where the largest residual still reaches about 200\,DN. The overall $\chi^2_\mathrm{red}$ has decreased to 4.1. Adding a third Gaussian component into the \ion{Si}{4} line would remove this discrepancy, and lower the $\chi^2_\mathrm{red}$. However, this would also mean that we require 9 free parameters to fit a single unblended line profile.

We note that in this fitting we did not assume the same value of $\lambda_0$ for both Gaussian components. Tying the $\lambda_0$ of both Gaussian components to a single value does not improve the goodness-of-fit in terms of the $\chi^2_\mathrm{red}$ for the double Gaussian fitting. Furthermore, the $\lambda_0$ obtained for the narrow and broader Gaussian do not differ by more than about a half of the \textit{IRIS} wavelength bin, i.e., $\approx$13 m\AA. This is not surprising, since the five TR lines are all symmetric (Sect. \ref{Sect:4.1}).

Finally, calculating the intensities of each Gaussian component (Eq. \ref{Eq:total_intensity_Gauss}), we found that the relative contribution of the broader second Gaussian to the total line intensity is about 30\% for \ion{Si}{4}, in agreement with the highest values reported by \citet{Peter01}, occurring in the middle transition region. However, for \ion{O}{4}, we find higher contributions of the second Gaussian, 39\%, 50\%, and 48\% for the \ion{O}{4} 1399.78\,\AA, 1401.16\,\AA, and 1404.82\,\AA~(bl), respectively. For the \ion{S}{4} line at 1406.06\AA, the second Gaussian contributes 58\% of its intensity. Such high values have not been reported in literature previously. Although the significance of these numbers is questionable for the weaker \ion{O}{4} and \ion{S}{4} lines because of the larger relative uncertainties in both the $I_\mathrm{G2}(\lambda_0)$ and $w_\mathrm{G2}$ compared to the narrow Gaussian, we find that this is not the case for the strongest \ion{O}{4} line at 1401.2\,\AA. Such a large contribution of the broad Gaussian to the total intensity, together with the intensity originating in closed transition-region loop (Fig. \ref{Fig:IRIS_sg}) casts doubt on the interpretation that this broader component originates in a coronal funnel \citep{Peter99,Peter00,Peter01}. Alternative interpretations for the two Gaussian components could however still be possible, but we do not engage in such speculations further.

\subsubsection{Fits with a $\kappa$-distribution}
\label{Sect:4.2.2}

We next fitted the lines with a $\kappa$-distribution (Eq. \ref{Eq:kappa_fit}). The results are shown in red color in the third column of Fig. \ref{Fig:Fits}. We see that the $\kappa$-distribution fits most of the profiles well, including the peaks and far wings. The total $\chi^2_\mathrm{red}$ = 3.1 is lower than for the double-Gaussian fit. This is in spite of the fact that a $\kappa$ distribution requires less free parameters than a double Gaussian fit. It is however not surprising given that the $\kappa$-distributions, or indeed any distributions broader than a Maxwellian, can be approximated by a sum of several Maxwellians \citep{Hahn15}. The lower $\chi^2_\mathrm{red}$ value comes in particular from a better match to the \ion{Si}{4} line center produced by the $\kappa$-fit. There, the highest residuals are about 150\,DN, and occur in fewer pixels than in the double Gaussian case. The $\chi^2_\mathrm{red}$ is still however larger than unity. In Appendix \ref{Sect:A}, we show that this is not due to optical thickness effects within the \ion{Si}{4} line, while in Appendix \ref{Sect:B} it is shown that the residuals in the \ion{Si}{4} line contribute only about 0.31 to the total $\chi^2_\mathrm{red}$, and that it is the pseudo-continuum that dominates the $\chi^2_\mathrm{red}$ value.

The values of $\kappa$ obtained from the line fitting are low, of about 1.9--2.1, see Table \ref{Table:2}. For the \ion{O}{4} 1401.2\,\AA~line, we obtain $\kappa$\,=\,2.35\,$\pm$\,0.08, higher than for other lines. The cause of this is uncertain. It is possible that the $\kappa$ and $w_\kappa$ are coupled, and we found that the $w_\kappa$ for this line is lower compared to other lines. To test this, we tried to arbitrarily restrict $w_\kappa$ to $w_\kappa \geq $\,0.14, i.e., to be larger than the lowest value found for other lines (Table \ref{Table:2}). Doing so, we found $\kappa$\,=\,2.09$\pm$\,0.07 for this \ion{O}{4} 1401.2\,\AA~line, and a slightly worse $\chi^2_\mathrm{red}$\,=\,3.2. Such value of $\kappa$ is in good agreement with the values obtained from other four TR lines. This exercise indicates that a restricted value of one of the $\kappa$ and $w_\kappa$ parameters could be mitigated by changes in the other one; however, at the expense of the $\chi^2_\mathrm{red}$. We note that the $\chi^2_\mathrm{red}$ is the quantity is minimized by the fitting procedure and we did not a priori enforce the same value of individual parameters for all five TR lines.

The values of $\kappa$ obtained here are low, located in the far-equilibrium thermodynamic region \citep[see][]{Livadiotis10,Livadiotis13}, i.e., near the extreme lower limit of $\kappa$\,$\to$\,3/2 for physically realizable $\kappa$-distributions. A similar value was found for a transient coronal loop by \citet{Dudik15}. \citet{Jeffrey17} obtained similarly low $\kappa$ values in flare loops, albeit they used the $\kappa$-distribution of the first kind, meaning that their $\kappa^*$ values correspond to our $\kappa$\,+\,1 (see Sect. \ref{Sect:2.1}). Such low $\kappa$\,$\approx$\,2 values are interesting, since if the line profile is given by the ion motions, then the energetic particles manifested in line wings would carry $\approx$80\% of total energy contained in the $\kappa$-distribution \citep[][Fig. 1b therein]{Oka13}. Finally we note that the discussion on the characteristic widths $w_\kappa$ obtained from the fitting, the contribution of non-thermal broadening, as well as the quantities derived from $w_\kappa$ are presented in Sect. \ref{Sect:7.3}.

%
\subsubsection{Inclusion of the \textit{IRIS} instrumental profile}
\label{Sect:4.2.3}

To verify whether the \textit{IRIS} instrumental profile influences the fitting results, we repeated the $\kappa$ fits by including the instrumental profile via its convolution with the $\kappa$ profile \citep[see Eq. 6 in][]{Jeffrey16}. Since only the FWHM$_\mathrm{instr}$ is known for the \textit{IRIS} instrument \citep[25.85\,m\AA~for the FUV channel,][]{DePontieu14}, but not its shape, we considered two extreme cases: Gaussian ($\kappa$\,$\to$\,$+\infty$) and Lorentzian (formally corresponding to $\kappa$\,=\,$-1$). We note however that the FWHM$_\mathrm{instr}$ is nearly the same as the size of a single wavelength pixel (with binning) in the present study; therefore, the influence of the instrumental profile is expected to be small.

This is indeed what we found. However, assuming either Gaussian or Lorentzian shapes of the instrumental profile does not improve upon the $\kappa$ fit. The Lorentzian leads to increased spread of $\kappa$ varying between 1.7--2.5 depending on the line, with overall $\chi^2_\mathrm{red}$\,=\,3.2. For the Gaussian instrumental profile, both the the $\chi^2_\mathrm{red}$ and the resulting values of $\kappa$ do not change appreciably compared to the purely $\kappa$-fit (Fig. \ref{Fig:Fits}, \textit{right column}). The only change are the decreased characteristic widths $w_\kappa$. However, since the true \textit{IRIS} instrumental profile is not known at present, in the remainder of this work, we use the purely $\kappa$-fit (Sect. \ref{Sect:5}) and the parameters derived from it (Sect. \ref{Sect:7} and Appendix \ref{Sect:A}). In Sect. \ref{Sect:5.2} we further investigate the effects of the instrumental profile providing additional evidence that it is unlikely to explain the observed non-Gaussian profiles.

%
\begin{figure*}[!t]
	\centering
	\includegraphics[width=0.75\textwidth]{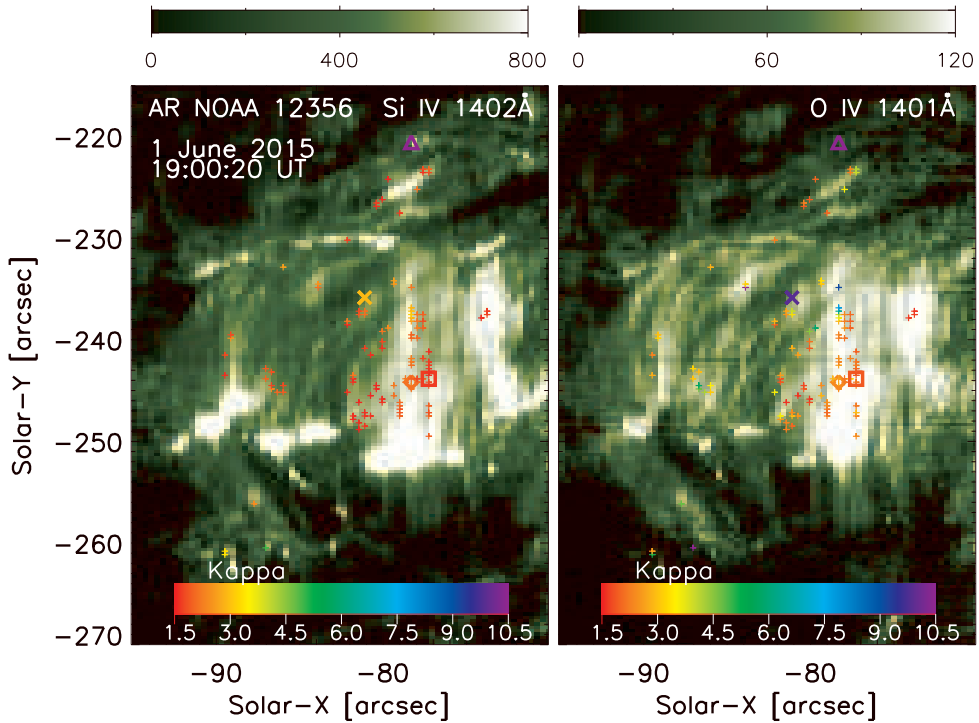}	
\caption{Spatial distribution of the $\kappa$ values obtained from the fitting of the \ion{Si}{4} 1402.8\,\AA~line (left) and the \ion{O}{4} 1401.2\,\AA~line (right). The small cross symbols overlaid on the Si IV and O IV images indicate the location of suitable pixels, with their color denoting the $\kappa$ values obtained from line profile fitting. The corresponding $\kappa$ color-bar is shown in the bottom part of each panel. The larger and thicker diamond, square, triangle and $\times$ symbols indicate the position of the pixels where we take the TR spectrum analyzed in Sect. \ref{Sect:5}. The corresponding spectra at these locations are shown in Fig. \ref{Fig:Fits_pixels}.
\label{Fig:IRIS_sg_fits}}
\end{figure*}
%
%
\begin{figure*}[!t]
	\centering
	\includegraphics[width=4.98cm,clip,bb= 0 52 487 340]{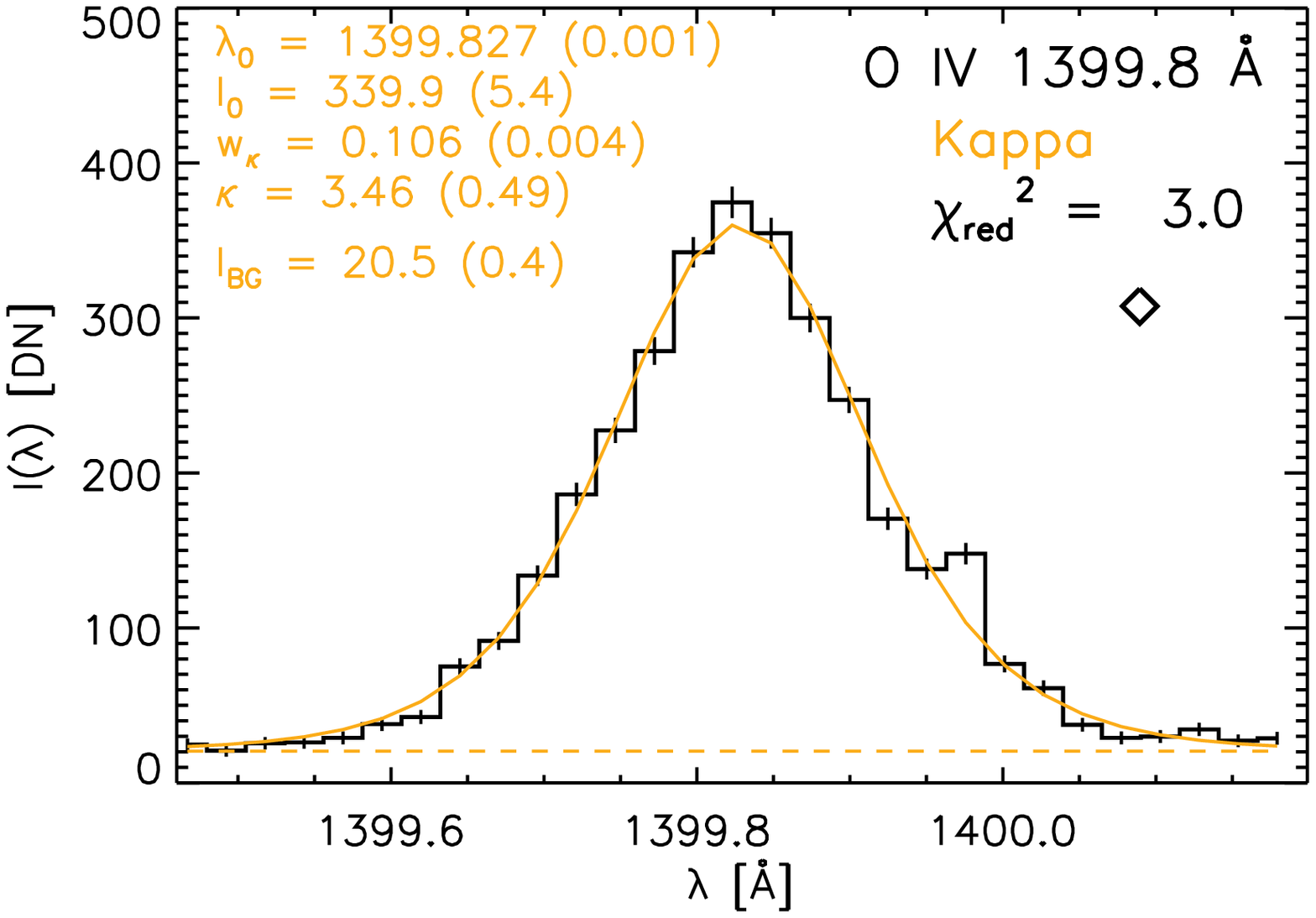}
	\includegraphics[width=4.21cm,clip,bb=75 52 487 340]{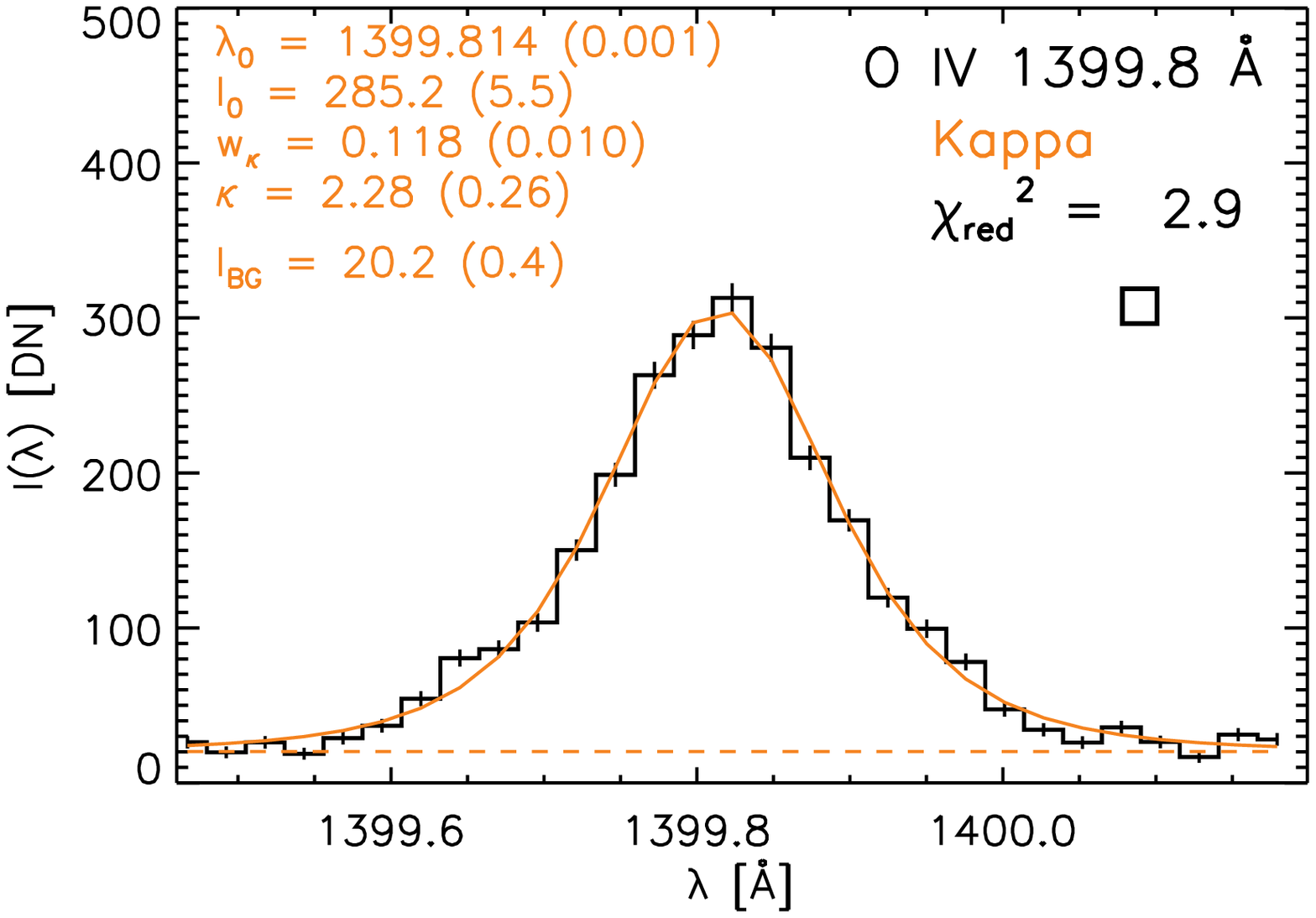}
	\includegraphics[width=4.21cm,clip,bb=75 52 487 340]{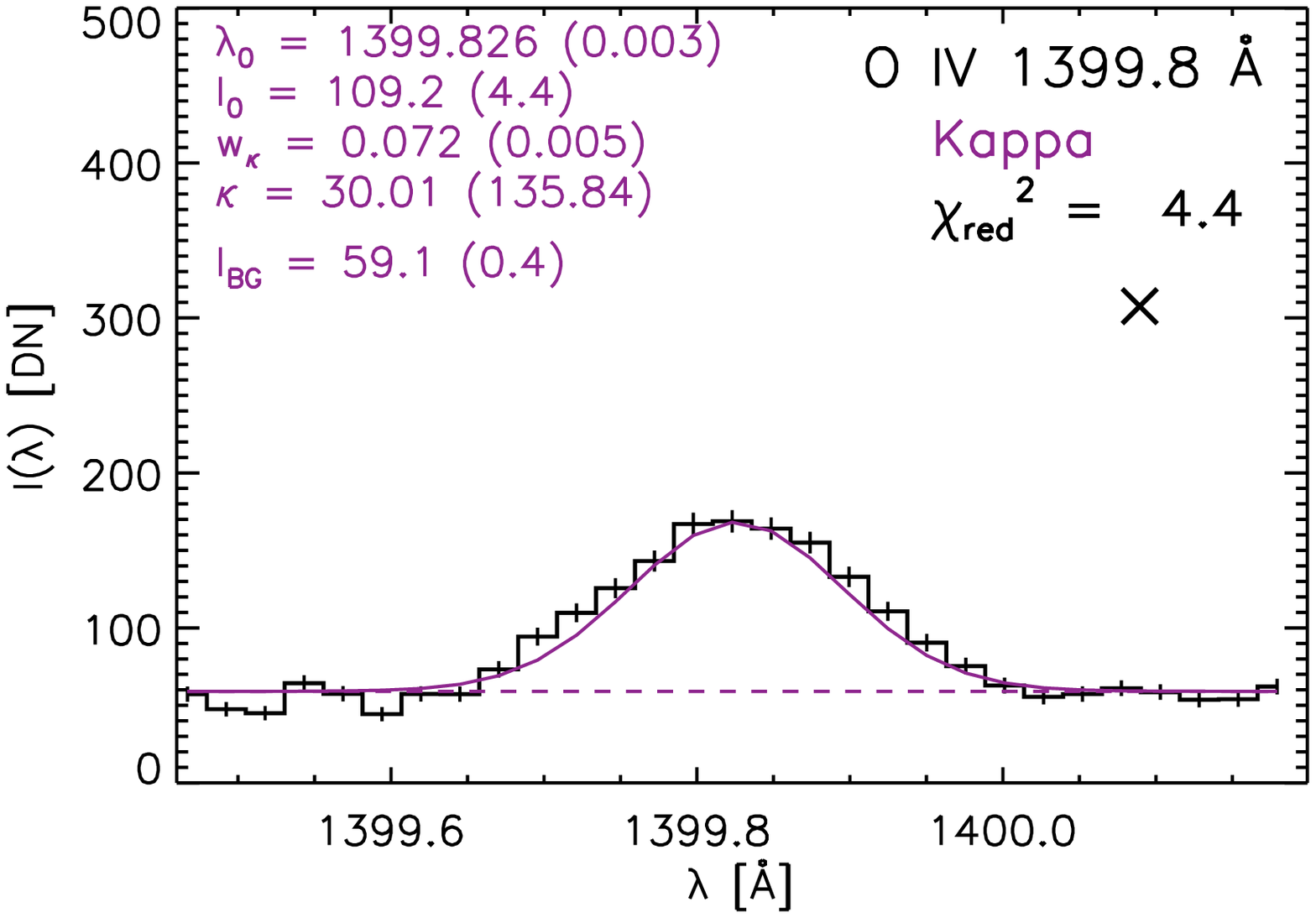}
	\includegraphics[width=4.21cm,clip,bb=75 52 487 340]{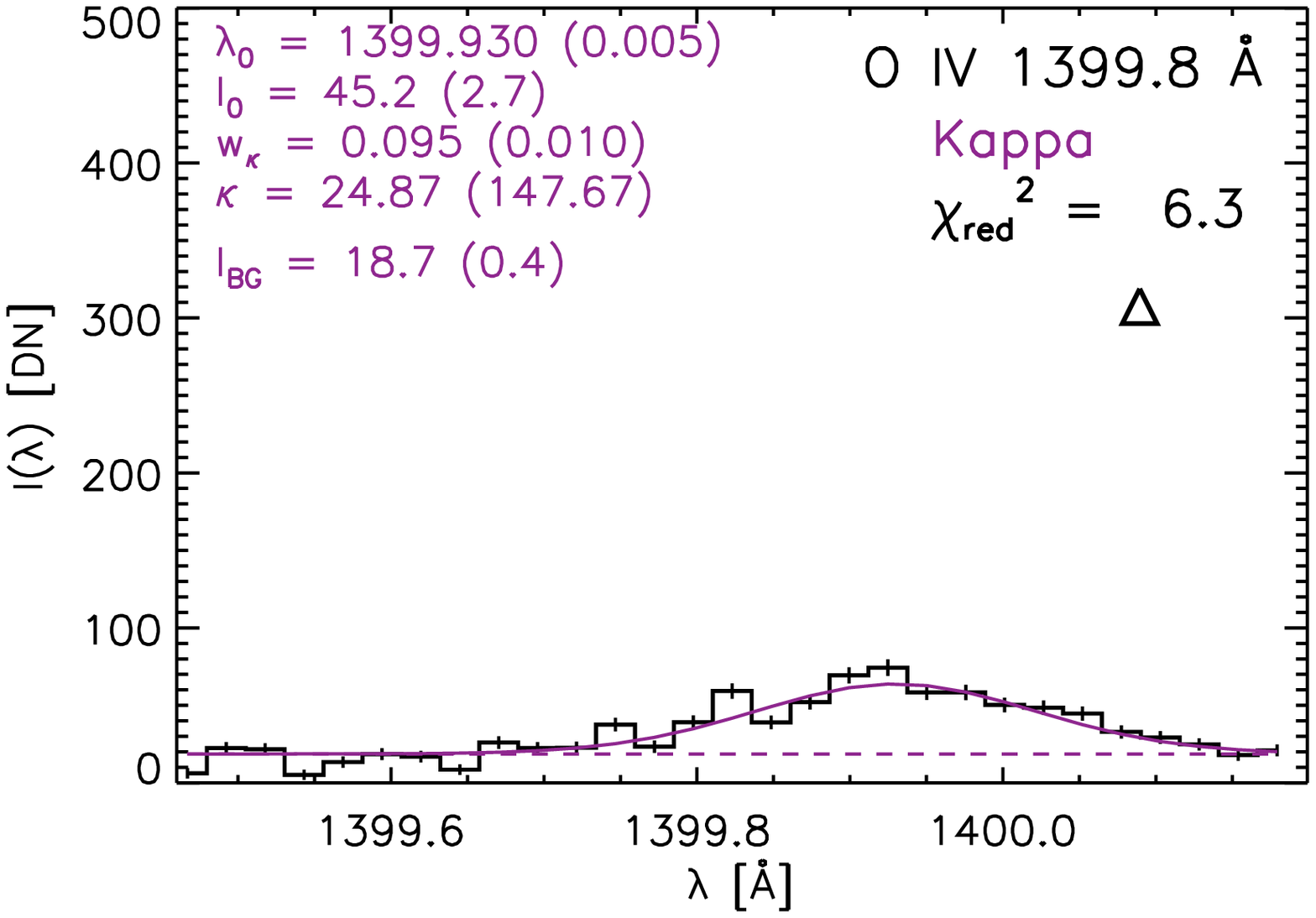}
	\includegraphics[width=4.98cm,clip,bb= 0  0 487 145]{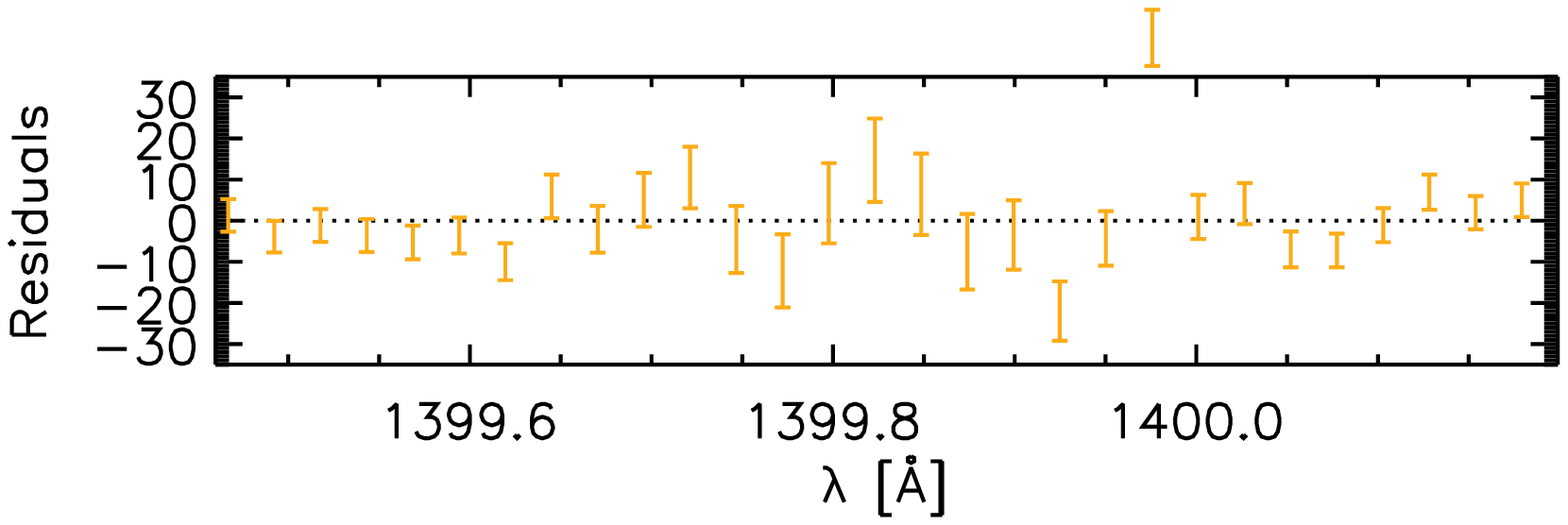}
	\includegraphics[width=4.21cm,clip,bb=75  0 487 145]{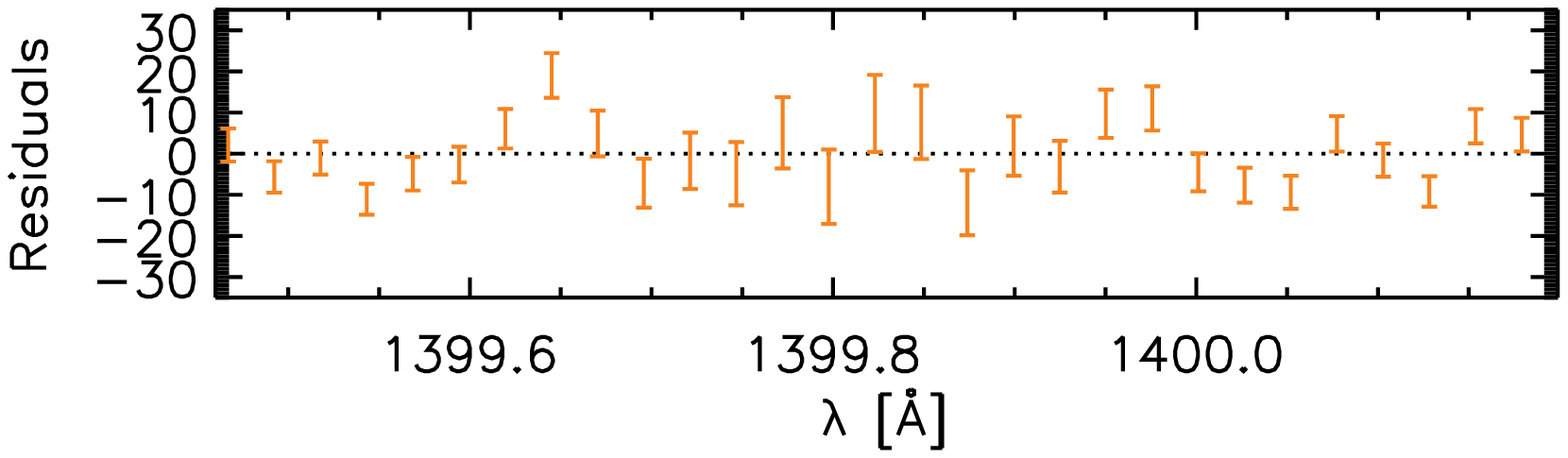}
	\includegraphics[width=4.21cm,clip,bb=75  0 487 145]{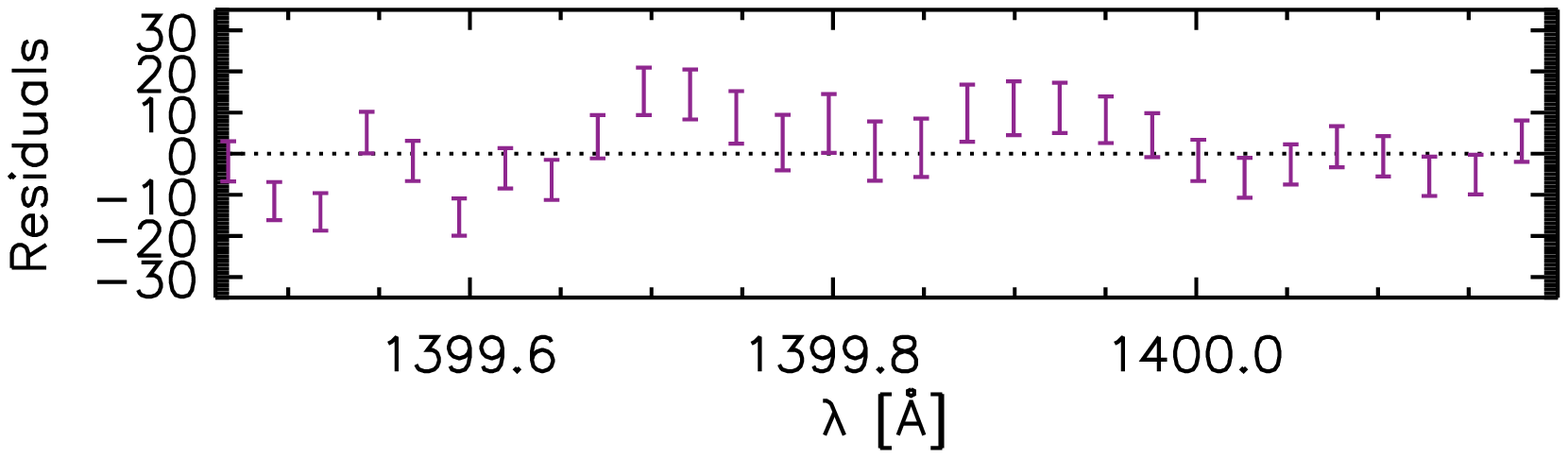}
	\includegraphics[width=4.21cm,clip,bb=75  0 487 145]{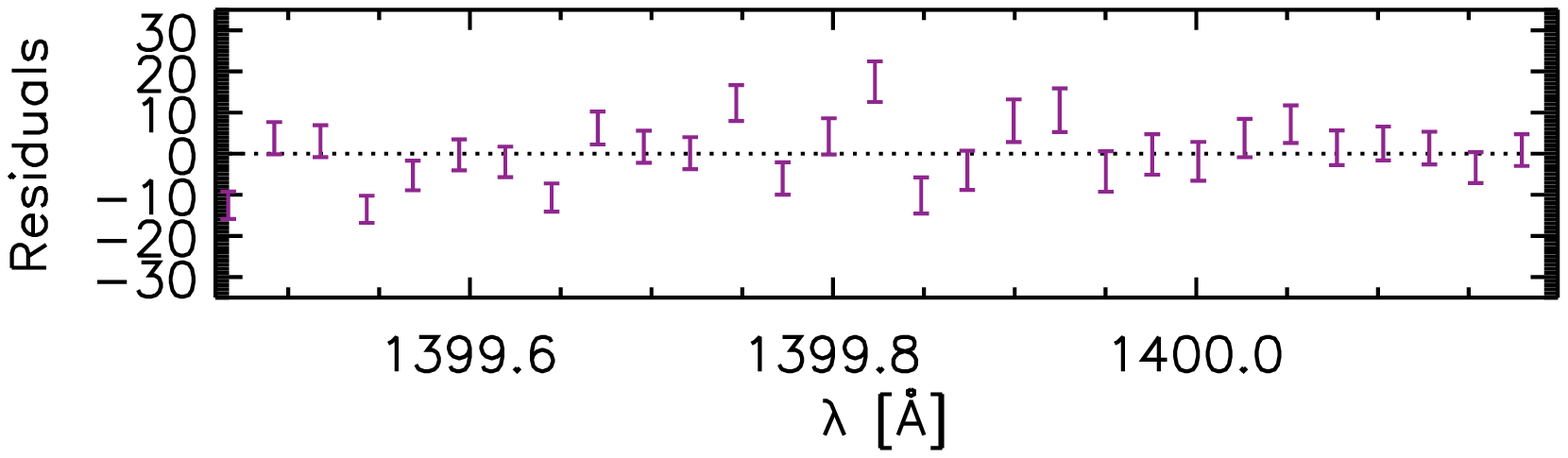}
	\includegraphics[width=4.98cm,clip,bb= 0 52 487 340]{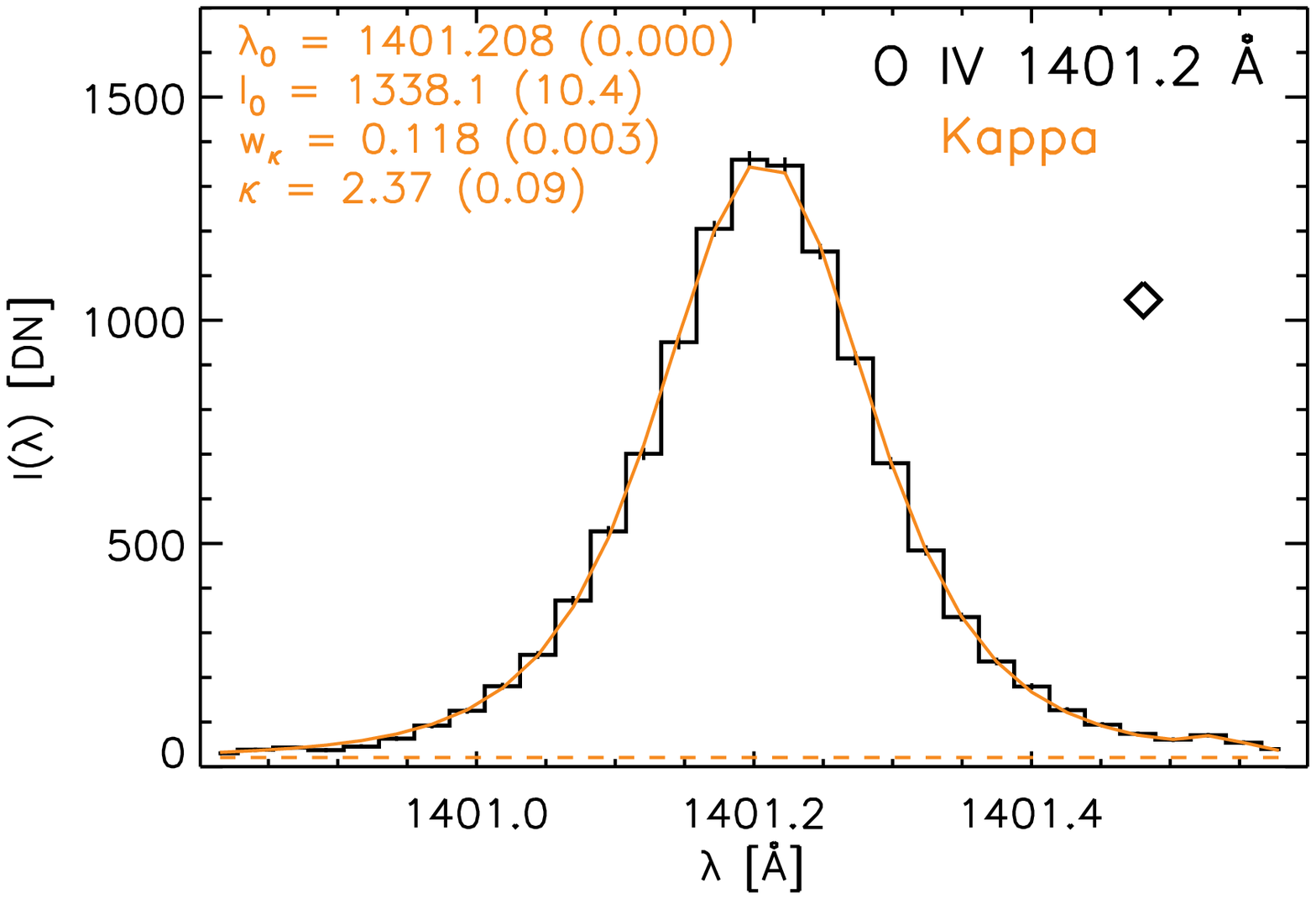}
	\includegraphics[width=4.21cm,clip,bb=75 52 487 340]{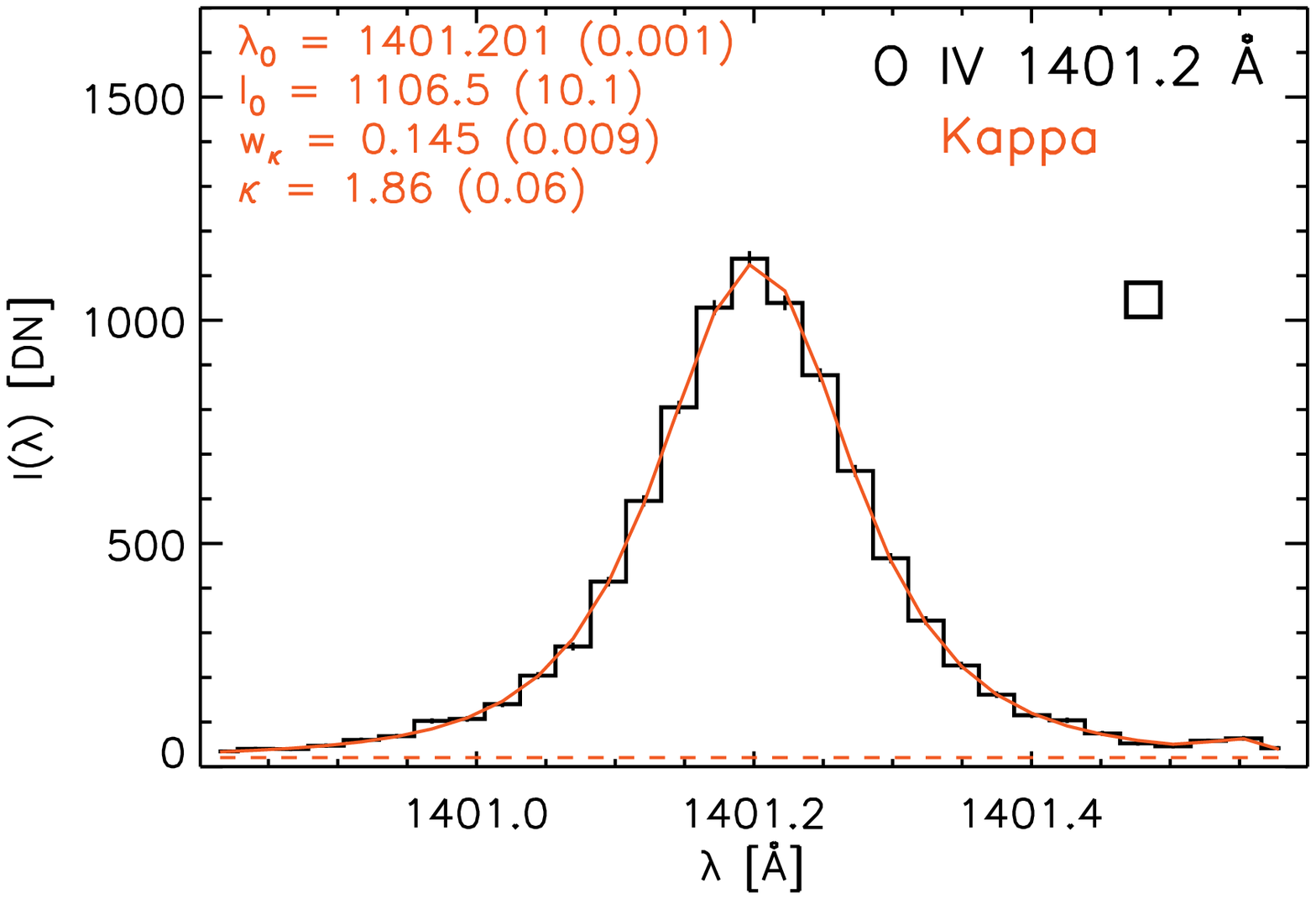}
	\includegraphics[width=4.21cm,clip,bb=75 52 487 340]{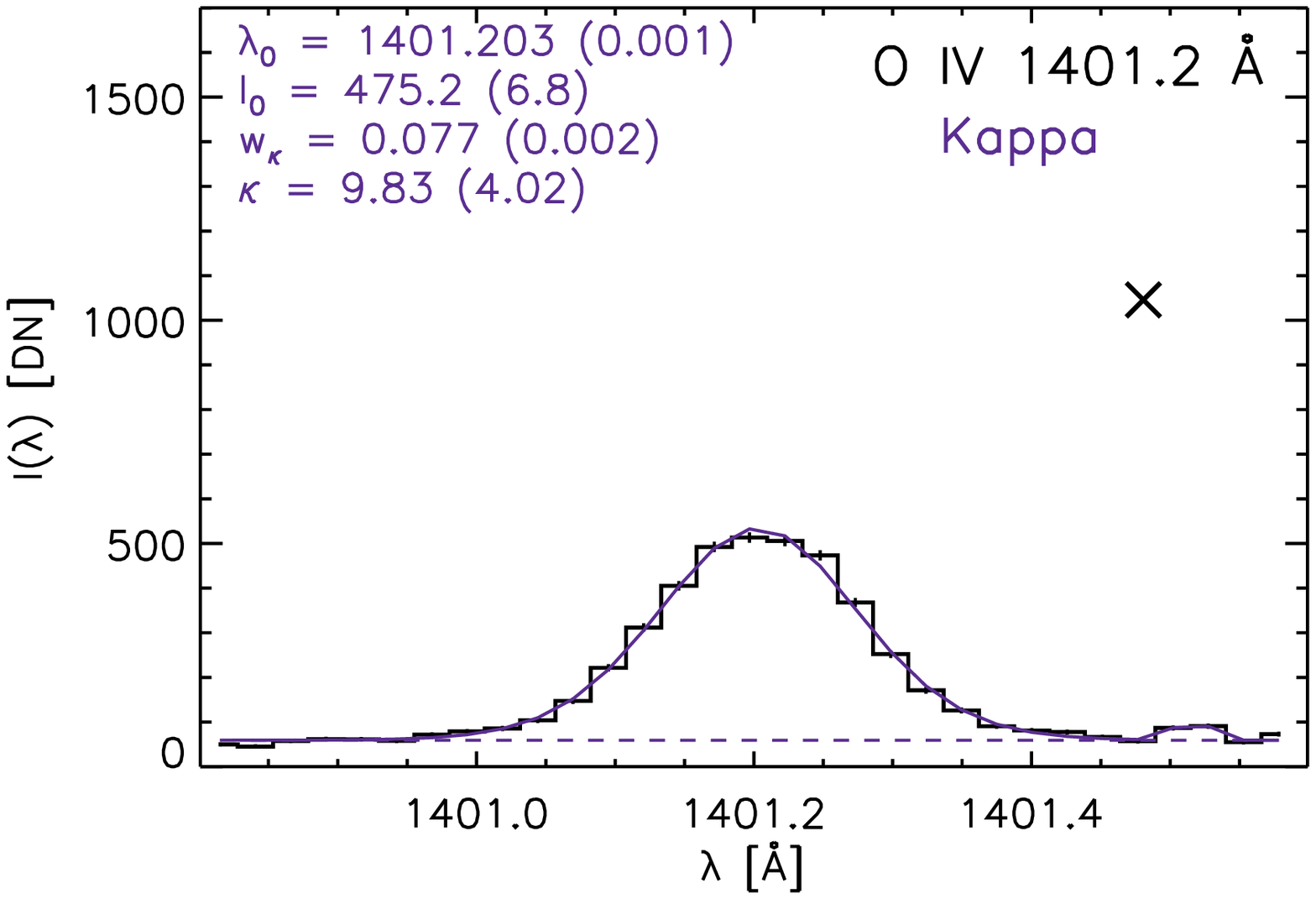}
	\includegraphics[width=4.21cm,clip,bb=75 52 487 340]{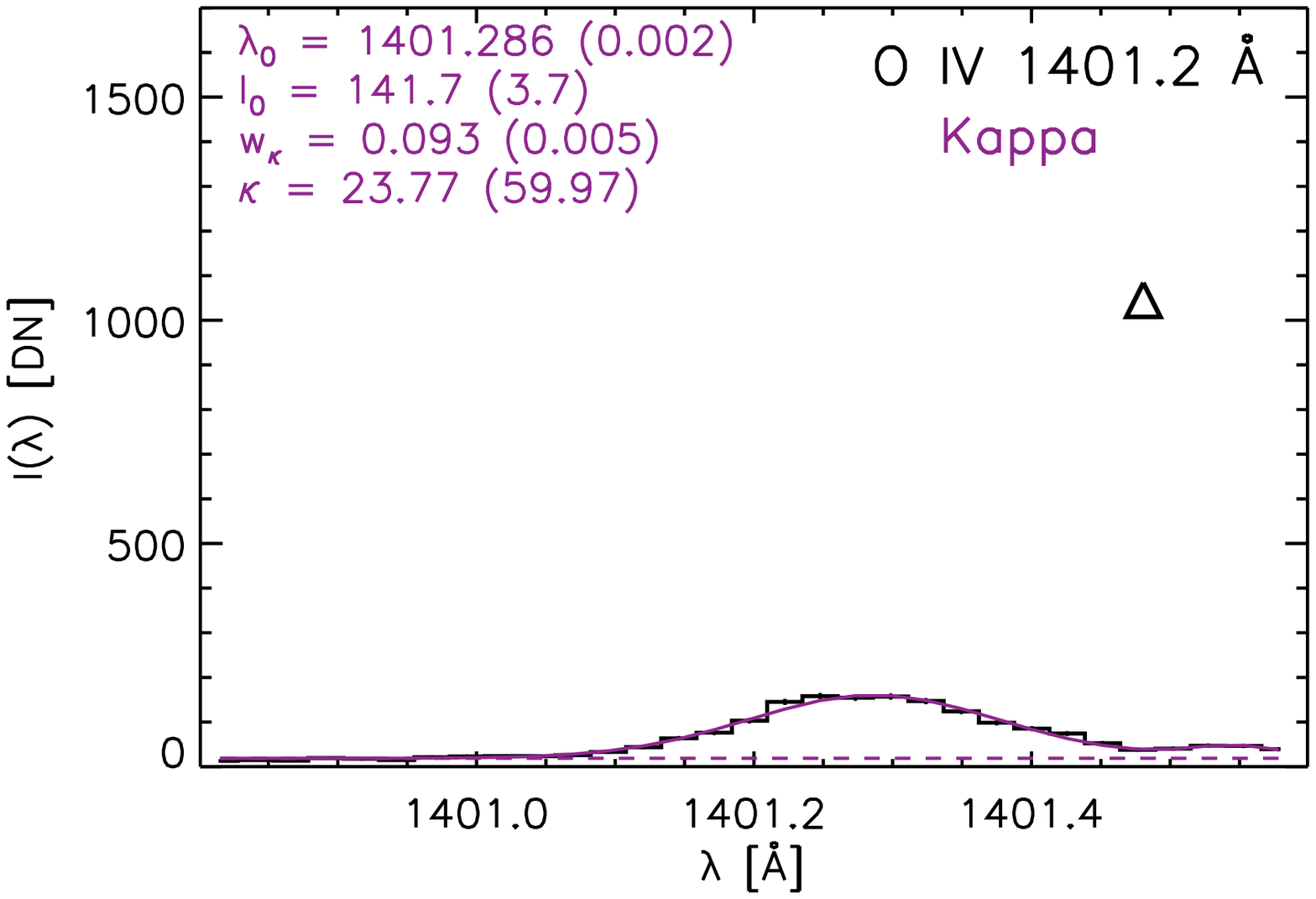}
	\includegraphics[width=4.98cm,clip,bb= 0  0 487 145]{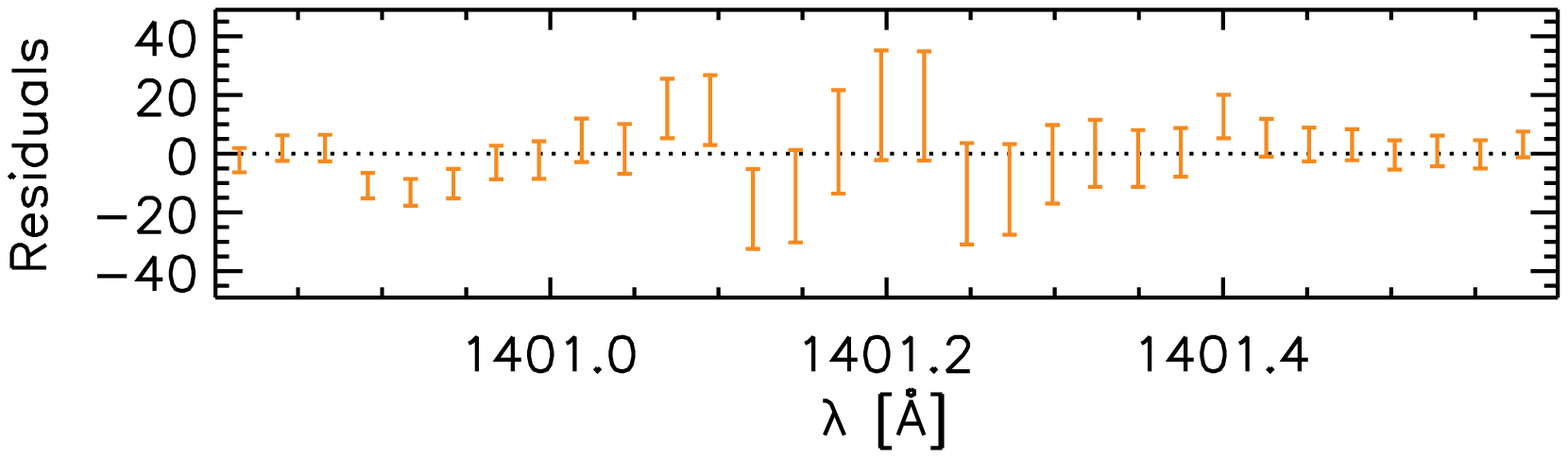}
	\includegraphics[width=4.21cm,clip,bb=75  0 487 145]{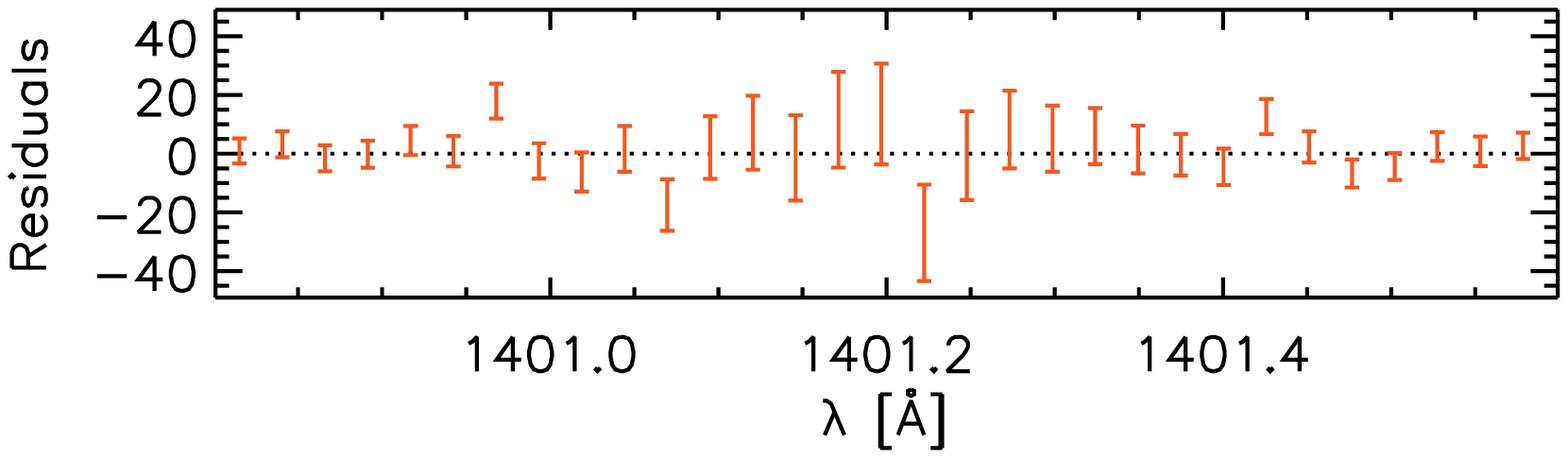}
	\includegraphics[width=4.21cm,clip,bb=75  0 487 145]{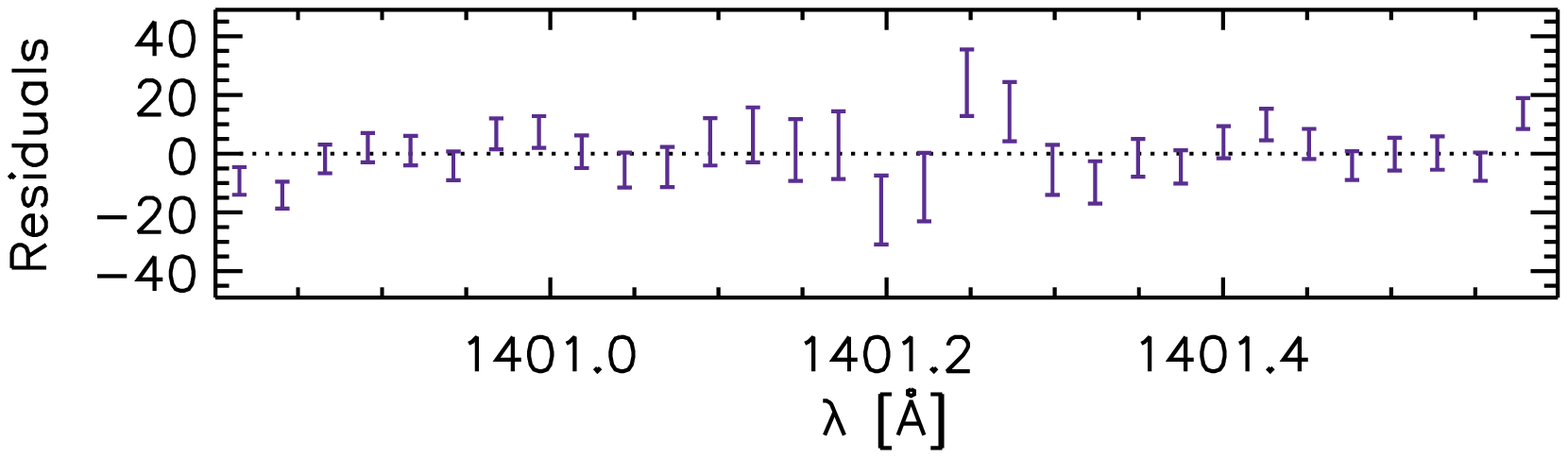}
	\includegraphics[width=4.21cm,clip,bb=75  0 487 145]{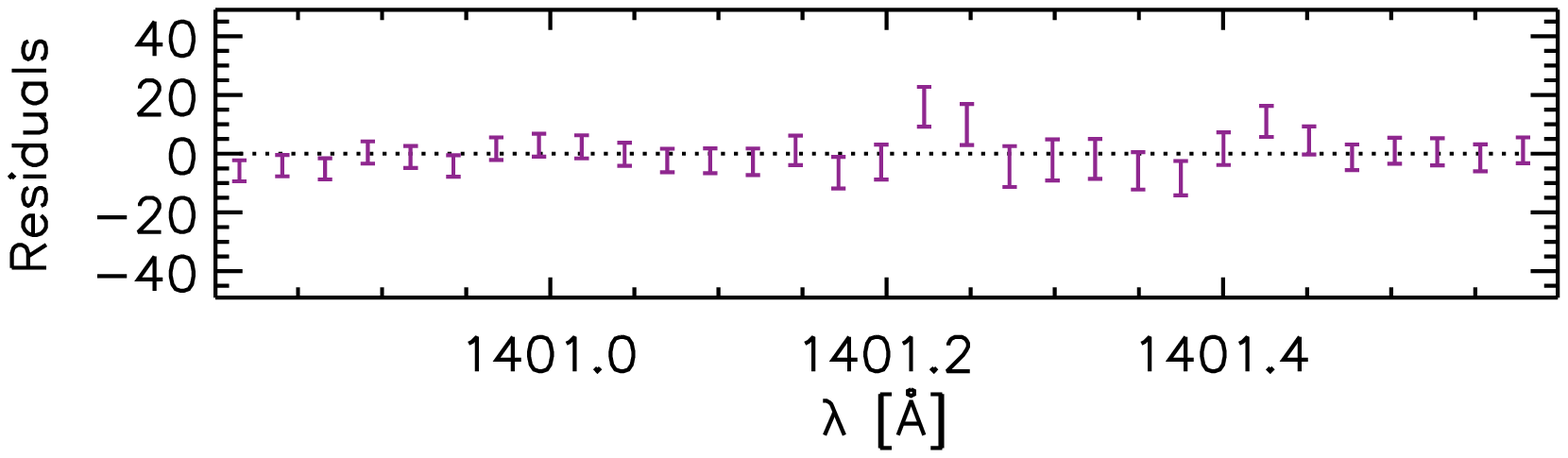}
	\includegraphics[width=4.98cm,clip,bb= 0 52 487 340]{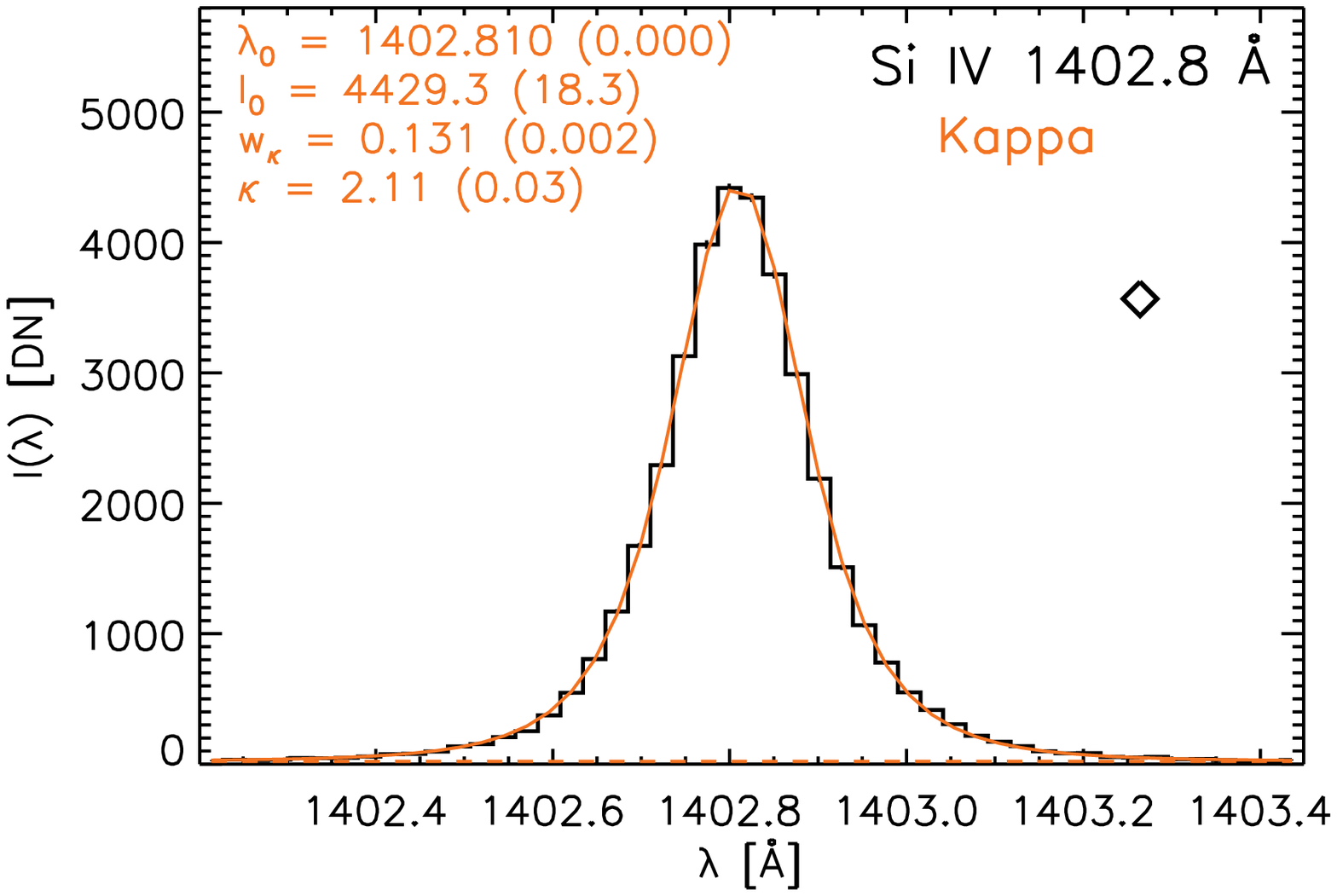}
	\includegraphics[width=4.21cm,clip,bb=75 52 487 340]{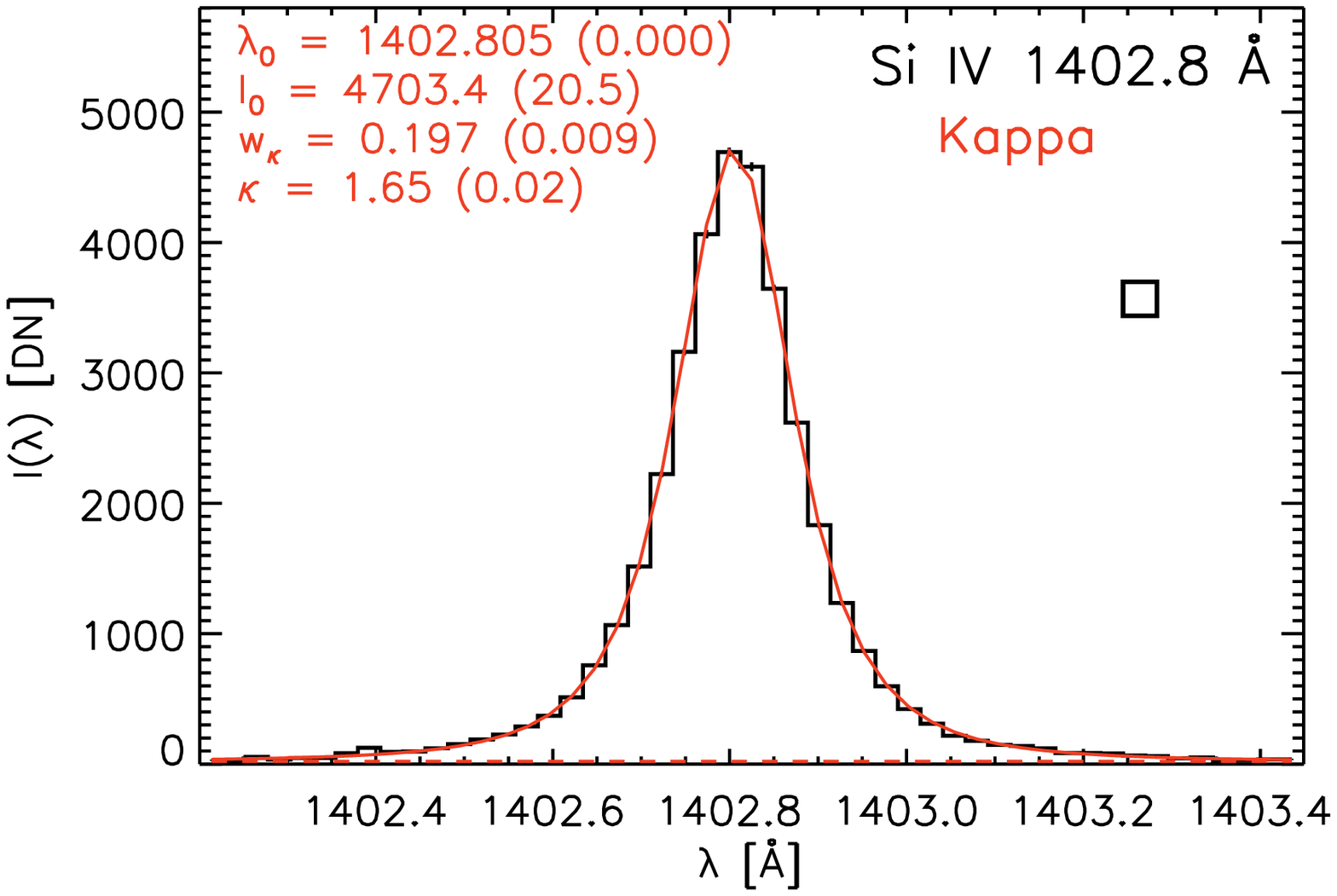}
	\includegraphics[width=4.21cm,clip,bb=75 52 487 340]{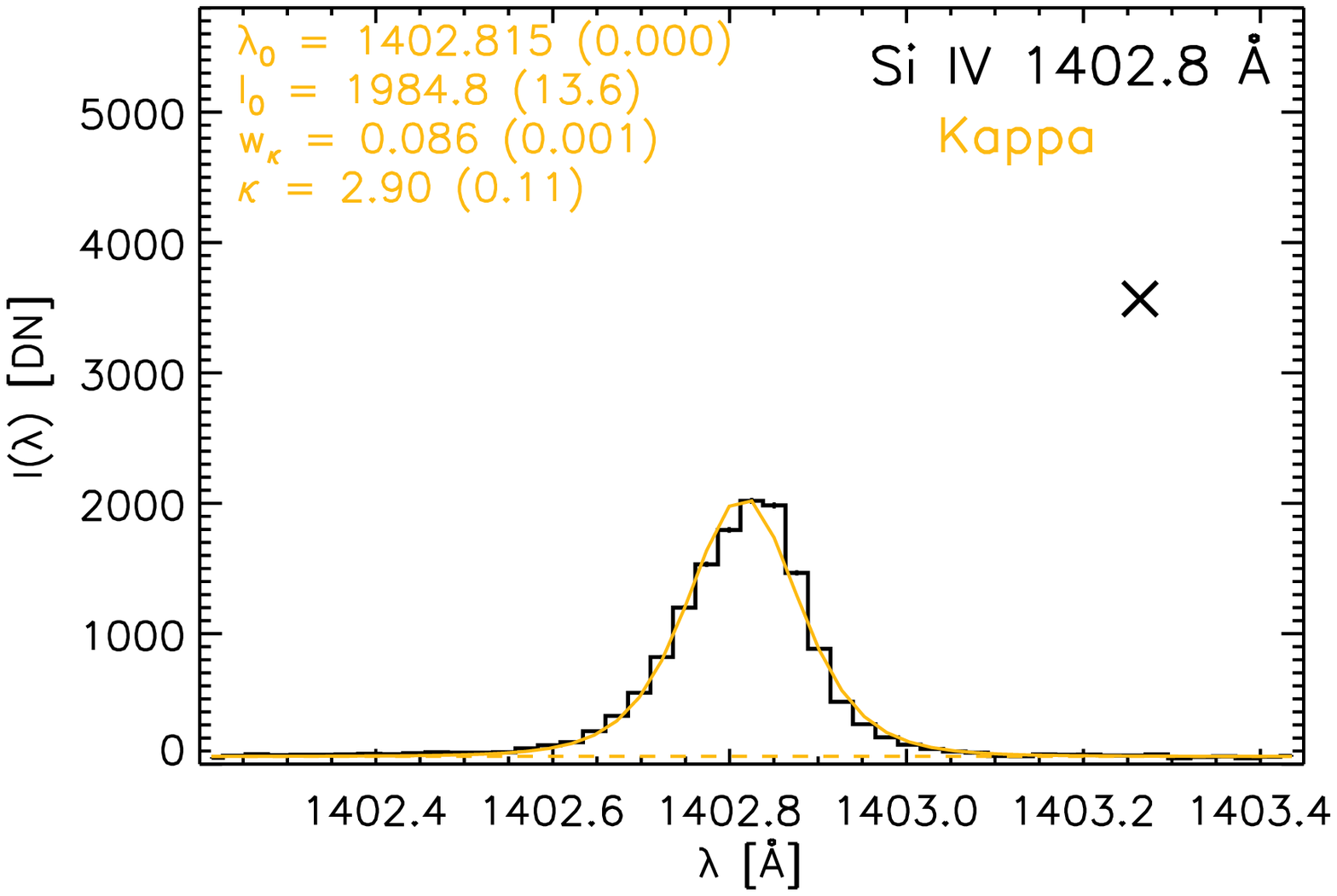}
	\includegraphics[width=4.21cm,clip,bb=75 52 487 340]{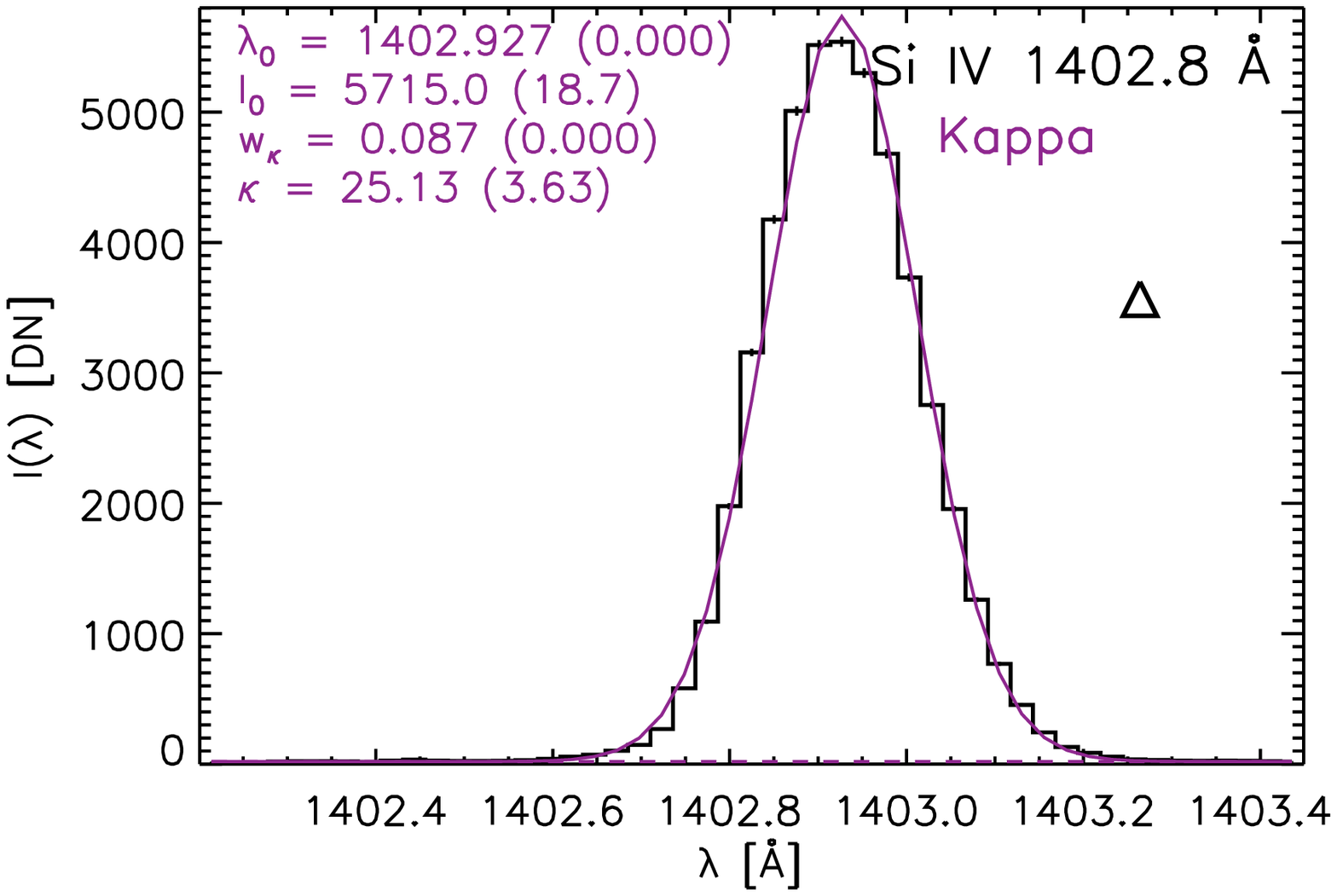}
	\includegraphics[width=4.98cm,clip,bb= 0  0 487 145]{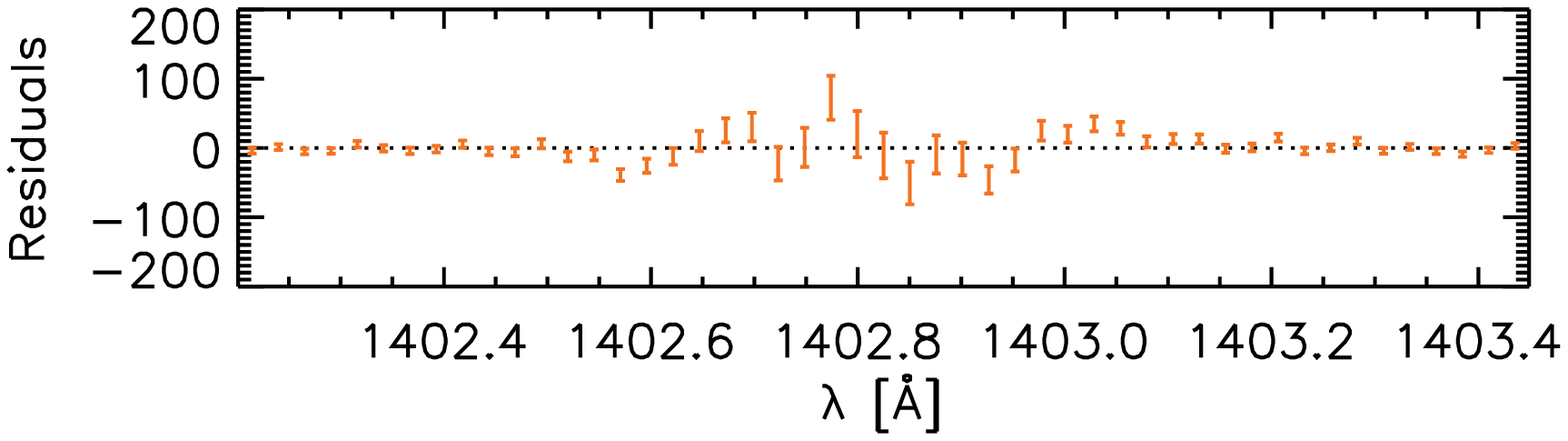}
	\includegraphics[width=4.21cm,clip,bb=75  0 487 145]{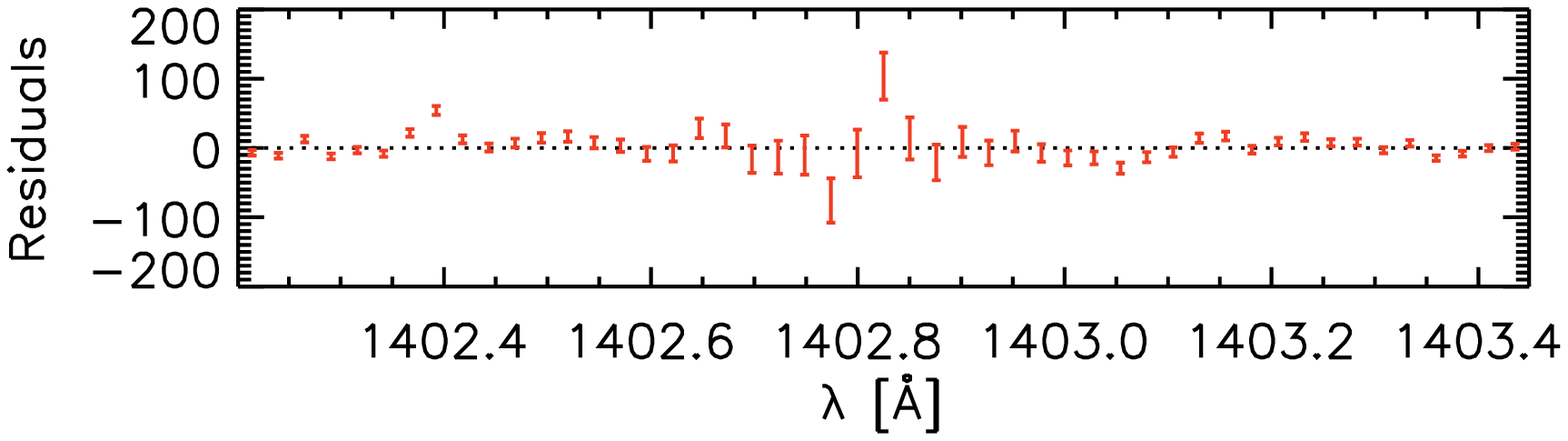}
	\includegraphics[width=4.21cm,clip,bb=75  0 487 145]{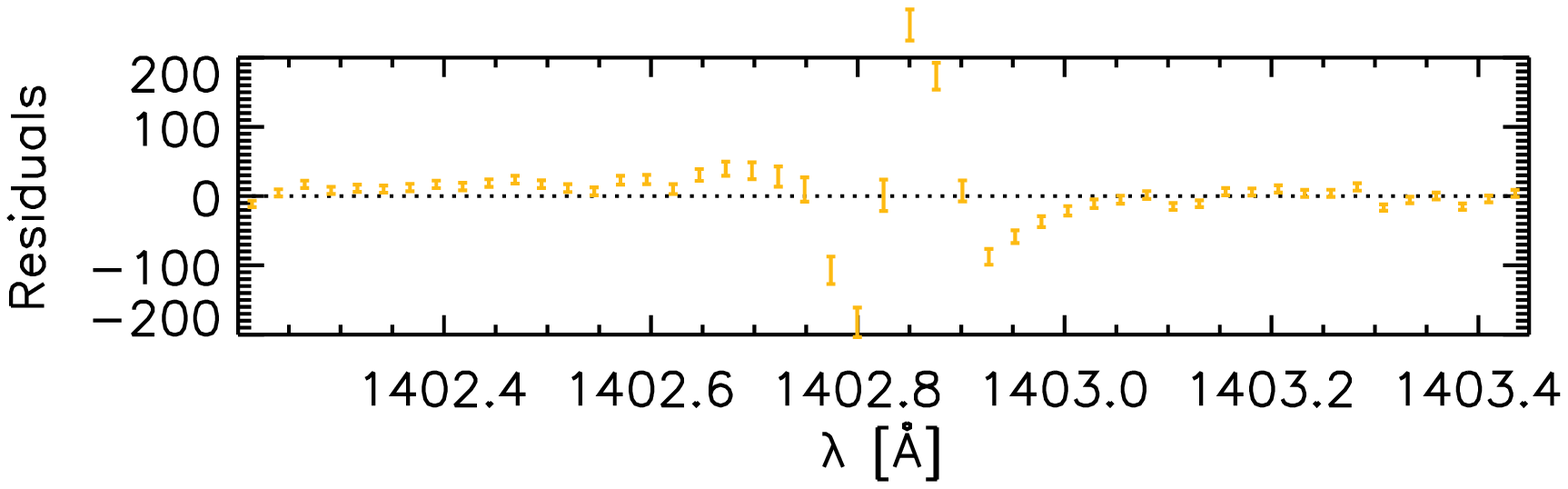}
	\includegraphics[width=4.21cm,clip,bb=75  0 487 145]{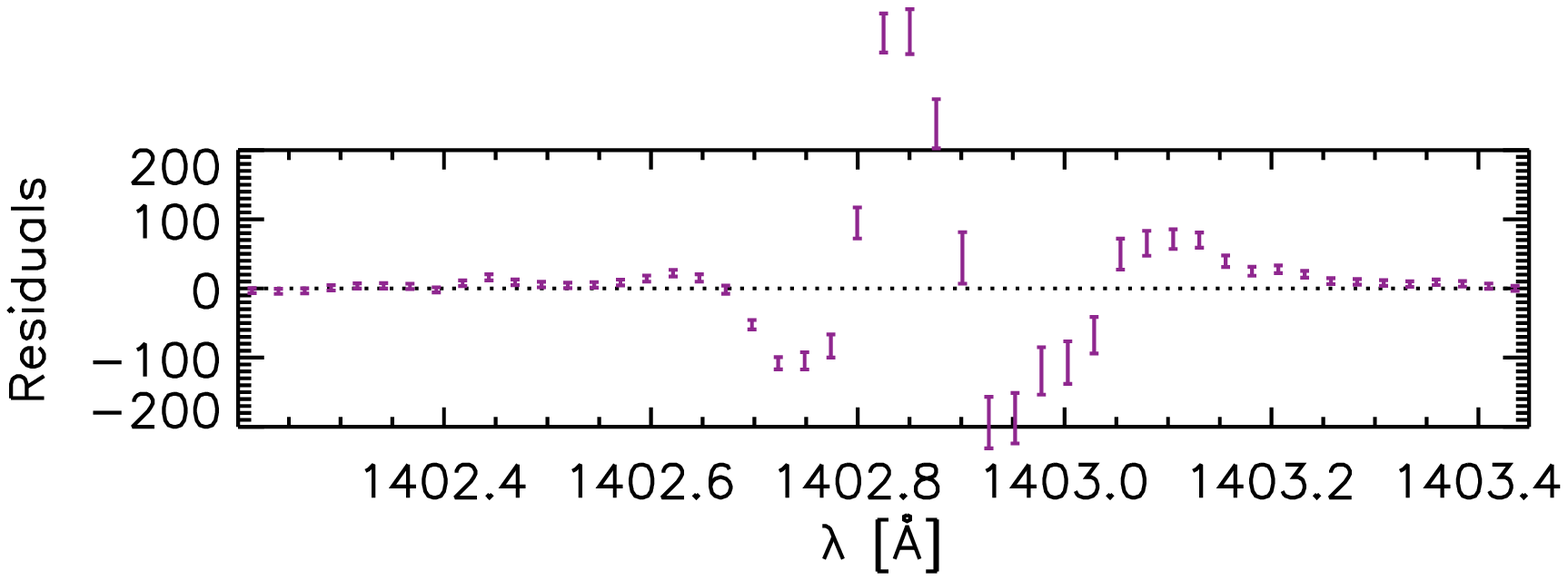}
	\includegraphics[width=4.98cm,clip,bb= 0 52 487 340]{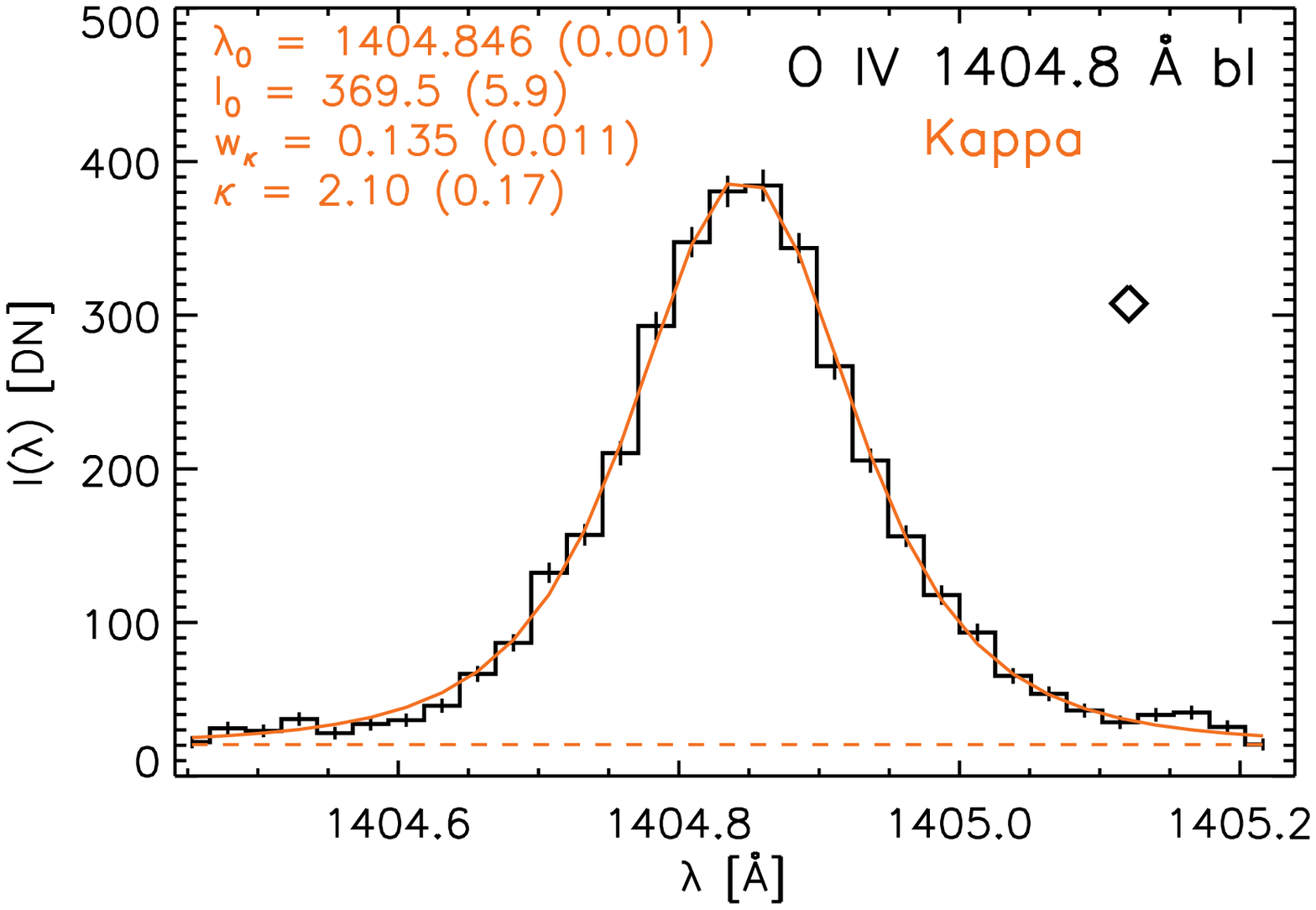}
	\includegraphics[width=4.21cm,clip,bb=75 52 487 340]{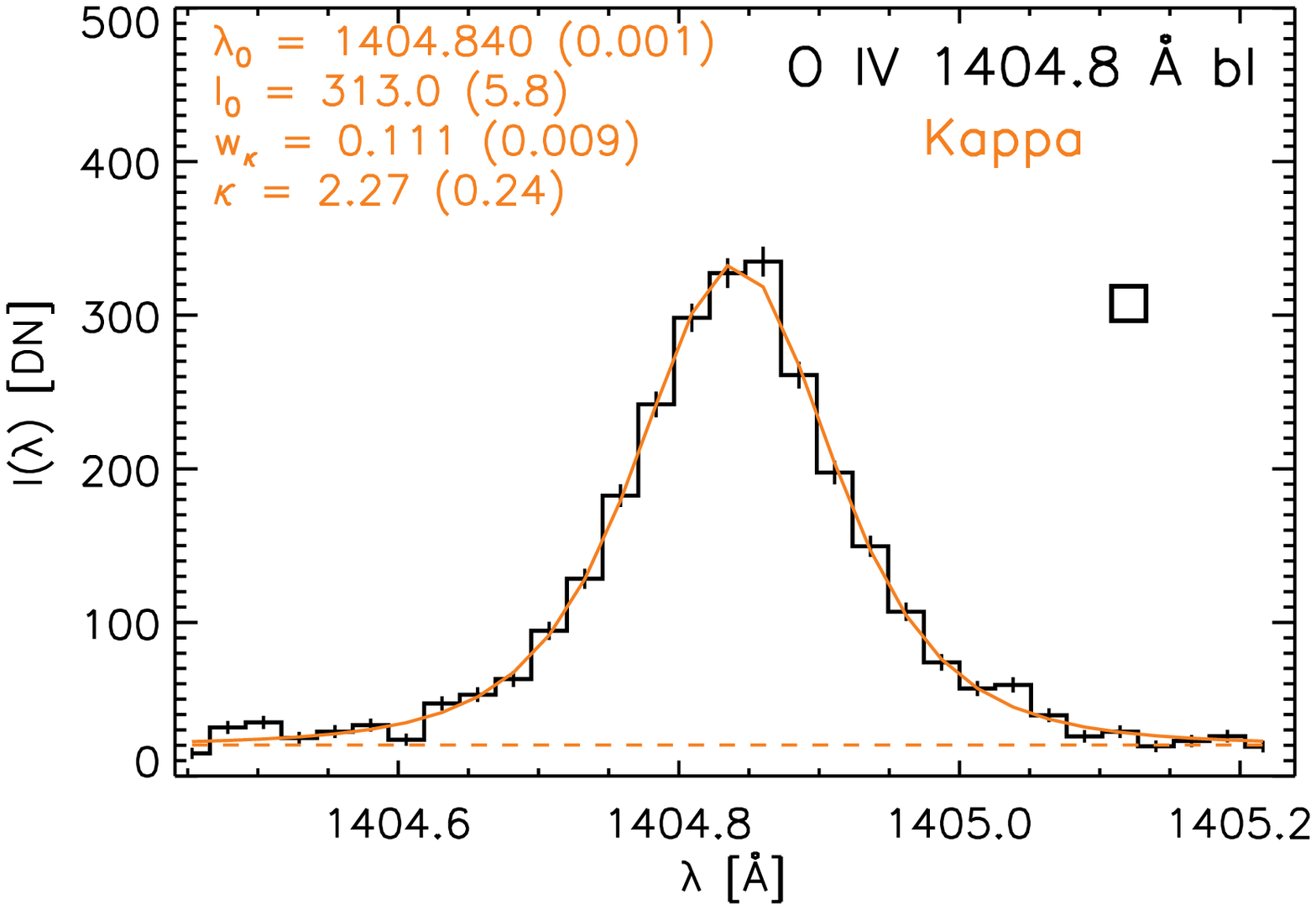}
	\includegraphics[width=4.21cm,clip,bb=75 52 487 340]{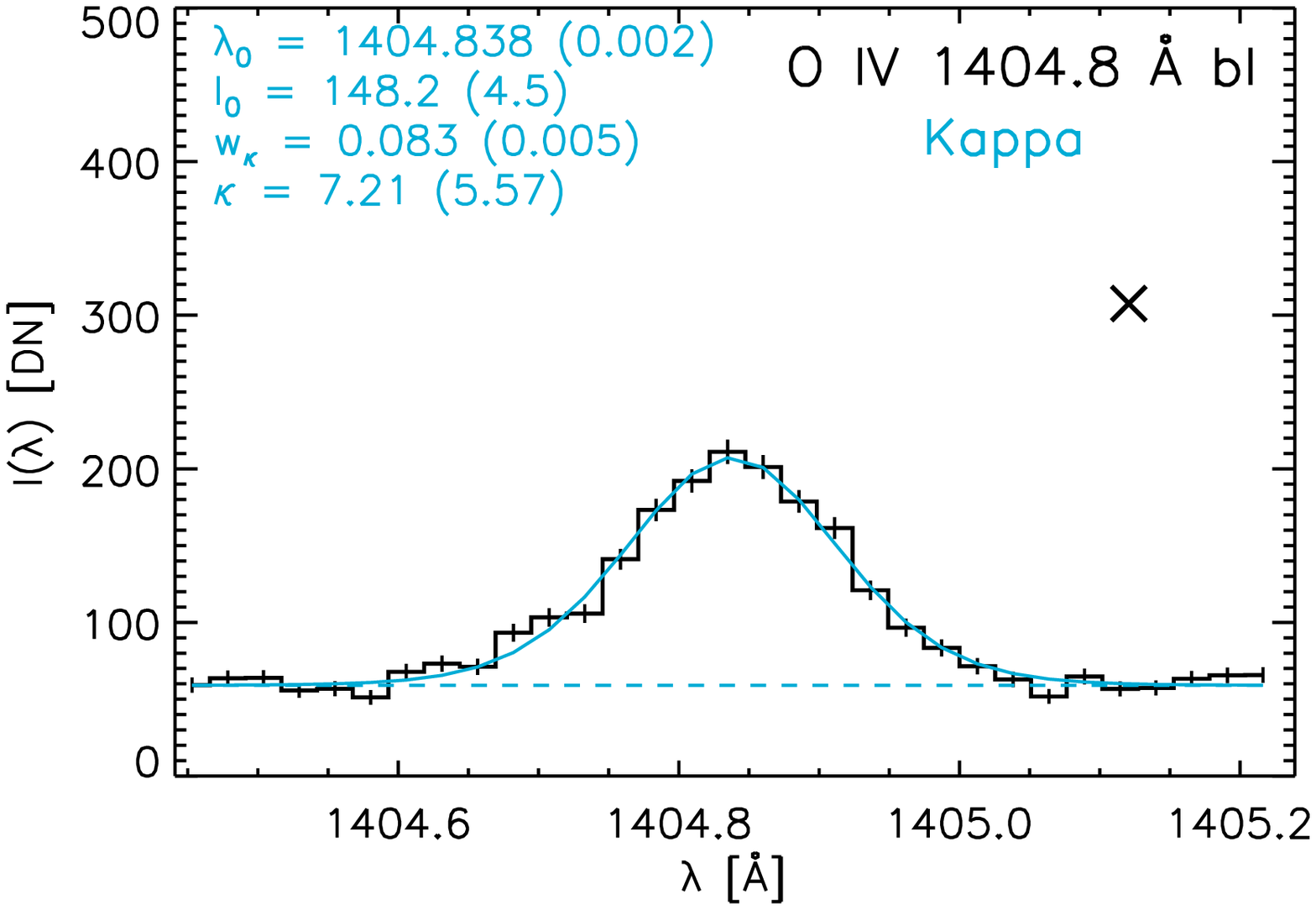}
	\includegraphics[width=4.21cm,clip,bb=75 52 487 340]{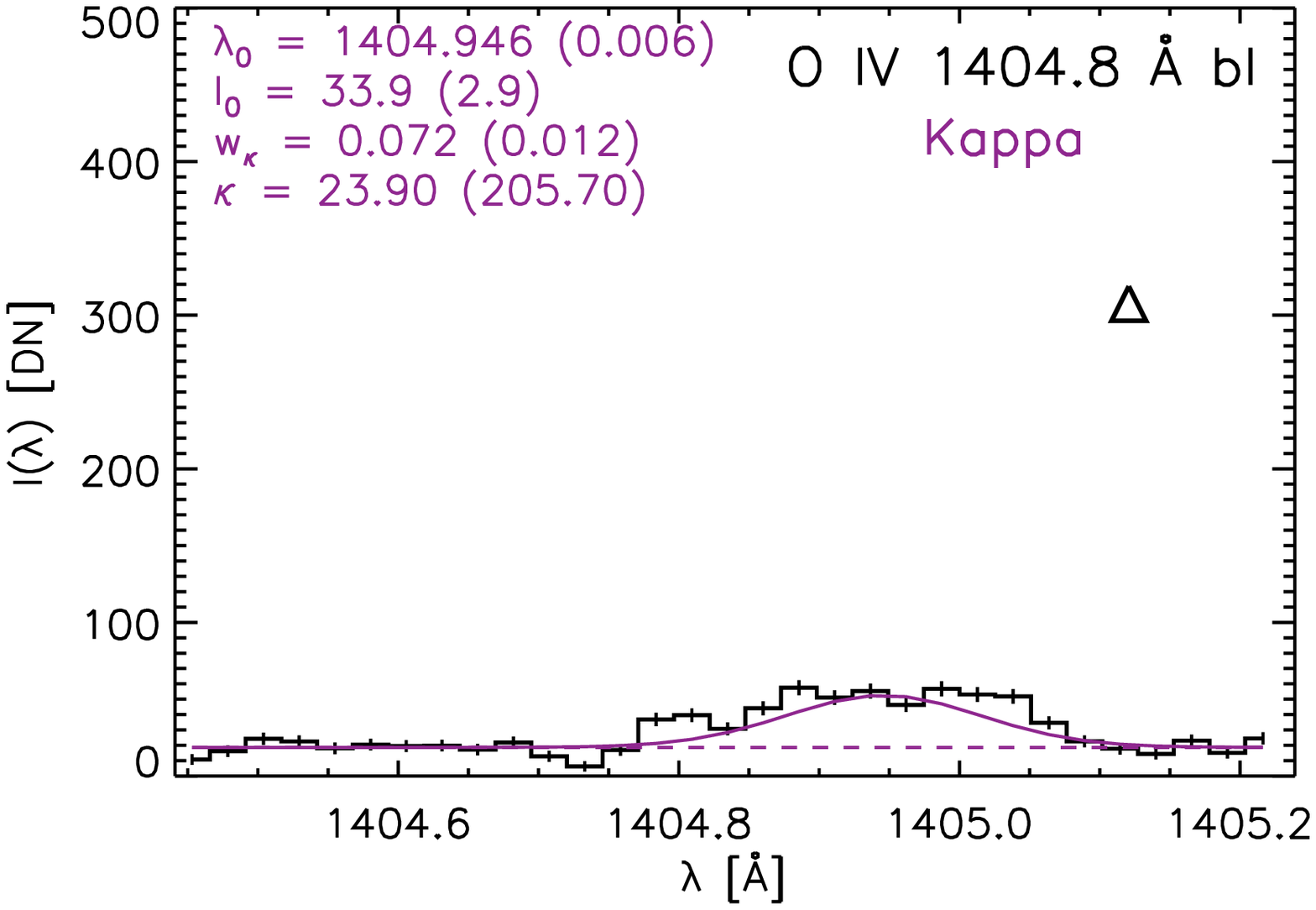}
	\includegraphics[width=4.98cm,clip,bb= 0  0 487 145]{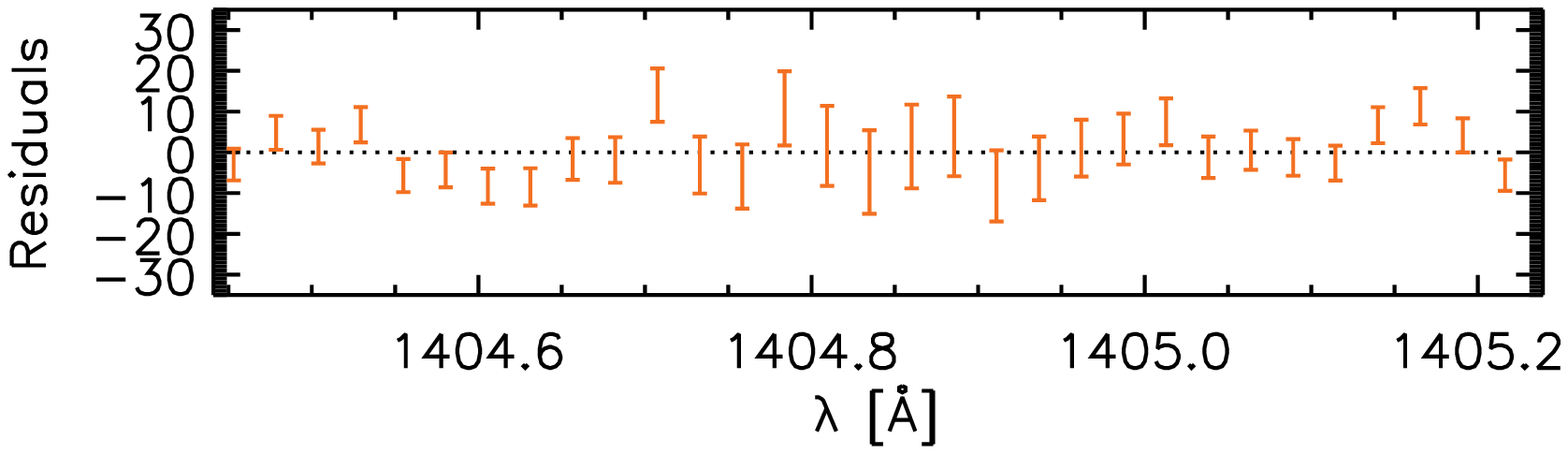}
	\includegraphics[width=4.21cm,clip,bb=75  0 487 145]{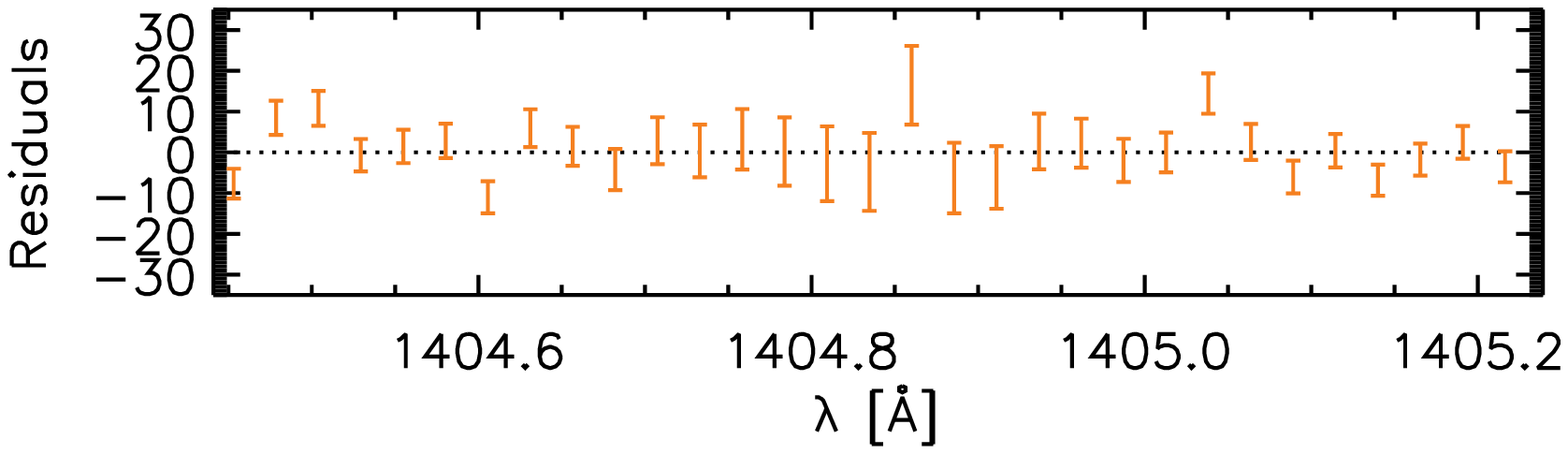}
	\includegraphics[width=4.21cm,clip,bb=75  0 487 145]{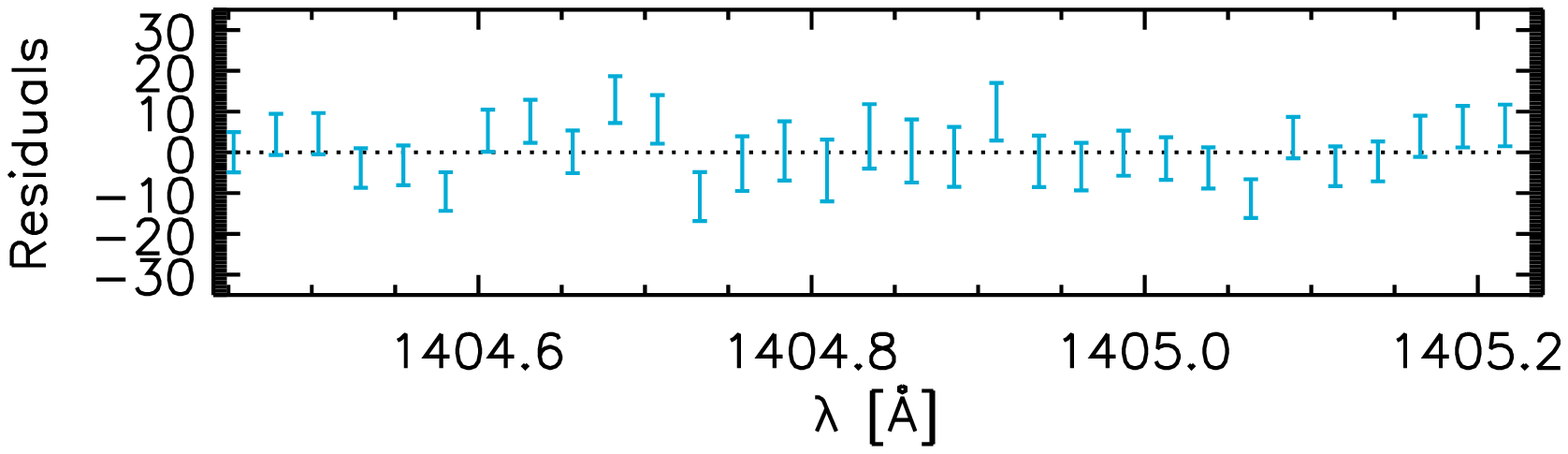}
	\includegraphics[width=4.21cm,clip,bb=75  0 487 145]{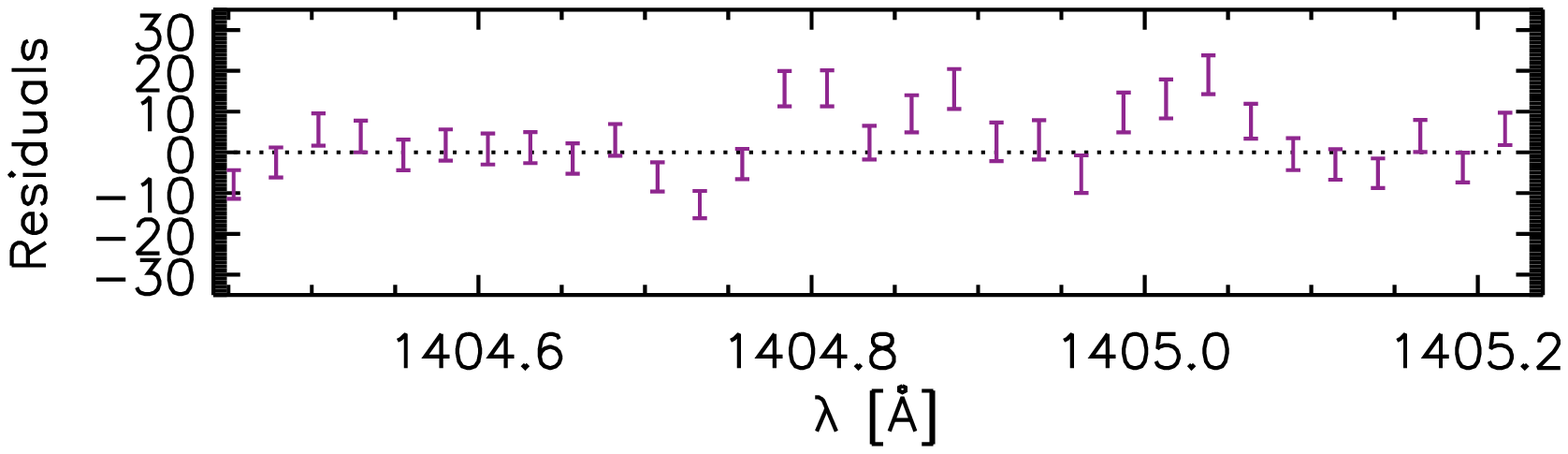}
	\includegraphics[width=4.98cm,clip,bb= 0 52 487 340]{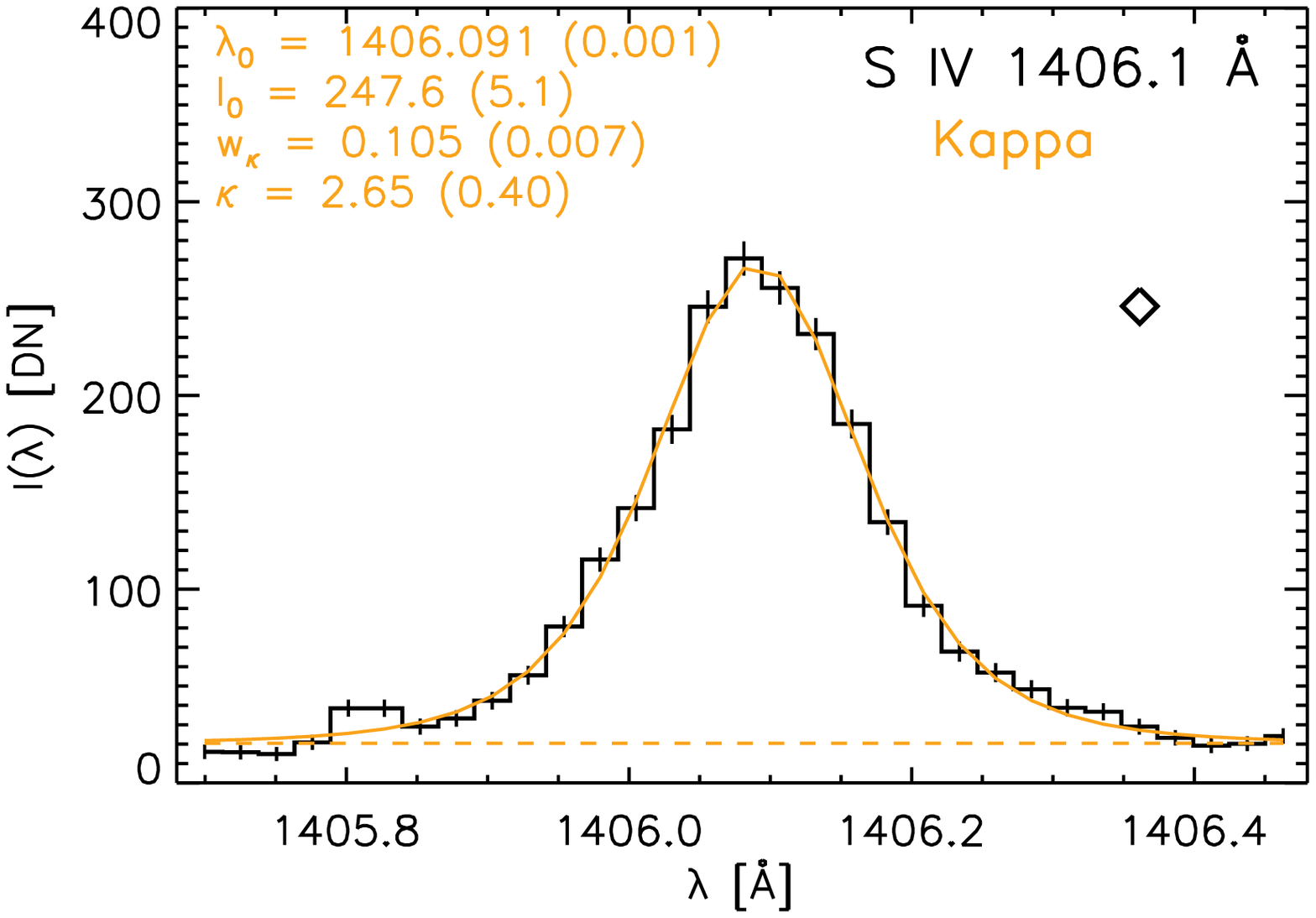}
	\includegraphics[width=4.21cm,clip,bb=75 52 487 340]{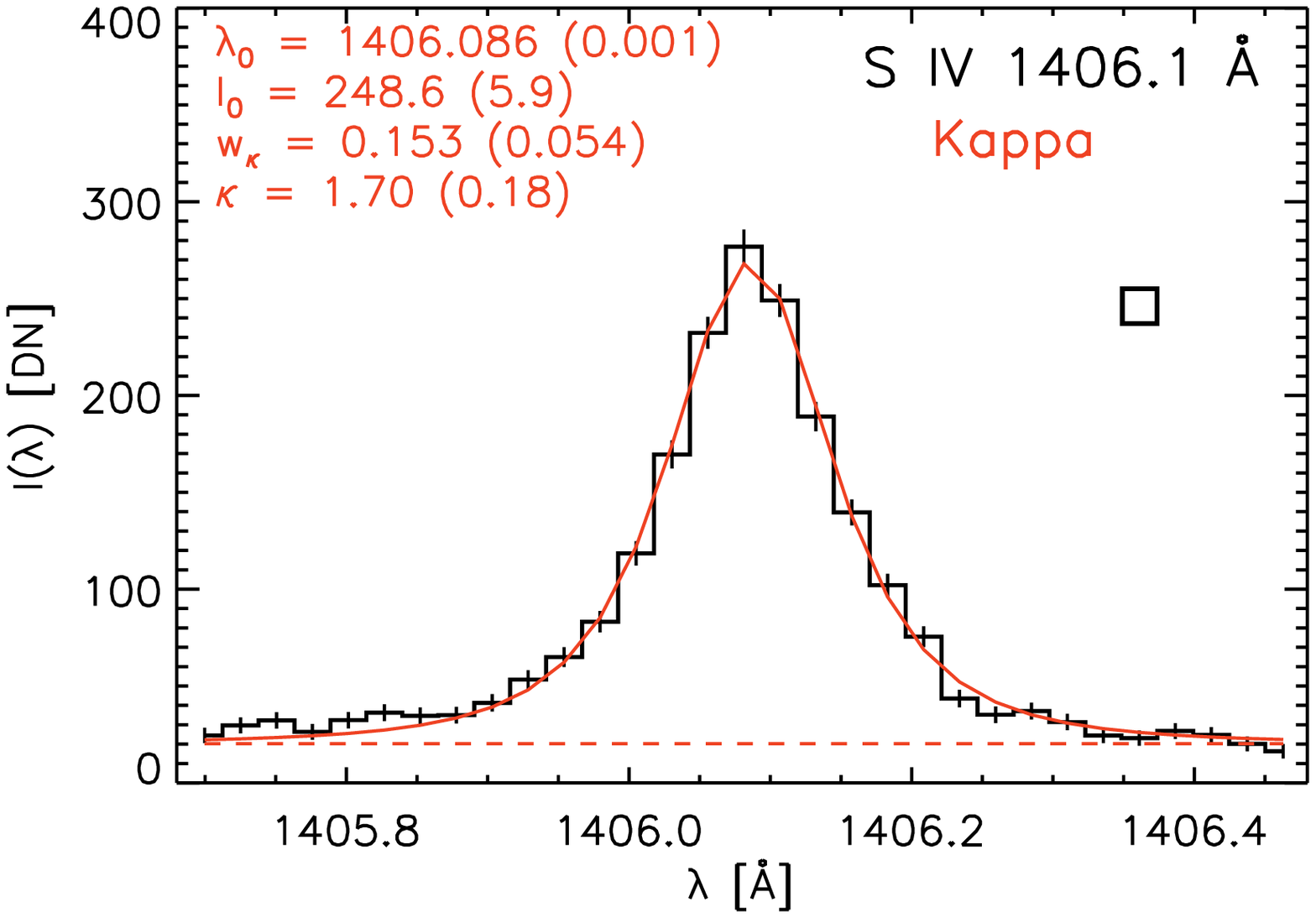}
	\includegraphics[width=4.21cm,clip,bb=75 52 487 340]{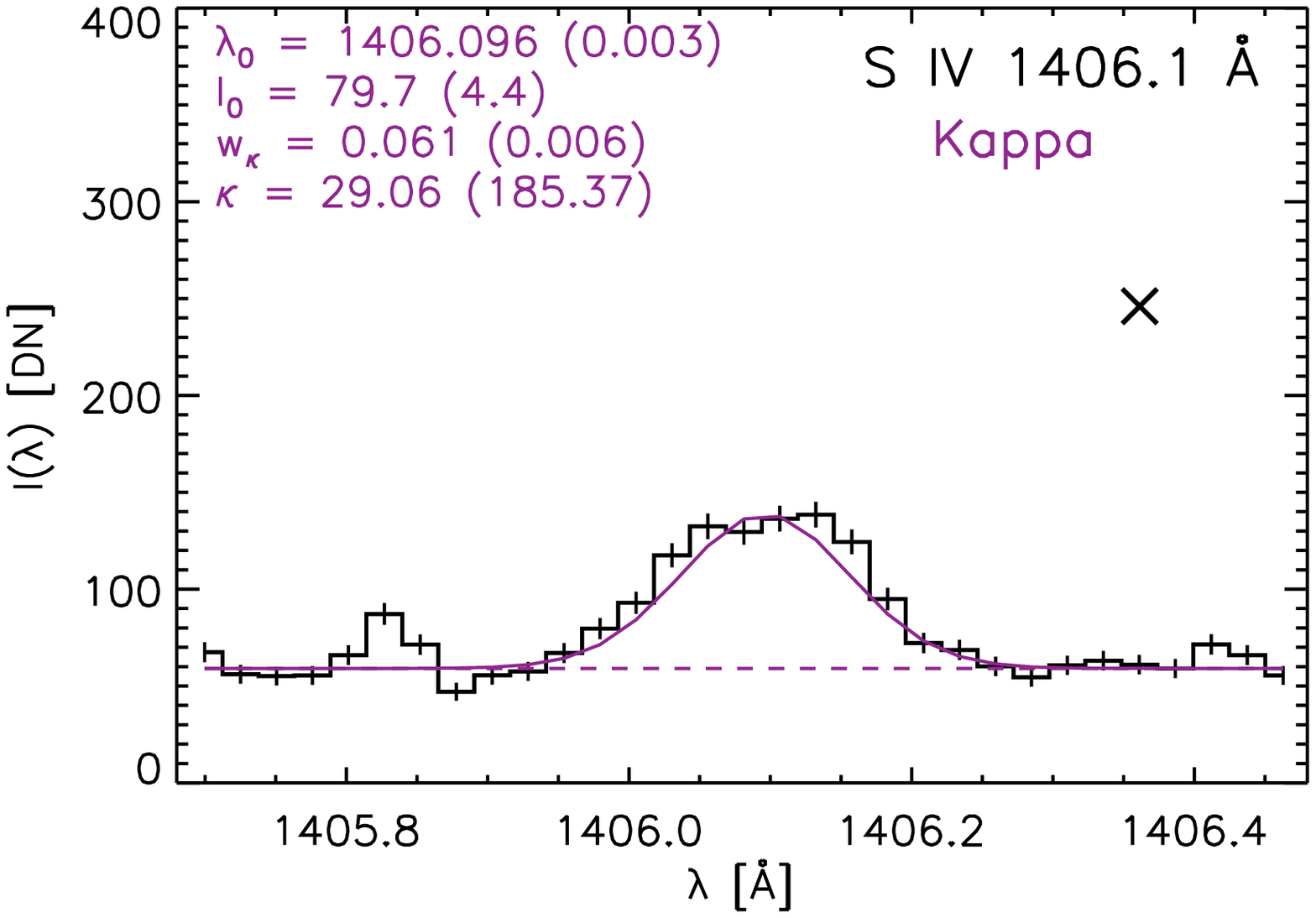}
	\includegraphics[width=4.21cm,clip,bb=75 52 487 340]{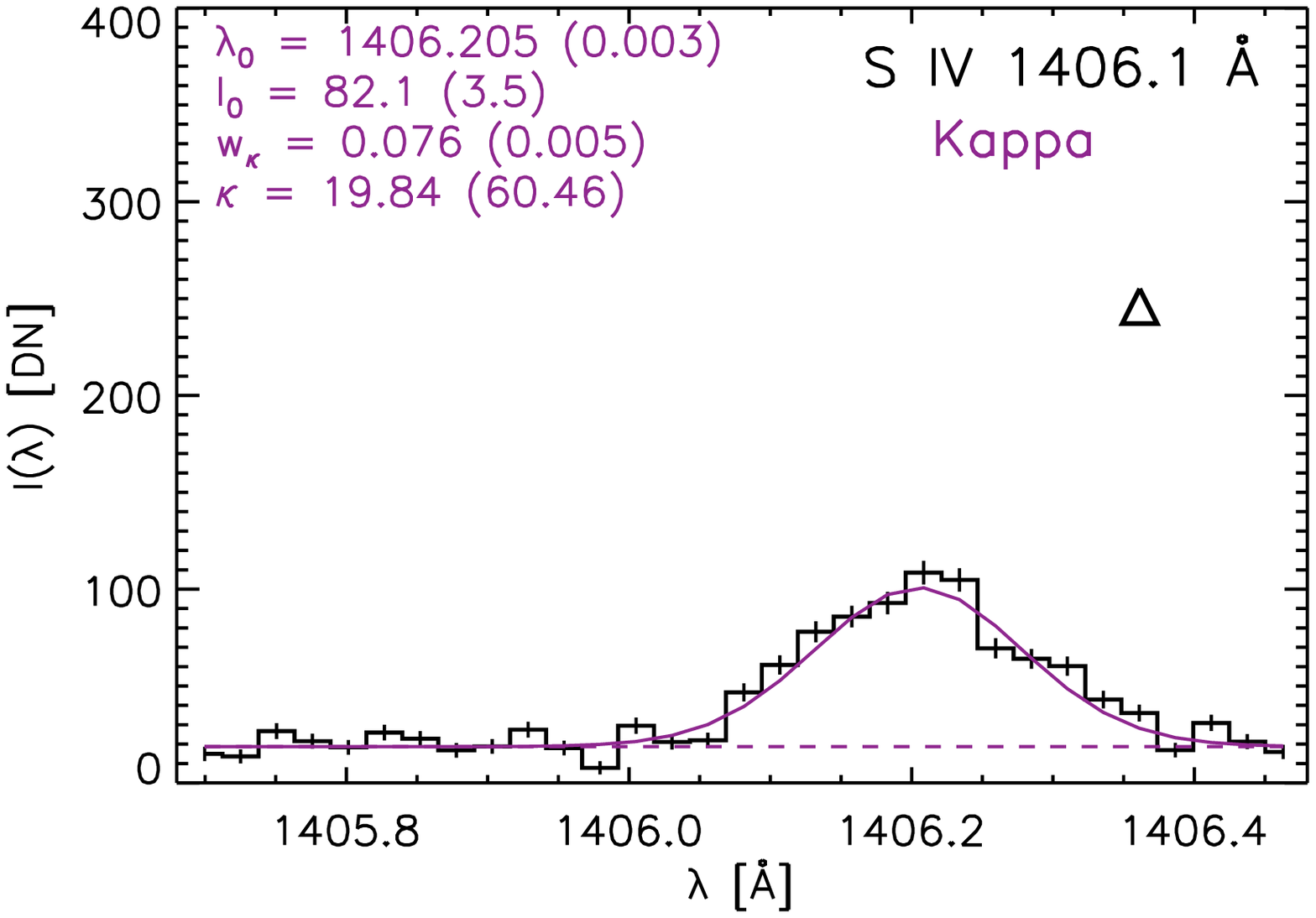}
	\includegraphics[width=4.98cm,clip,bb= 0  0 487 145]{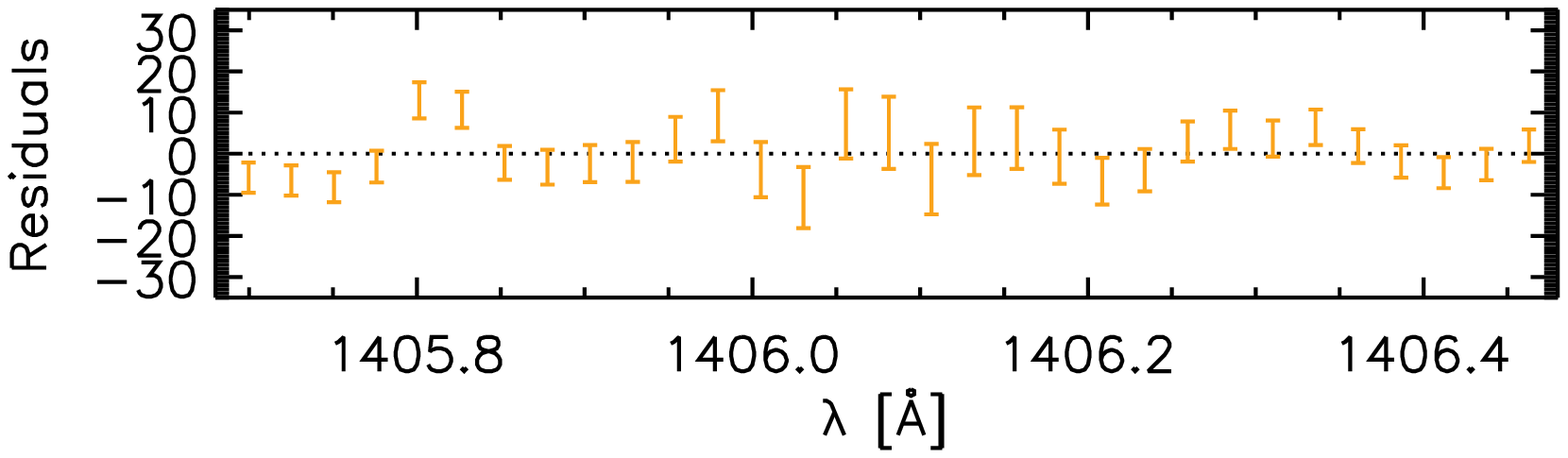}
	\includegraphics[width=4.21cm,clip,bb=75  0 487 145]{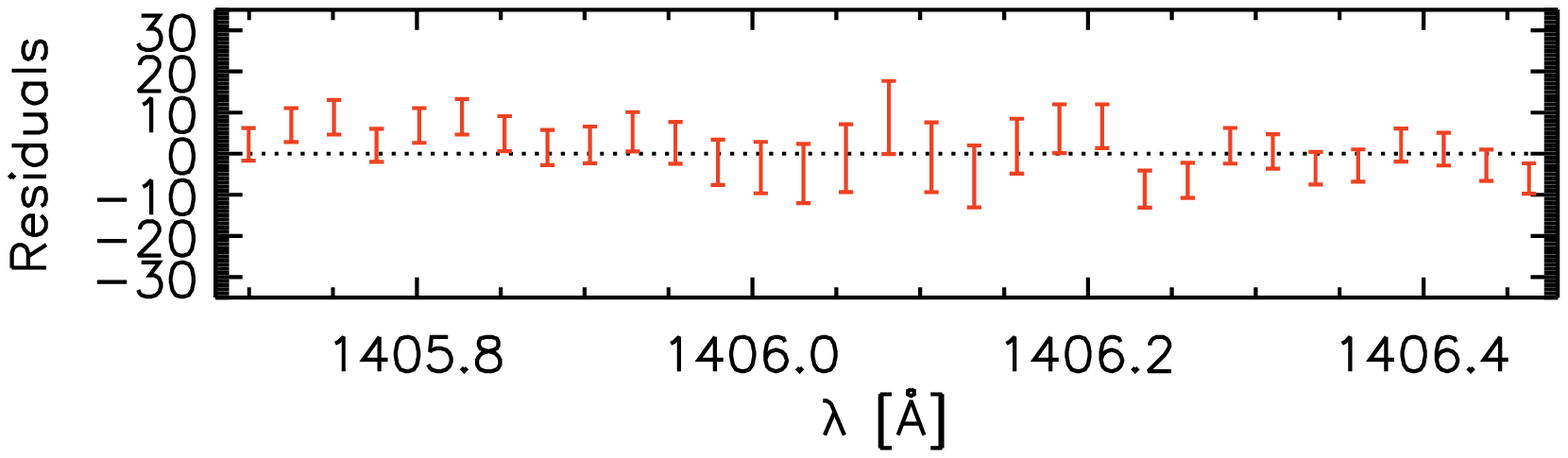}
	\includegraphics[width=4.21cm,clip,bb=75  0 487 145]{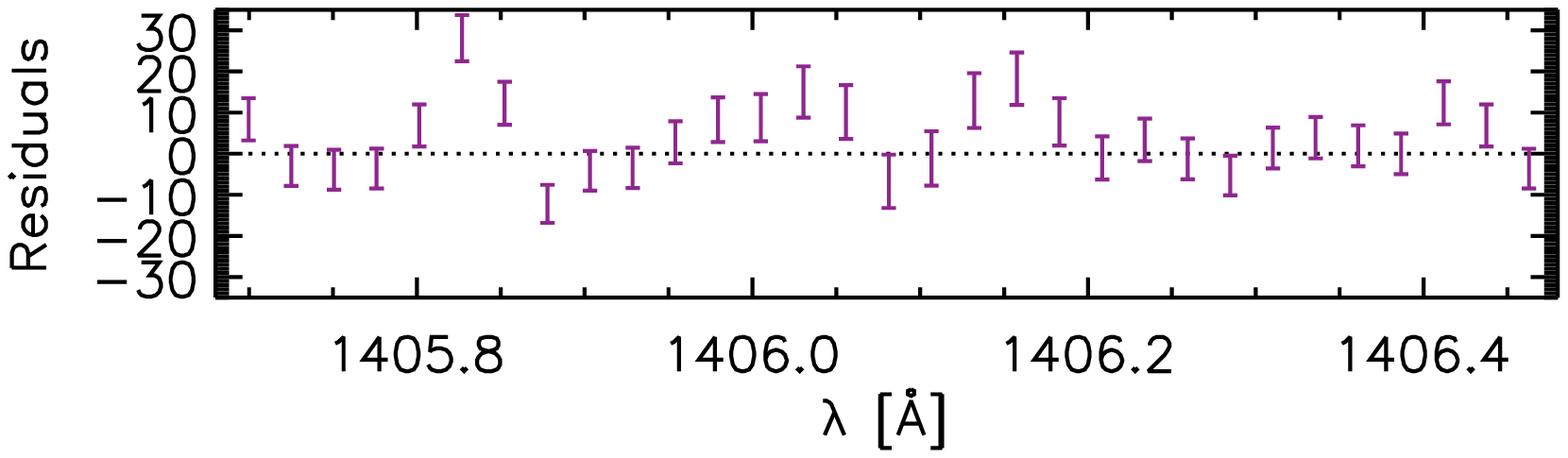}
	\includegraphics[width=4.21cm,clip,bb=75  0 487 145]{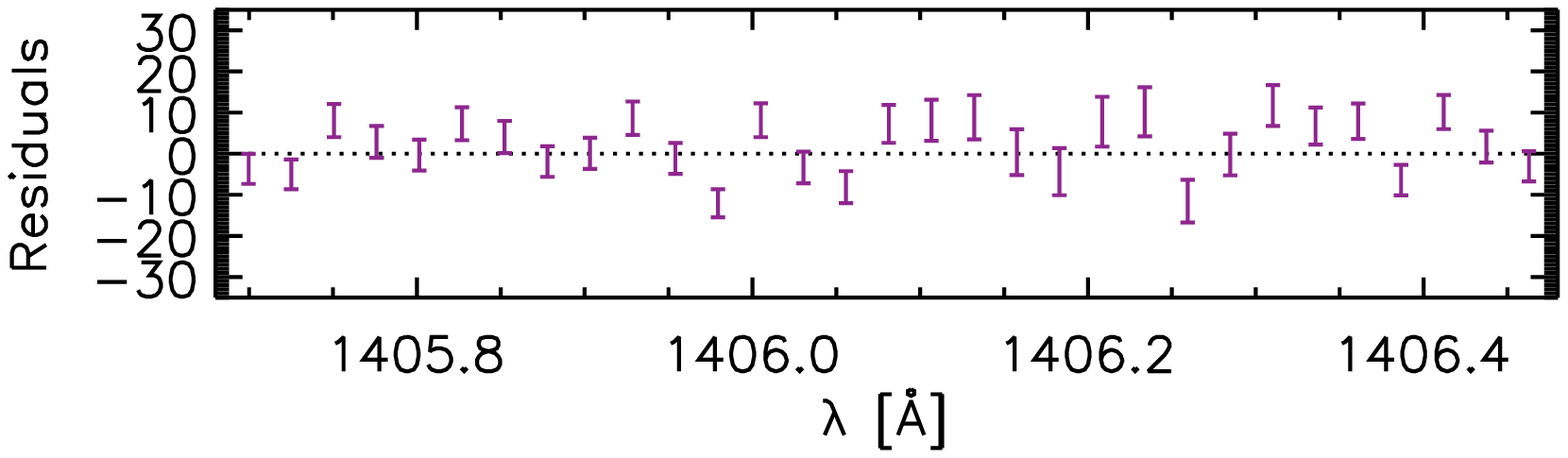}
\caption{Example spectra in four pixels spatial pixels discussed in Sects. \ref{Sect:5.1} and \ref{Sect:5.2}. The pixels are denoted by the diamond, square, cross, and triangle symbols. The location of these pixels is shown in Fig. \ref{Fig:IRIS_sg_fits}.
\label{Fig:Fits_pixels}}
\end{figure*}
%
\begin{figure*}[!t]
	\centering
	\includegraphics[width=6.46cm,clip,bb= 0 55 655	340]{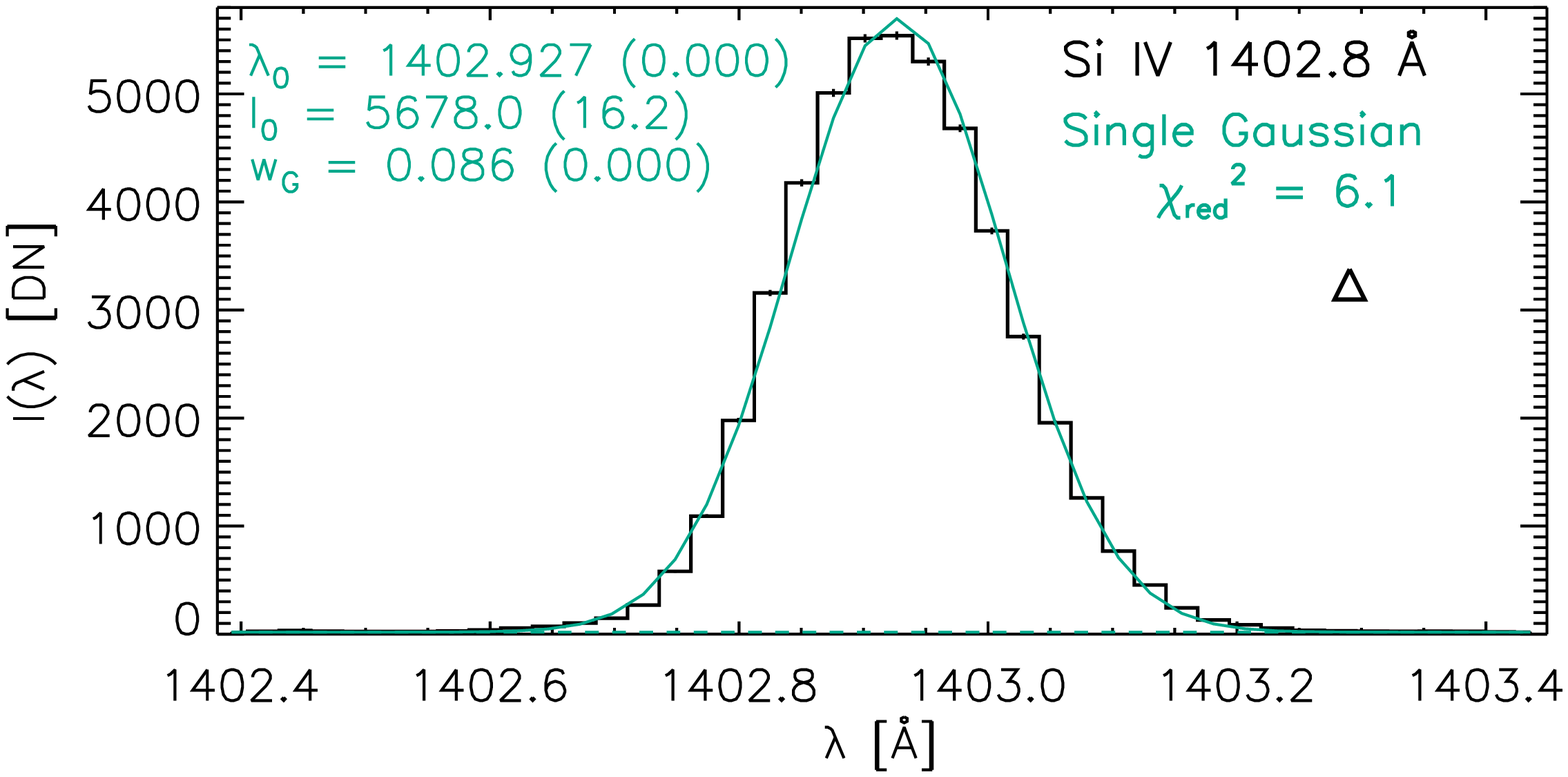}
	\includegraphics[width=5.57cm,clip,bb=90 55 655 340]{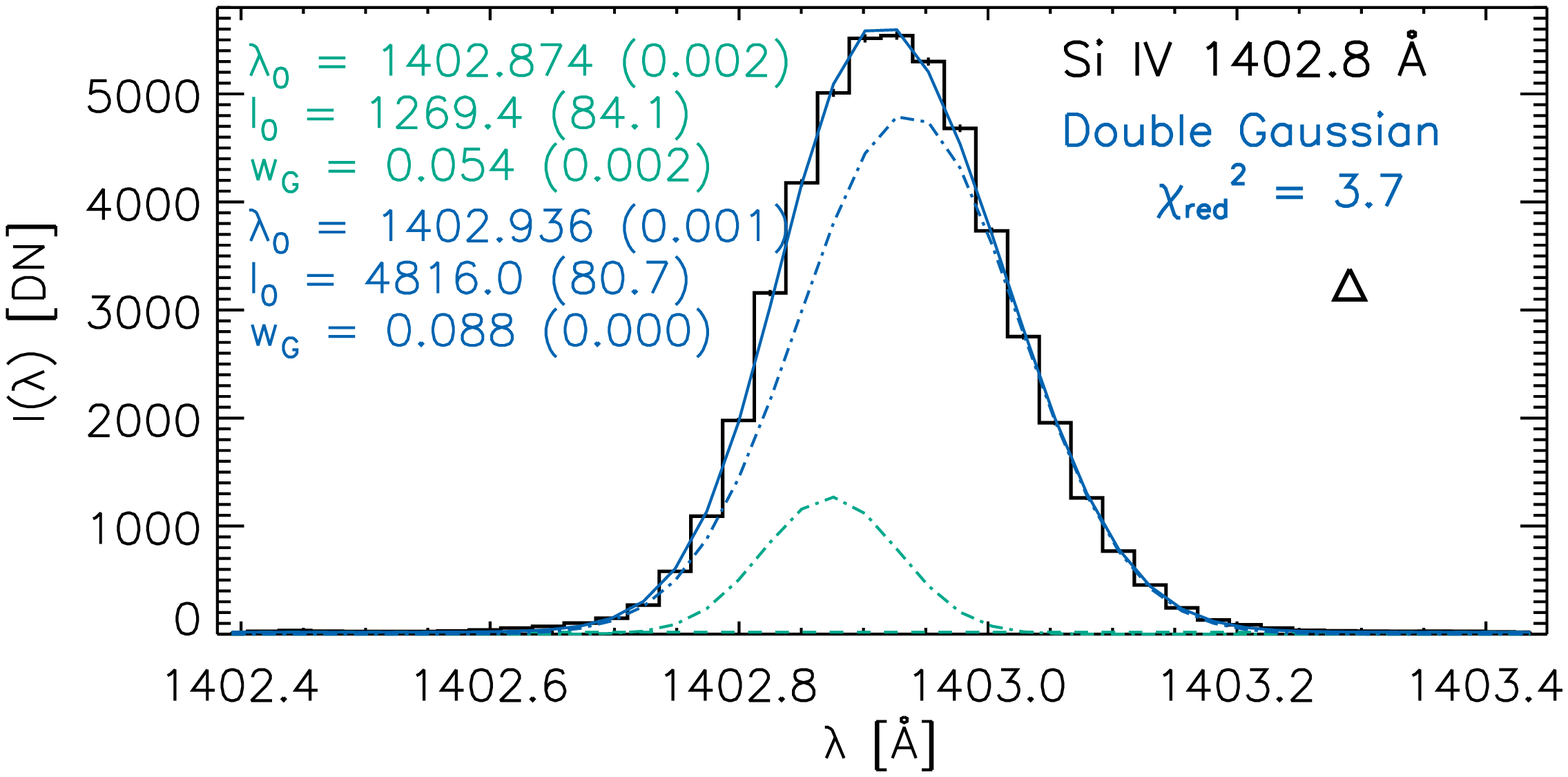}
	\includegraphics[width=5.57cm,clip,bb=90 55 655 340]{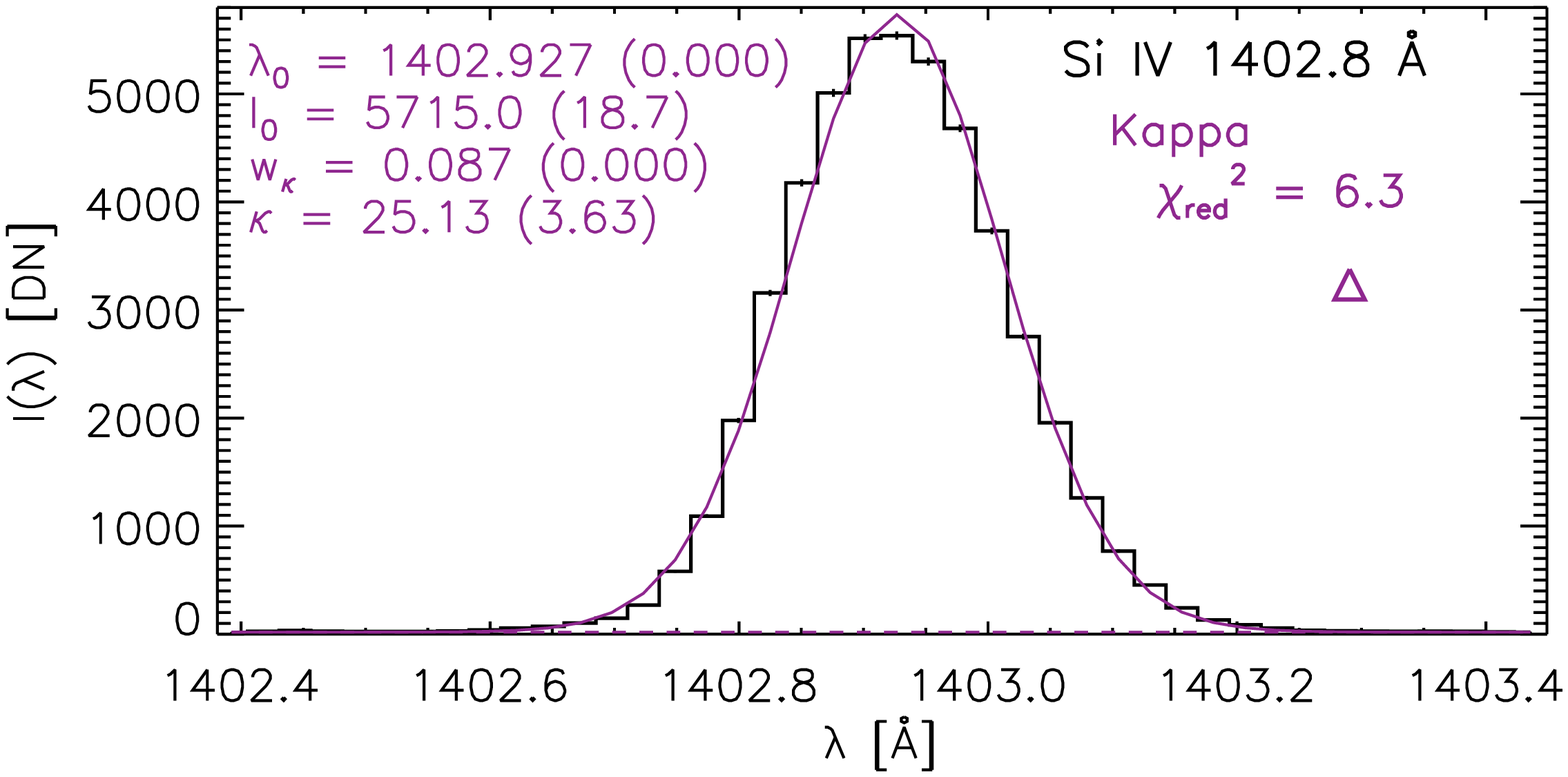}
	\includegraphics[width=6.46cm,clip,bb= 0  0 655 310]{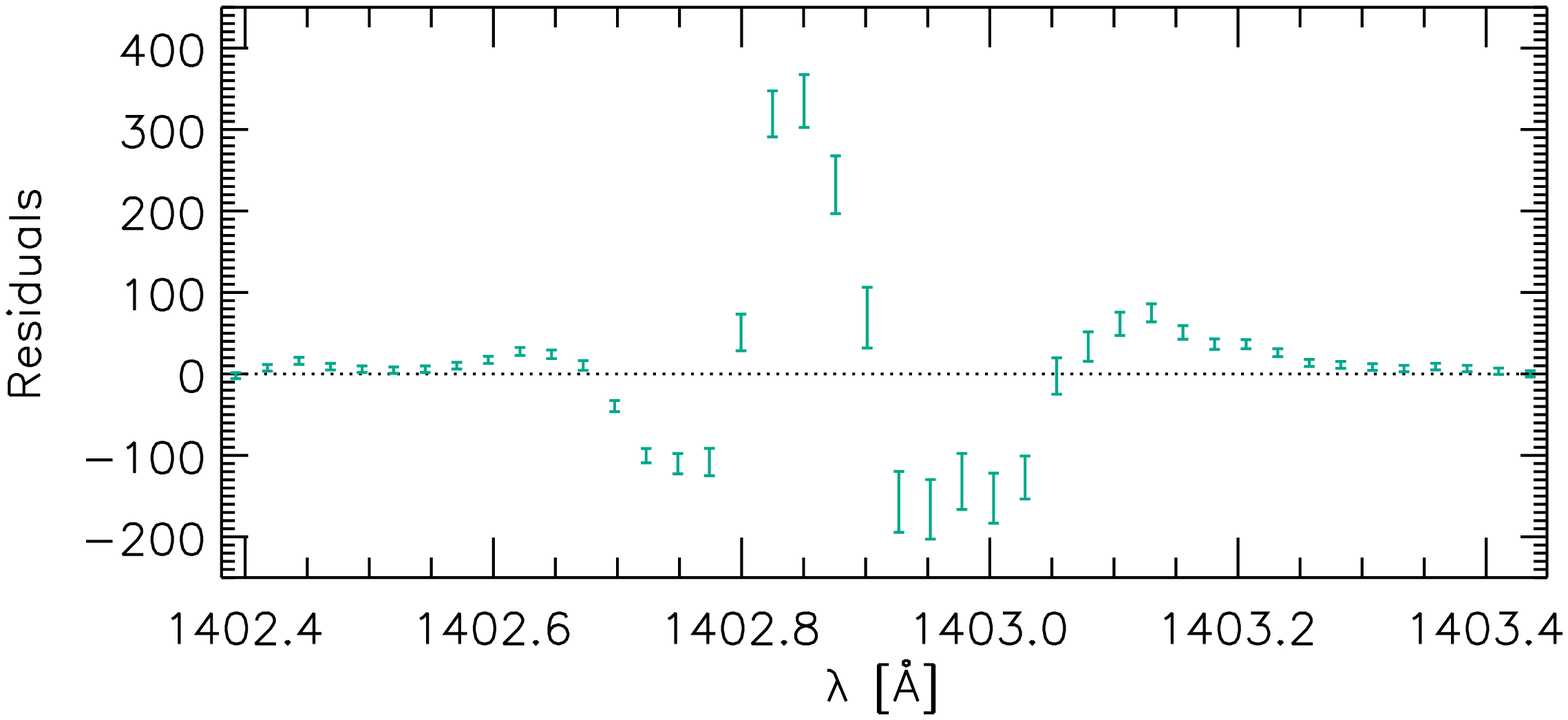}
	\includegraphics[width=5.57cm,clip,bb=90  0 655 310]{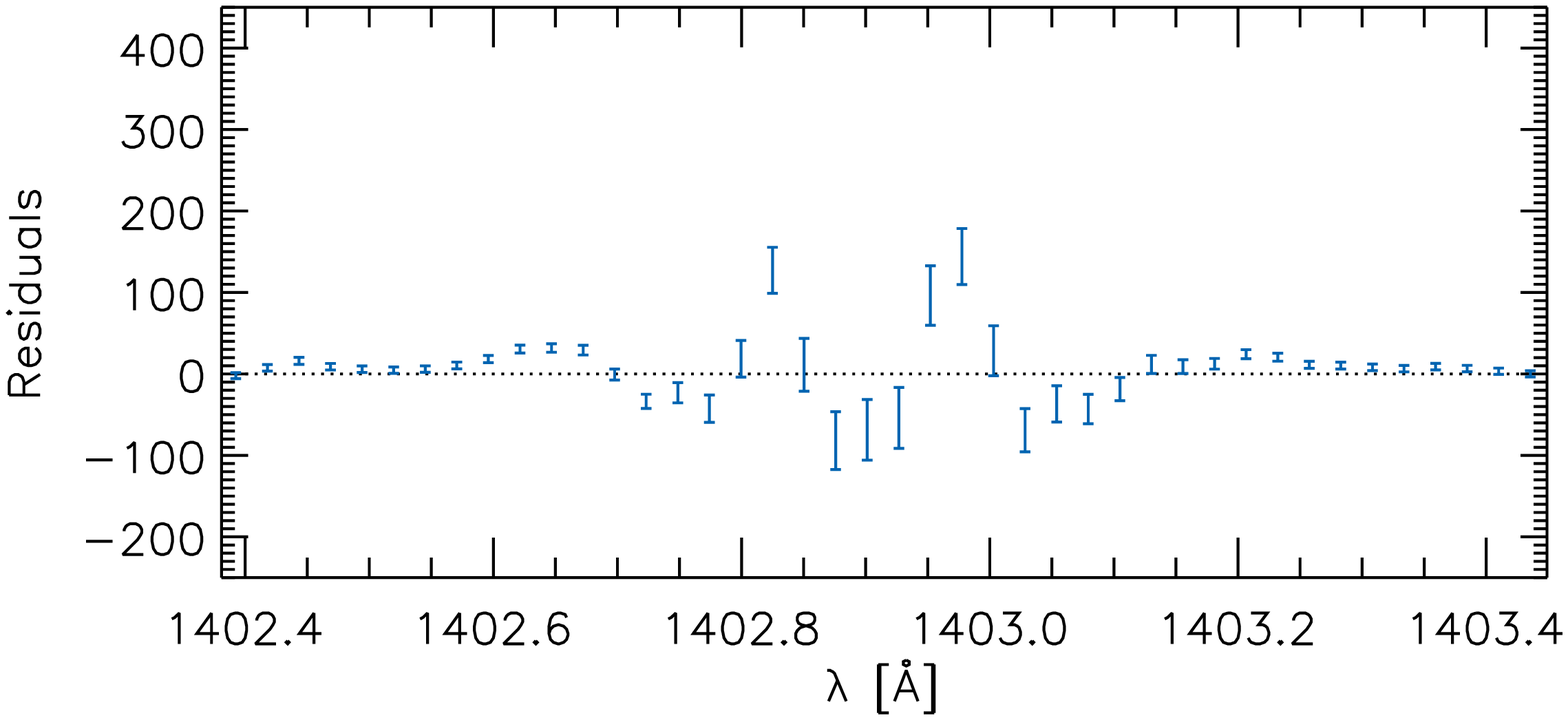}
	\includegraphics[width=5.57cm,clip,bb=90  0 655 310]{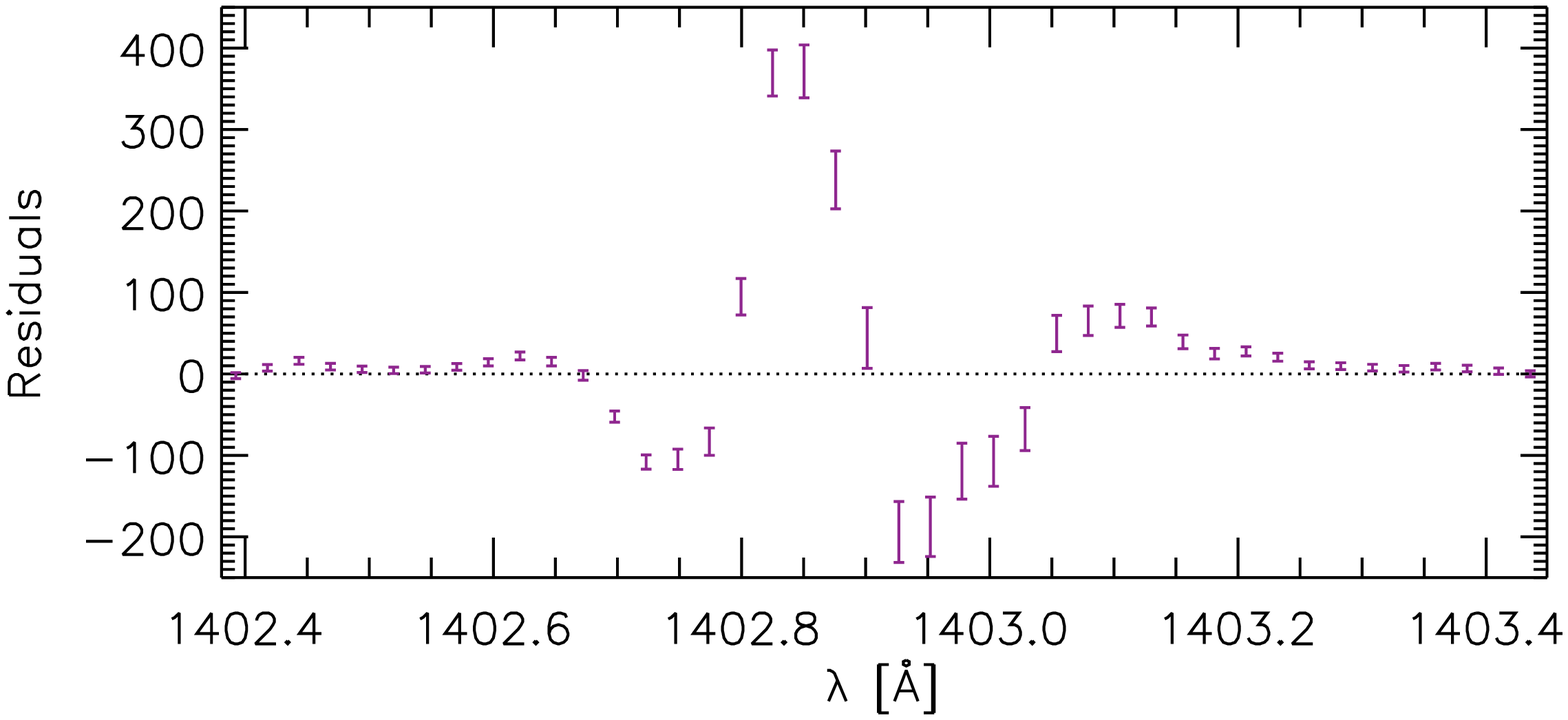}
\caption{Fitting of the \ion{Si}{4} 1402.8\,\AA~line without strong wings. Fits with a single Gaussian (left), double Gaussian (middle) and $\kappa$ components with $\kappa$\,=\,25.1 (right) are shown. See Sect. \ref{Sect:5.2} for details.
\label{Fig:Gauss_pixel}}
\end{figure*}
%
\section{Spatial distribution of $\kappa$ profiles}
\label{Sect:5}

We next performed the fitting in each suitable pixel within the field of view of Fig. \ref{Fig:IRIS_sg}. A pixel is defined as suitable if the \ion{Si}{4} 1402.8\,\AA~line satisfies $|S|$\,$\leq$\,0.1 and $I(\lambda_0)$\,$\geq$\,10$^3$\,DN. There are 291 pixels satisfying these criteria. We do not invoke such constraints on other \ion{O}{4} or \ion{S}{4} lines, since their intensities vary relatively to the \ion{Si}{4} 1402.8\,\AA~one. Furthermore, these lines can be blended, such as \ion{O}{4} 1401.2\,\AA~with \ion{S}{1} 1401.5\AA~in the far wing, which can distort the value of $|S|$ if the blend is relatively strong.

The 291 pixels however still contain spectra where the \ion{Si}{4} is strong but obviously asymmetric, such as saturated, extremely wide lines, or lines with closely spaced double peaks, for which $|S|$\,$<$\,0.1 is still satisfied. Such lines obviously cannot be fitted well with a $\kappa$-distribution. Therefore, we present the results of the $\kappa$-fits only in pixels where an additional constraint of $\chi^2_\mathrm{red}$\,$\,\leq$\,7 is met. This value was found empirically upon reviewing the 291 fit results. This third constraint removes badly fitted spectra and results in 120 pixels where a satisfactory fit is performed. We note that the three constraints do not introduce a preference towards spectra with a $\kappa$-distribution as opposed to a more Gaussian ones.

The 120 suitable pixels are located in the TR loops, as well as in the plage region located at about Solar $Y$ = $-227\arcsec$ \citep[Fig. \ref{Fig:IRIS_sg_fits}; cf.,][Fig. 4 therein]{Polito16b}. A single suitable pixel is located further north, in a point-like bright dot (triangle in Fig. \ref{Fig:IRIS_sg_fits}). An inventory of the $\kappa$-values obtained by fitting of the \ion{Si}{4} 1402.8\,\AA~and \ion{O}{4} 1401.2\,\AA~lines is presented in Fig. \ref{Fig:IRIS_sg_fits}. There, the suitable pixels are shown as plus symbols, whose color depends on the $\kappa$-value in the given line. Four pixels including the bright dot are denoted by different symbols. The spectra observed in these pixels are shown in Fig. \ref{Fig:Fits_pixels} and discussed in Sects. \ref{Sect:5.1} and \ref{Sect:5.2}.

Overall, we see that a vast majority of the suitable \ion{Si}{4} profiles have $\kappa$\,$\leq$\,2.5. Values as low as $\kappa$\,$\approx$\,1.7 are the most common, being present in 21 pixels. An example of such spectrum is discussed in Sect. \ref{Sect:5.1.2}. Contrary to that, the $\kappa$ values found from the \ion{O}{4} 1401.2\,\AA~profiles are typically higher, $\kappa$\,=\,2.0--2.5, with a peak at 2.3 (17 pixels). This is probably at least in part due to the lower intensities of the \ion{O}{4} lines, and thus lower S/N ratio. We however did not find a correlation ($r$\,$<$\,0.4) between $\kappa$ and $I_0$. Furthermore, there are several pixels where the $\kappa$ derived from \ion{O}{4} 1401.2\,\AA~line is significantly higher than those from \ion{Si}{4} 1402.8\,\AA, up to a factor of several. This suggests that in some pixels, the $\kappa$ values could indeed differ among the two lines in some cases. One such case is discussed in Sect. \ref{Sect:5.1.3}.

%
\subsection{Example spectra with strong $\kappa$ profiles}
\label{Sect:5.1}

We now present the spectra and their fitting in 3 additional pixels, corresponding to the first three columns of Fig. \ref{Fig:Fits_pixels}. The first one (denoted by a diamond) is located along the same loop bundle as the spectrum analyzed in Sect. \ref{Sect:4}. The second one (asterisk) is a spectrum from a neighboring loop bundle, and the third one ($\times$ symbol) is an example of a spectrum with different shapes of the \ion{Si}{4} and \ion{O}{4} profiles.

\subsubsection{Loop bundle}
\label{Sect:5.1.1}

Figure \ref{Fig:IRIS_sg_fits} shows that the example spectrum analyzed in Sect. \ref{Sect:4} occurs in a cluster of suitable pixels located along the same loop bundle. Here, we discuss a spectrum located in the center of the cluster, 1.33$\arcsec$ (2 spatial pixels) northward of the one analyzed in Sect. \ref{Sect:4}. The five TR line profiles are shown in the first column Fig. \ref{Fig:Fits_pixels}. There, the values of $\kappa$ are nearly the same as those obtained in Sect. \ref{Sect:4.2.2}. In particular $\kappa_\mathrm{Si\,IV}$\,=\,2.11\,$\pm$\,0.03 and $\kappa_\mathrm{O\,IV}$\,=\,2.37\,$\pm$\,0.09. These two strong lines are very well approximated with the $\kappa$-fits, and we obtain $\chi^2_\mathrm{red}$\,=\,3.0, lower than in Sect. \ref{Sect:4.2.2}.

The other lines have similar $\kappa$ values, except the \ion{O}{4} 1399.8\,\AA~one, whose $\kappa$\,=\,3.5\,$\pm$\,0.5 is higher, and only consistent with the other ones within three times its uncertainty. A possible contributor is a single spurious higher-intensity (by about $\approx$\,40\,DN) wavelength bin in its red wing, which could correspond to a known \ion{Fe}{2} 1399.97\,\AA~blend. Since the chromospheric lines are typically very narrow, with widths of only a few wavelength bins, we cannot reliably verify the presence of such blend from a single spurious wavelength bin. Adding a narrow Gaussian at this wavelength to the fitting procedure would lower the $\kappa$\,=\,3.1\,$\pm$\,0.4 and produce a better fit of the \ion{O}{4} 1399.8\,\AA~line, as well as an overall $\chi^2_\mathrm{red}$\,=\,2.8. We note that the value of $\kappa$ can be further lowered to values consistent with the \ion{Si}{4} line by restricting the width $w_\kappa$ similarly as in the exercise mentioned in Sect. \ref{Sect:4.2.2}, again however at the expense of a somewhat higher $\chi^2_\mathrm{red}$.

\subsubsection{Neighboring loop}
\label{Sect:5.1.2}

Intense TR lines with $\kappa$ line profiles can also be found in a neighboring bright TR loop, located further $\approx$1$\arcsec$ westward. The spectra of this loop are suitable for fitting in many neighboring pixels located along the north-south direction of the loop. We present an example spectrum in second column of Fig. \ref{Fig:Fits_pixels}. The values of $\kappa$ obtained there are among the lowest in the 120 pixels where fitting was performed, with $\kappa$\,=\,1.65\,$\pm$\,0.02 for \ion{Si}{4} and $\kappa$\,=\,1.86\,$\pm$\,0.06 for the strongest \ion{O}{4} line. The weaker \ion{O}{4} lines have higher $\kappa$\,$\approx$\,2.3, but again lower $w_\kappa$ (see discussion in Sect. \ref{Sect:4.2.2}). The overall $\chi^2_\mathrm{red}$\,=\,2.9 is among the lowest found for the 120 suitable pixels.

\subsubsection{Example spectrum with different \ion{Si}{4} and \ion{O}{4} profiles}
\label{Sect:5.1.3}

As an example of a spectrum with different types of profiles obtained from fitting of \ion{Si}{4} and \ion{O}{4} lines, we discuss the spectrum obtained in the pixel denoted by the symbol $\times$, i.e., pixel [41, 128] of the raster. The profiles of the five TR lines are shown in the third column of Fig. \ref{Fig:Fits_pixels}. The \ion{Si}{4} line has about 1950\,DN in its peak, i.e., it is more than a factor of 2 weaker than the examples studied in Sects. \ref{Sect:4} and \ref{Sect:5.1.1}--\ref{Sect:5.1.2}. Its profile is somewhat asymmetric in the peak, but still having $|S|$\,=\,$-0.05$. The line nevertheless can be fitted with a $\kappa$\,=\,2.90\,$\pm$\,0.11, with a maximum residual of about 200\,DN.

Contrary to that, for the \ion{O}{4} 1401.2\,\AA~and 1404.8\,\AA~blend we obtain $\kappa$\,=\,9.8\,$\pm$\,4.0 and 7.2\,$\pm$\,5.6, respectively, which are much higher than for the \ion{Si}{4} line. We again tried restricting the $w_\kappa$ to that of the \ion{Si}{4} line, but this does not lower the $\kappa$ to a value consistent with the \ion{Si}{4} profile, indicating that in this case, the $\kappa$ value can indeed differ for \ion{O}{4} and \ion{Si}{4}. Finally, the two weak \ion{O}{4} and \ion{S}{4} lines appear asymmetric, and have indeterminable $\kappa$, with uncertainties of more than $\pm$130. 

%
\subsection{Are the $\kappa$ profiles an instrumental effect?}
\label{Sect:5.2}

Since we found prevalence of small $\kappa$, especially derived from the \ion{Si}{4} line, is it possible that such profiles are an instrumental effect? Although the influence of the instrumental profile is small (as discussed in Sect. \ref{Sect:4.2.3}), it could still be conceivable that the wings are created somehow by the instrumental profile itself, e.g., if it has very large wings. Alternatively, the observed line wings could be due to stray light (i.e., the point-spread function) within the \textit{IRIS} instrument. We therefore searched for presence of a strong line with a Gaussian line profile, since the presence of such observed line profile would rule out the instrumental effects.

We indeed found one single pixel among the 120 suitable ones, denoted by triangle in Fig. \ref{Fig:IRIS_sg_fits}. This pixel is located in a bright dot-like feature. There, the \ion{Si}{4} line is both nearly Gaussian, as well as very strong. With $I_0$\,=\,5715\,$\pm$\,19\,DN, it is among third strongest \ion{Si}{4} line within the 120 suitable spectra. The corresponding spectrum of the five TR lines is shown in the fourth column of Fig. \ref{Fig:Fits_pixels}. The \ion{Si}{4} 1402.8\,\AA~line does not show presence of large wings. We obtain $\kappa$\,=\,25.1\,$\pm$\,3.6, which indicates that the shape is almost indistinguishable from a Gaussian \citep[compare Fig. \ref{Fig:kappa_v}, right; see also][]{Jeffrey16}. We have verified this by fitting the line with a single Gaussian (Fig. \ref{Fig:Gauss_pixel}), and found that the single Gaussian fit is indeed similar, but slightly better than a fit with $\kappa$\,$\approx$\,25. The detection of a Gaussian \ion{Si}{4} profile, moreover in direct neighborhood of an even brighter pixel (with $I_0$\,$\approx$\,7000\,DN), means that the strong wings in the line profiles, approximated by a $\kappa$-distribution, and not present in this spectrum, do not arise as a result of an instrumental effect. 

We note that although the \ion{Si}{4} line can be fitted with a single Gaussian, the corresponding residuals are asymmetric, which can indicate either a double-component line or presence of optical thickness effects. The line could indeed be fitted well with two Gaussian components (overall $\chi^2_\mathrm{red}$\,=\,3.7 instead of 6.3, Fig. \ref{Fig:Gauss_pixel}), with the stronger Gaussian being more red-shifted. However, the likely presence of large electron densities, of the order of 10$^{11}$\,cm$^{-3}$ or higher indicated by the weak \ion{O}{4} lines and the \ion{S}{4} 1406\,\AA~being stronger than the \ion{O}{4} 1404.8\,\AA~blend \citep[cf.,][Figs. 9, 11, B.1, and B.4 therein]{Polito16b} means that optically thick effects cannot be dismissed for this \ion{Si}{4} line. For further discussion of optically thick effects see Appendix \ref{Sect:A}.

The other \ion{O}{4} and \ion{S}{4} lines are very weak, with peaks less than 150\,DN, and have correspondingly indeterminable shape with large uncertainties in $\kappa$ (last column of Fig. \ref{Fig:Fits_pixels}).

%
%
%
%
\begin{figure*}
	\centering
	\includegraphics[width=17.6cm,bb= 0 52 997 374,clip]{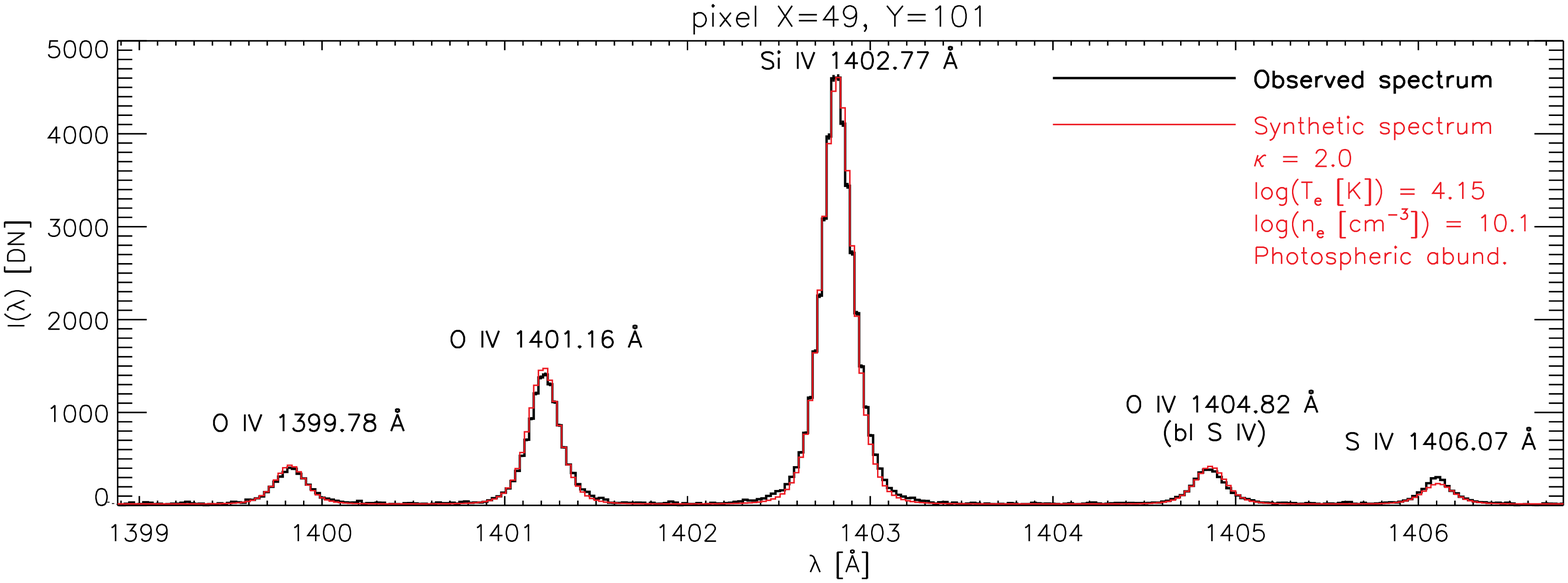}
	\includegraphics[width=17.6cm,bb= 0  0 997 347,clip]{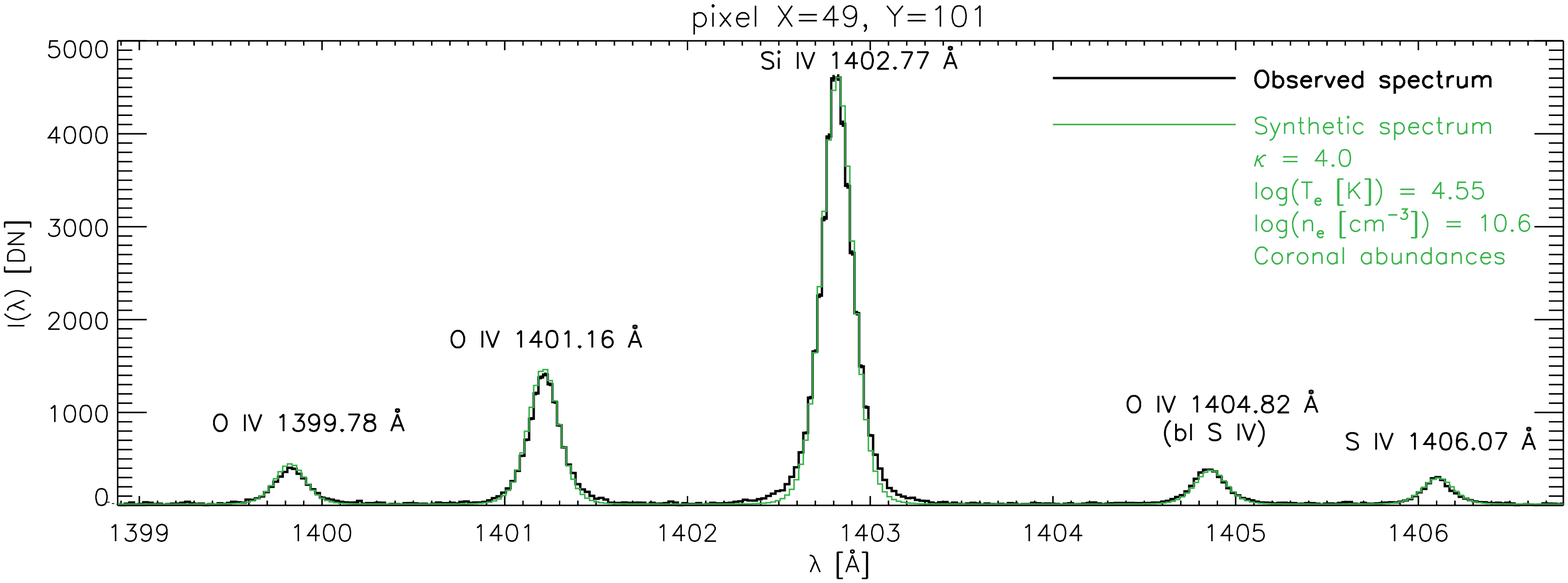}
\caption{Best approximation of the observed spectrum with a synthetic one obtained for photospheric (red) and coronal abundances (green). The parameters of the distribution are listed in each panel. Black line represents the observed \textit{IRIS} spectrum analyzed in Sect. \ref{Sect:4}.
\label{Fig:Spectrum_obs_synt}}
\end{figure*}
%

\section{Line intensities}
\label{Sect:6}

The $\kappa$-distributions also influence intensities of the emission lines studied \citep[e.g.,][]{Dudik14a,Dudik15,Dzifcakova15}. Unlike the line profiles, the line intensities are influenced by distribution of electron energies, since the ionization, recombination, and excitation processes all occur dominantly via electron-ion collisions \citep[e.g.,][]{Phillips08}. Although the line intensities of \textit{IRIS} transition-region lines depend on $\kappa$ \citep{Dudik14a}, the lines are too close in wavelength (excitation energy) to offer unique diagnostics of $\kappa$ from observations using the line ratio-ratio method \citep{Dudik14b,Dudik15}.

Nevertheless, the fact that the $\kappa$-distributions influence the \ion{Si}{4}\,/\,\ion{O}{4} ratios can be used to constrain the value of $\kappa$, if additional assumptions on the nature of the emitting region are made. Only five TR lines are observed, which provides four intensity ratios. Therefore, these additional assumptions are indispensable to restrict the number of free parameters in the calculations of synthetic intensities. Since Si is an element with low ionization potential, unlike oxygen, these assumptions necessarily involve elemental abundances. Further assumptions on the thermal structure of the emitting region are also required, such as on the differential emission measure. Since we observe a bright transition-region loop, we invoke an isothermal and iso-density assumption. This assumption is coupled with the assumption of the collisional ionization equilibrium common in calculation of synthetic spectra for $\kappa$-distributions \citep{Dzifcakova13a,Dzifcakova15,Dudik14a}. We note that possible transient ionization effects are discussed in Sect. \ref{Sect:7.2}.

Under these assumptions, we compare a grid of synthetic intensities with the observed ones in all 120 pixels analyzed in Sect. \ref{Sect:5}. We use the intensities relative to the \ion{Si}{4} 1402.8\,\AA, calculated for log($T_\mathrm{e}$\,[K])\,=\,4.0--6.0 with a step of 0.05, log$(N_\mathrm{e}$\,[cm$^{-3}$])\,=\,9--12 with a step of 0.1, and $\kappa$\,=\,1.6, 1.7, 1.8, 1.9, 2, 3, 4, 5, 7, 10, 15, 25, 33, and Maxwellian, for which the corresponding ionization equilibrium files are available in the KAPPA database \citep{Dzifcakova15}. The line intensities are calculated using the method of \citet{Dudik14a}. In doing so, we use the atomic data of \citet{Liang09}, \citet{Liang12}, and \citet{DelZanna16} for \ion{Si}{4}, \ion{O}{4}, and \ion{S}{4}, respectively. We note that for the transitions from levels 3--20 in \ion{O}{4}, we use the $A_{ij}$ values from \citet{Correge04} instead of \citet{Liang12}. This is for consistency with the CHIANTI database, version 8 \citep{DelZanna15b}. Details on the \ion{O}{4} and \ion{S}{4} atomic data used can be found in Appendix A of \citet{Polito16b}.

In summary, we invoke three free parameters, $T_\mathrm{e}, N_\mathrm{e}$, and $\kappa$, to approximate four  observed line intensity ratios. This is done for each of the 120 suitable pixels. The best approximation is found by minimizing the quantity $\chi^2 = \sum_i(O_i-C_i)^2/C_i$, where $O_i$ and $C_i$ are four observed and calculated line intensity ratios, respectively.

We found that the observed intensity ratios are typically best approximated with $\kappa$\,=\,1.9--2 (12.5\% and 78.3\% of the suitable pixels, respectively) if photospheric abundances of \citet{Asplund09} are assumed. In Fig. \ref{Fig:Spectrum_obs_synt} top, we plot in black an example of the spectrum from a single pixel analyzed in Sect. \ref{Sect:4}. The synthetic spectrum calculated for $\kappa$\,=\,2, log($T_\mathrm{e}$\,[K])\,=\,4.15, and log$(N_\mathrm{e}$\,[cm$^{-3}$])\,=\,10.1 is shown in red. The synthetic spectrum has been shifted by 0.05\,\AA~to compensate for the ubiquitous redshift of TR lines, and we assumed the widths derived from line fitting (Sect. \ref{Sect:4.2}). The synthetic spectrum is a good match to the observed one, despite the rather simplifying assumptions. The $\chi^2$ obtained for this approximation is about 2--times lower than the minimum $\chi^2$ if $\kappa$\,=\,3, and about 14.8--times lower than for a Maxwellian. Discrepancies however still occur in the \ion{O}{4} 1401.2\,\AA~and \ion{S}{4} 1406.1\,\AA~lines, which are over-estimated and underestimated by about 100\,DN, respectively. The reason for this discrepancy is not clear; we suspect it may be related to the discrete $\kappa$ values used, which e.g. do not contain a value of $\kappa$\,=\,2.1 (Sect. \ref{Sect:4.2.2}). We tried using a different set of photospheric abundances of \citet{Caffau11}, but this resulted in a worse match.

If the abundances are assumed to be coronal, the best approximation is obtained for $\kappa$\,=\,3--4 (78.3\% and 15.8\% of the suitable pixels, respectively). In the case of the single pixel analyzed in Sect. \ref{Sect:4}, the best match obtained is a synthetic spectrum with $\kappa$\,=\,4, log($T_\mathrm{e}$\,[K])\,=\,4.55, and log($N_\mathrm{e}$\,[cm$^{-3}$])\,=\,10.6, shown in green color in the bottom panel of Fig. \ref{Fig:Spectrum_obs_synt}. This synthetic spectrum fits the \ion{S}{4} line better; however, the \ion{O}{4} 1401.2\,\AA~remains overestimated similarly as in the case of photospheric abundances.

Finally, if the synthetic line intensity calculations are restricted by a further assumption of only a Maxwellian distribution, the closest match to the observations is found for coronal abundances, log($T$\,[K])\,=\,5.2, and log($N_\mathrm{e}$\,[cm$^{-3}$])\,=\,12.9. Even so, the \ion{O}{4} lines at 1399.8\,\AA~and 1401.2\,\AA~are over- and under-estimated by about 150 and 170\,DN, respectively. This agreement is significantly worse than for the $\kappa$-distributions. Furthermore, such extremely high densities are not realistic, see \citep{Polito16b} and \citet{Judge15}. Under the assumption of photospheric abundances, no good approximation to the observed line intensity ratios can be found.

%
\begin{table*}[ht]
\caption{Thermal and non-thermal widths, $w^\mathrm{(th)}$ and $w^\mathrm{(nth)}$, derived for the Maxwellian and $\kappa$\,=\,2 distributions, respectively. The $T_\mathrm{max}$ represent electron temperatures corresponding to the peak of the relative ion abundance in ionization equilibrium. The corresponding $w^\mathrm{(th)}$ are derived using Eq. (\ref{Eq:w_kappa}) with $T$\,=\,$T_\mathrm{max}$. The non-thermal widths are derived using Eq. (\ref{Eq:nonth_widths}) from the observed $w_\kappa$ and $w^\mathrm{(th)}$.}
\label{Table:3}
  $$ 
  \centering
  \begin{tabular}{lccccccccc}
	\hline
	\hline
	\noalign{\smallskip}
	Line 						& $w_\kappa$ [\AA]	& & log($T_\mathrm{max,Maxw}$ [K])	& $w^\mathrm{(th)}_\mathrm{Maxw}$	& $w^\mathrm{(nth)}_\mathrm{Maxw}$ & & log($T_{\mathrm{max},\kappa=2}$ [K])	& $w^\mathrm{(th)}_{\kappa=2}$	& $w^\mathrm{(nth)}_{\kappa=2}$	\\
	\noalign{\smallskip}
	\hline
	\noalign{\smallskip}
	\ion{O}{4}  1399.78 \AA~			&  0.143 $\pm$ 0.010	& & 5.15			& 0.040			& 0.137				& & 4.45			& 0.018				& 0.141				\\
	\ion{O}{4}  1401.16 \AA~			&  0.127 $\pm$ 0.003	& & 5.15			& 0.040			& 0.121				& & 4.45			& 0.018				& 0.126				\\
	\ion{Si}{4} 1402.77 \AA~			&  0.136 $\pm$ 0.001	& & 4.90			& 0.023			& 0.134				& & 4.10			& 0.009				& 0.136				\\
	\ion{O}{4}  1404.82 \AA~(bl \ion{S}{4})		&  0.163 $\pm$ 0.018	& & 5.15			& 0.040			& 0.158				& & 4.45			& 0.018				& 0.162				\\
	\ion{S}{4}  1406.06 \AA~			&  0.144 $\pm$ 0.021	& & 5.05			& 0.025			& 0.141				& & 4.20			& 0.009				& 0.143				\\
	\noalign{\smallskip}
	\noalign{\smallskip}
	\hline
  \end{tabular}
  $$ 
\end{table*}
%
%
%
%
\section{Discussion}
\label{Sect:7}

\subsection{On the consistency of electron and ion distributions}
\label{Sect:7.1}

It is not obvious that the electron and ion velocity distributions should be the same. 
The classical relaxation time theory predicts that both electrons and ions should thermalize quickly at the high densities typical of the TR (see, e.g., Chapter 5 of \citeauthor{Spitzer62} \citeyear{Spitzer62}, or Chapter 3.2.4 of  \citeauthor{Goedbloed04} \citeyear{Goedbloed04}). For example, using the values of $T_\mathrm{e}$ and $N_\mathrm{e}$ obtained in Sect. \ref{Sect:6} together with Eq. (3.50) of \citet{Goedbloed04}, the estimated relaxation timescale to an electron fluid is about\,4\,$\times$10$^{-6}$\,s. The corresponding electron-ion equipartition time, i.e., the timescale for both ions and electrons to reach Maxwellians at the same $T$, is longer by a factor of $m_\mathrm{i}/2m_\mathrm{e}$. For oxygen, it is 0.1\,s. 

These classical relaxation timescales are however applicable only for particles with $v$\,$\lesssim$\,$\theta$. That means that they are not valid for the case of $\kappa$\,$\approx$\,2, where $\approx$35\% of particles have $v$\,$>$\,$\theta$ \citep{Oka13}. Furthermore, since the collision frequency of the high-energy particles scales as $v^{-3}$, such particles become increasingly collisonless. Derivation of the corresponding equilibration timescales for $\kappa$-distributions is however out of the scope of this work.

\subsection{Notes on transient ionization effects}
\label{Sect:7.2}

The relative line intensities, especially the ratios of \ion{Si}{4}\,/\,\ion{O}{4}, can exhibit departures from equilibrium values due to transient ionization effects, i.e., in situations where the plasma is ionizing or recombining \citep[e.g.,][]{Judge12,Doyle13,Olluri13b,Olluri15,DePontieu15,Martinez-Sykora16}. \citet{Smith10} calculated the ionization equilibration timescales for astrophysically important elements as a function of temperature. Using their Fig. 1, which assumes a Maxwellian distribution, known $T$, and no flows, we find that the typical ionization equilibration timescales for Si and O can be of the order of 10\,s or higher at electron densities of the order of 10$^{10}$\,cm$^{-3}$, which we obtained in Sect. \ref{Sect:6}.

For the $\kappa$-distributions, such ionization timescales will be shorter, since the total ionization rate is enhanced by orders of magnitude, while the total recombination rate is enhanced by a factor of about 2 for low $\kappa$ values compared to the Maxwellian \citep{Dzifcakova13a}. However, detailed hydrodynamic or magnetohydrodynamic modeling of TR loops, which is beyond the scope of this work, would be required to study the presence of non-equilibrium ionization. Without it, we cannot exclude the presence of transient ionization effects. We however note that the non-equilibrium ionization effects would primarily impact the line intensity ratios. Their influence on the line profiles occurs only through the distribution of temperatures over which the individual ions exist when out of ionization equilibrium. In this regard, the observed shape of the line profiles in a chosen single pixel could serve as a strong constraint on the transient ionization simulations.

\subsection{Line widths and temperatures for the $\kappa$-fit}
\label{Sect:7.3}

Having obtained the fit parameters (Fig. \ref{Fig:Fits} and Table \ref{Table:2}), we calculated the FWHM$_\kappa$ and the corresponding ion temperatures $T_\mathrm{i}$ for the example spectrum studied in Sect. \ref{Sect:4}. To do this, we used Eqs. (\ref{Eq:w_kappa}). In doing so, the uncertainties of the fit parameters are propagated to obtain the uncertainties of the resulting FWHM$_\kappa$ and $T_\mathrm{i}$. The derived values are also listed in Table \ref{Table:2}. The large uncertainties for the \ion{O}{4} 1404.8\,\AA~and \ion{S}{4}\,1406.0\,\AA~lines come from the corresponding uncertainties in both $\kappa$ and $w_\kappa$; however, visual inspection of the observed line profiles (Fig. \ref{Fig:Fits}) show that the derived value of FWHM$_\kappa$\,$\approx$\,0.2 is in accordance with the observations despite its large calculated uncertainty.

We point out that the FWHM$_\kappa$\,$\approx$\,0.2 is the same for all five lines, including \ion{Si}{4}. We note that the allowed \ion{Si}{4} line is often reported to have different width compared to the intercombination \ion{O}{4} lines \citep[e.g.,][]{Kjeldseth-Moe77,Doschek77,Doschek78,Doschek04,Feldman77,Akiyama05,Doschek16}. Our result indicates that the observed spectrum is unlikely to contain structures with wildly different electron densities.

The interpretation of the line profile as being given by the ion $\kappa$-distribution does not remove the need for non-thermal broadening\footnote{Here, the term ``non-thermal'' does not refer to non-Maxwellian; rather, it refers to broadening processes other than ion thermal motions, i.e., unresolved motions such as waves.}. This is obvious from the ion temperatures $T_\mathrm{i}$ derived from the line widths (Table \ref{Table:2}), which are all above 1.4 MK and varying among the lines. This variation is likely being caused by the factor $(\kappa-3/2)w_\kappa^2$ being present in the denominator of Eq. (\ref{Eq:Kappa_profile}), which makes the $w_\kappa$ not completely independent from $\kappa$ (see discussion in Sect. \ref{Sect:4.2.2}) even if FWHM$_\kappa$ of the line is a well-defined quantity. Low uncertainties on both $\kappa$ and $w_\kappa$ are found only for the strongest lines; the lower the $I_0$, the larger the corresponding uncertainty on these fit parameters (see Table \ref{Table:2}).

The non-thermal broadening can be for the case of $\kappa$-distributions derived analogously as for the Maxwellian. Recalling Eq. (\ref{Eq:w_kappa}) and assuming that the thermal and non-thermal broadening have the same $\kappa$, we can write
\begin{equation}
	w_\kappa^2 = \frac{1}{2} \frac{\lambda_0^2}{c^2} (\theta^2 + (\theta^\mathrm{(nth)})^2) = (w^\mathrm{(th)}_\kappa)^2 + (w^\mathrm{(nth)}_\kappa)^2 \,.
	\label{Eq:nonth_widths}
\end{equation}
where $w_\kappa$ is now the observed width, and $w_\kappa^\mathrm{(th)}$ and $w_\kappa^\mathrm{(nth)}$ are the thermal and non-thermal contributions. We note that these numbers differ for each $\kappa$ due to the shift of the electron temperature $T_\mathrm{max}$ at which the relative ion abundance has its peak. For $\kappa$-distributions and TR ions, the shift is towards lower log($T_\mathrm{max}$\,[K]) for smaller $\kappa$ \citep{Dzifcakova13a,Dudik14a}, with the corresponding decrease of $w^\mathrm{(th)}$ if the ions are assumed to have the same temperature as electrons. These numbers are shown in Table \ref{Table:3} for the Maxwellian and $\kappa$\,=\,2 distributions, respectively, together with the resulting $w_\kappa^\mathrm{(nth)}$. It is obvious that the $\kappa$-distributions with low $\kappa$, detected by fitting the line profiles in this work, lead to a small increase of the non-thermal characteristic widths. 

There is however no reason to assume that the thermal and non-thermal broadening components have the same $\kappa$, as we have done in Eq. (\ref{Eq:nonth_widths}). For example, the ion distribution could be a Maxwellian at temperatures close to $T_\mathrm{max}$, while it could be only the broadening component (e.g. due to turbulence) that has a $\kappa$-distribution. The resulting line profile would then be a convolution of a Gaussian and a $\kappa$ profile (cf., \ref{Sect:4.2.3}). We have attempted to fit such convolved profiles to the observed TR lines, assuming Gaussian FWHM$_\mathrm{G}$ corresponding to $w_\mathrm{Maxw}^\mathrm{(th)}$ (Table \ref{Table:3}) and a $\kappa$ component. The resulting fits have somewhat lower $\kappa$ values than a purely $\kappa$ fit. For the example spectrum studied in Sect. \ref{Sect:4.2.2} we obtained $\kappa$\,=\,2.00 $\pm$0.08 for \ion{O}{4} 1401.2\,\AA, while for \ion{Si}{4} 1402.8\,\AA~we got $\kappa$\,=\,2.06 $\pm$0.03. These lower of $\kappa$ values are not surprising: If the Gaussian component of the convolution has non-negligible width, the convolved $\kappa$-Gaussian profile has lower wings than a purely $\kappa$ one. Therefore, the fit has to decrease the fitted $\kappa$ in the convolved profile in order to match the observed line profiles.

%
\section{Summary}
\label{Sect:8}

We analyzed the \textit{IRIS} FUV observations of an active region containing closed, bright TR loops and a plage. Spectra containing strong and symmetric spectral lines, belonging to \ion{O}{4}, \ion{Si}{4}, and \ion{S}{4}, were fitted using $\kappa$-distributions. An example spectrum of a closed TR loop was also fitted using single and double-Gaussian fits. The single-Gaussian fits failed to properly account for the line profiles, which showed both pronounced peaks and wings. Two Gaussians fitted the profiles much better, but the fit required the relative intensities of the two Gaussians to be nearly the same, especially for weaker lines of \ion{O}{4} and \ion{S}{4}. 

All five TR lines can be fitted with a $\kappa$-distribution of ion velocities equally well, or in the case of \ion{Si}{4} better than the double-Gaussian fits. The $\kappa$-fit however contains less free parameters. The values of $\kappa$ obtained from the line profile fitting are low, about 2, and are typical especially for the strongest 1402.8\,\AA~line of \ion{Si}{4}. Similar values of $\kappa$ were found for all five lines in the majority of pixels, but there were some pixels where the \ion{O}{4} lines can have higher values of $\kappa$. A single spectrum from a dot-like bright point however exhibited a nearly Gaussian \ion{Si}{4} profile, which allowed us to rule out the instrumental profile or the point-spread function as a cause of the large line wings. In addition to the line profile analysis, the line intensities relative to the \ion{Si}{4} one can also be well-fitted with an electron $\kappa$-distribution with a similar value of $\kappa$\,=\,2, if photospheric abundances are assumed.

We however found that the $\kappa$-distributions do not remove the non-thermal broadening component. If anything, they increase the non-thermal width, since for $\kappa$-distributions the TR ions are expected to exist at lower $T_\mathrm{e}$ than for the Maxwellian case. We also found that all five TR lines can have the same FWHM, of about 0.2\,\AA, irrespective of whether the line is an allowed or an intercombination one. This is contrary to most reports within the literature (although see \citeauthor{Polito16b} \citeyear{Polito16b} as well as Fig. 8 of \citeauthor{Doschek16} \citeyear{Doschek16}), and suggests that the allowed and intercombination lines in our case do not form in regions with highly different densities \citep[cf.,][]{Doschek84,Doschek16}. Furthermore, since the optical thickness of a line depends inversely on its width, we found that the observed non-thermal width, which is about an order of magnitude larger than the thermal one, is sufficient to make the \ion{Si}{4} 1402.8\,\AA~line optically thin. This is because for the large non-thermal width, the potential absorbers are spread throughout the wavelength range of the line, decreasing the number of absorbers at any given $\lambda$.

These results mean that, at least in the cases where the FWHM and $\kappa$ are the same for all five lines, the mechanism creating the line profile has to act in the same way on all TR ions observed, \ion{Si}{4}, \ion{O}{4}, and \ion{S}{4}, i.e., through at least the range of temperatures where these ions are formed. However, similar values of $\kappa$ have been obtained for a transient coronal loop observed by \textit{Hinode}/EIS in \ion{Fe}{11}--\ion{Fe}{12} emission \citep{Dudik15} and for flare loops emitting in \ion{Fe}{16} and \ion{Fe}{23} by \citet{Jeffrey16,Jeffrey17}. This could mean that the range of temperatures is not limited to the transition region investigated here, but could be present throughout the outer solar atmosphere.

Although the values of $\kappa$ obtained here from the line profile fitting and the line intensity ratios are consistent, we finally note that the present data do not provide unambiguous evidence for the presence of $\kappa$-distributions of ion and electron velocities in the solar transition region. A unique diagnostics can only be obtained from ratios of lines originating from levels with widely different excitation energies within the same ion \citep{Dudik14b,Dudik15}, but such lines cannot be observed by \textit{IRIS} due to its limited wavelength range. In this regard, a coordinated \textit{IRIS} observations together with the future SPICE instrument on board the \textit{Solar Orbiter} could be helpful, since both instruments are designed to observe \ion{O}{4} lines, but at wavelength ranges different by a factor of two.

\acknowledgments
The authors thank the anonymous referee for comments that helped to improve the manuscript. The authors are also grateful to H. E. Mason, P. Heinzel, H. Tian, and N. Jeffrey for useful discussions. J.D. and E.Dz. acknowledge Grants No. 17-16447S and 16-18495S of the Grant Agency of the Czech Republic, as well as institutional funding RVO:67985815 from the Czech Academy of Sciences. J.D. and G.D.Z. also acknowledges support from the Royal Society via the Newton Fellowships Alumni Programme. G.D.Z. acknowledges STFC funding through the DAMTP astrophysics grant. 
IRIS is a NASA small explorer mission developed and operated by LMSAL with mission operations executed at NASA Ames Research center and major contributions to downlink communications funded by the Norwegian Space Center (NSC, Norway) through an ESA PRODEX contract.
CHIANTI is a collaborative project involving the NRL (USA), the University of Cambridge (UK), and George Mason University (USA).
\facilities{IRIS.}



\bibliographystyle{aasjournal}
\bibliography{IRIS_TR_Profiles}

%
\begin{figure*}[!ht]
	\centering
	\includegraphics[width=6.14cm,bb=10  0 350 355,clip]{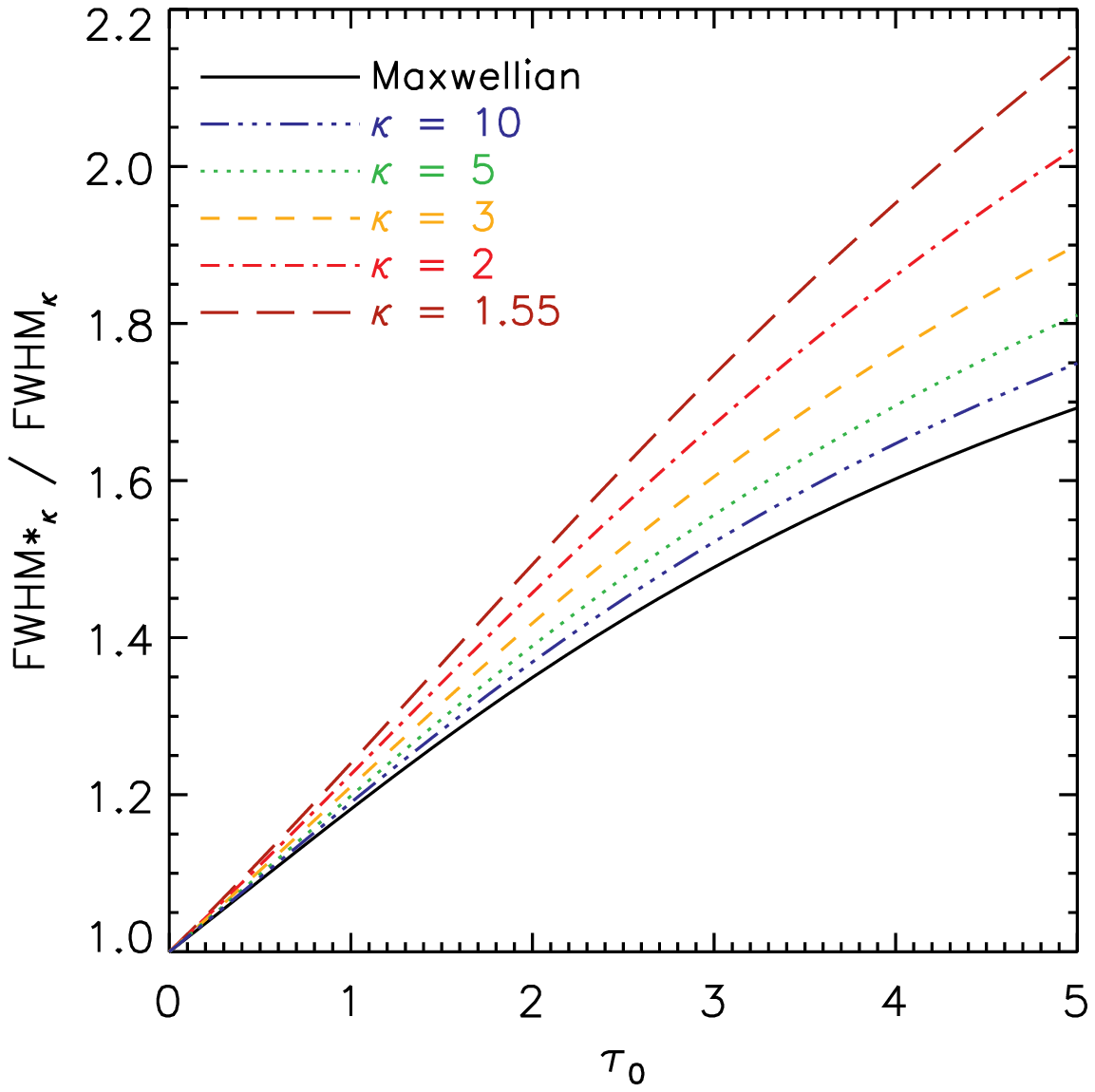}
	\includegraphics[width=6.32cm,bb=00  0 350 355,clip]{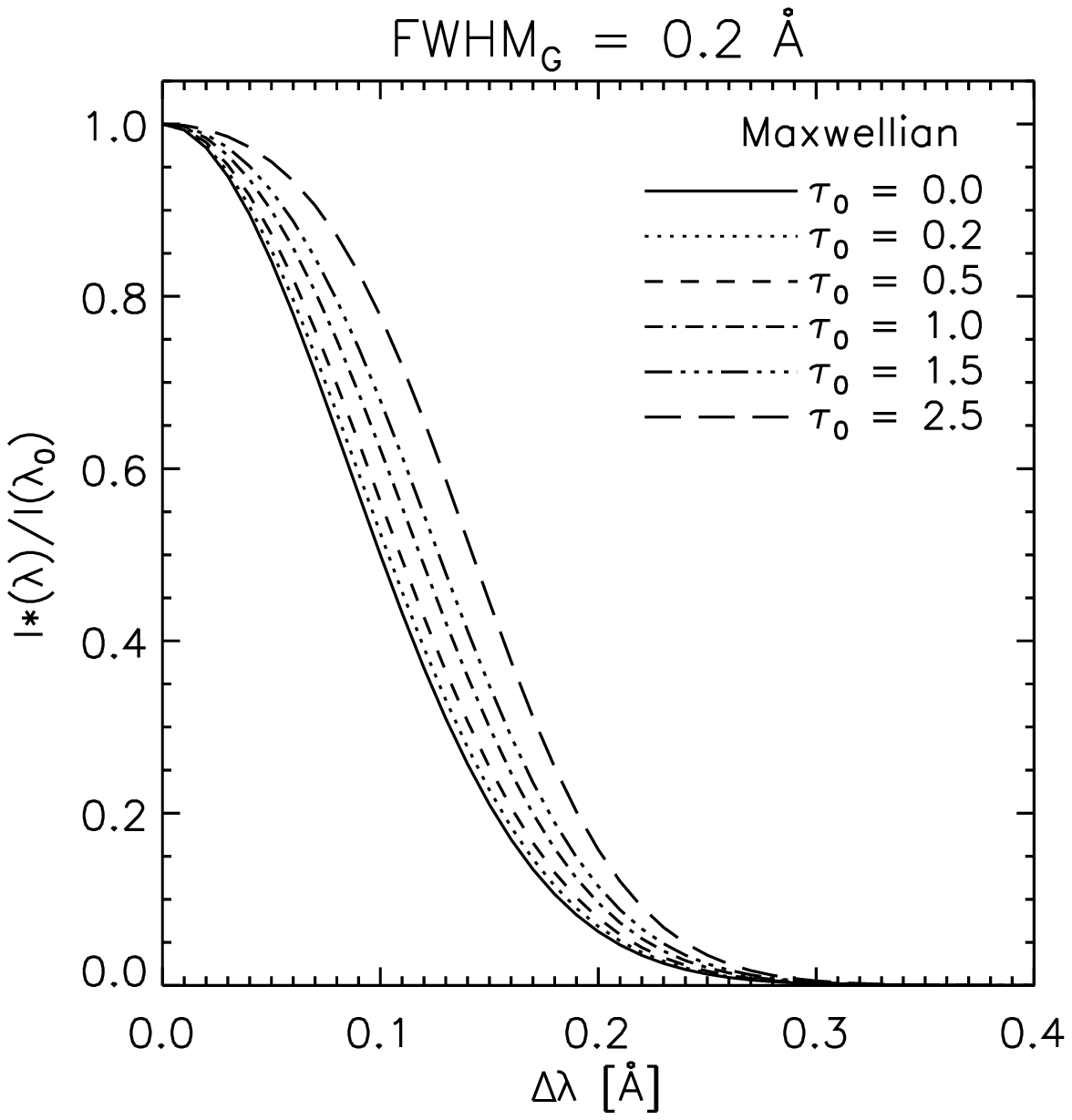}
	\includegraphics[width=5.14cm,bb=65  0 350 355,clip]{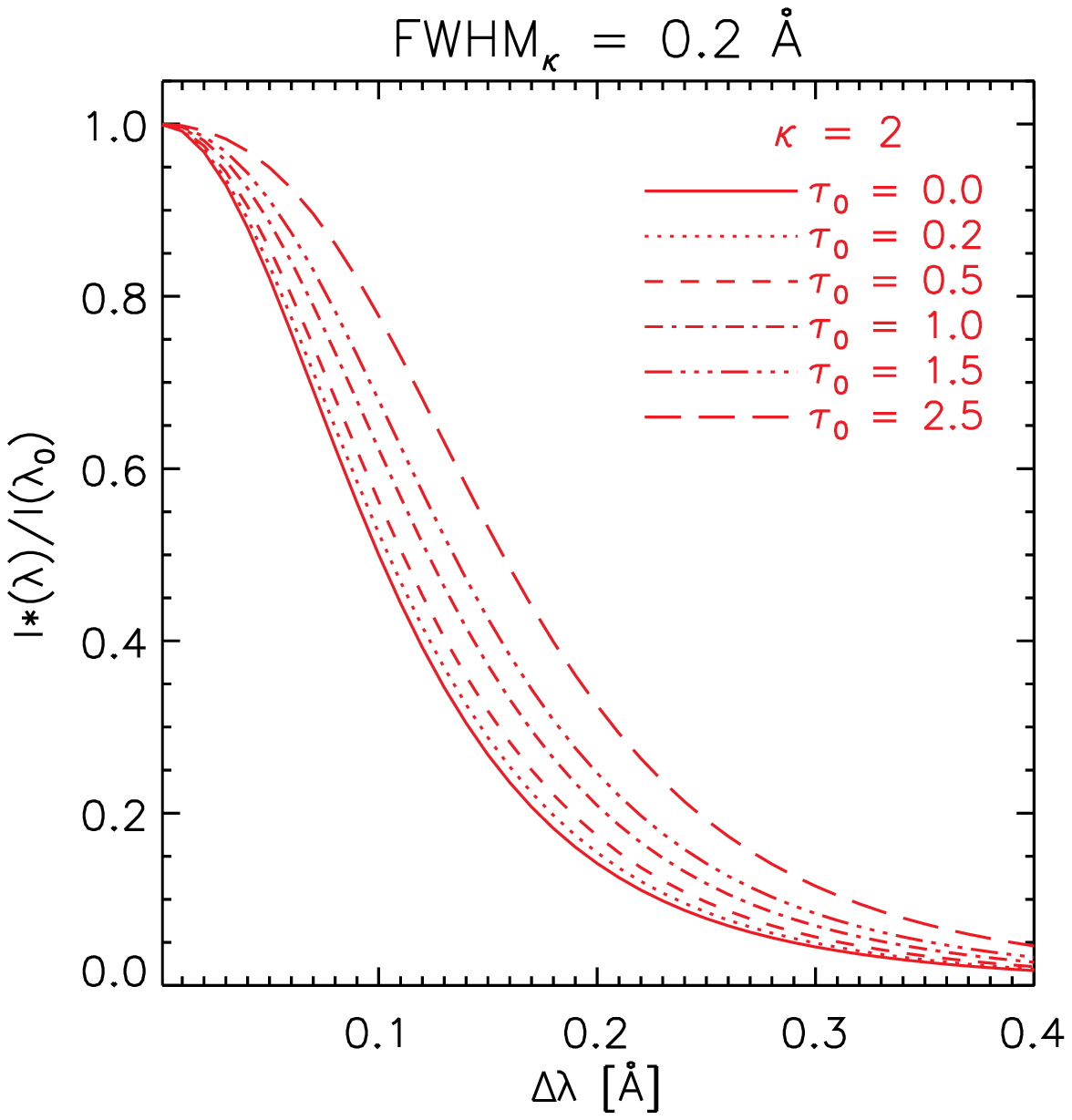}

\caption{Left: Ratio of the FWHM for an optically thick line to the optically thin case. Middle and right: Corresponding line profiles for a Gaussian and a $\kappa$\,=\,2 cases, respectively.
\label{Fig:FWHM_kappa_tau}}
\end{figure*}
%
%

\appendix
\section{Is the \ion{Si}{4} 1402.8\,\AA~line optically thick?}
\label{Sect:A}

In Sects. \ref{Sect:4.2} and \ref{Sect:5} we found that the $\chi^2_\mathrm{red}$ is always higher than 1, with the lowest values found being of about 3 despite the $\kappa$-distribution providing good approximation of all five TR line profiles. We have also reported that the largest residuals occur within the \ion{Si}{4} profile and that the $\chi^2_\mathrm{red}$ does not depend on the small ``lumps and bumps'' above the pseudo-continuum, such as the \ion{S}{1} or \ion{Fe}{2} lines. 

A possible explanation of these high residuals is that the \ion{Si}{4} 1402.8\,\AA~line is not entirely optically thin. Although we do not observe any self-absorption features \citep{Yan15}, weak optically thick effects could still lead to flattening of the peak of the line. Since the \textit{IRIS} observation analyzed here does not contain the other Si doublet line at 1393.8\,\AA~\citep[see, e.g., Sect. 5.1 in][]{DelZanna02}, we resort to an estimate the optical depth $\tau$ in the \ion{Si}{4} 1402.8\,\AA~line. To do that, we use the classical formula \citep[e.g.,][Eq. (15) therein]{Buchlin09}
\begin{equation}
	\tau(\lambda) = \tau_{0}(\lambda_0) \Phi(\lambda) = \frac{\lambda_0^4 A_{ij}\Phi(\lambda)}{4\pi^{3/2} c \Delta\lambda_\mathrm{D}}  \frac{N(\mathrm{Si}^{+3})}{N(\mathrm{Si})} A(\mathrm{Si}) \frac{N_\mathrm{H}}{N_\mathrm{e}} \left<N_\mathrm{e}\right> \Delta s \,,
	\label{Eq:tau_Si4}
\end{equation}
where $\Delta\lambda_\mathrm{D}$ is the (Doppler) width of the absorption profile $\Phi(\lambda)$, which is normalized to $\Phi(\lambda_0)$\,=\,1, $N(\mathrm{Si}^{+3})/N(\mathrm{Si})$ is the relative ion abundance of \ion{Si}{4}, $A($Si) is the abundance of Si, which we take to be photospheric (see Sect. \ref{Sect:6}), $\left<N_\mathrm{e} \right>$ is the average electron density in the emitting source, and $\Delta s$ is the path length along the line of sight through the source.

Taking for simplicity $\Delta\lambda_\mathrm{D} = w_\mathrm{Maxw}^\mathrm{(th)}$ corresponding to Gaussian thermal profile (Table \ref{Table:3}), $N(\mathrm{Si}^{+3})/N(\mathrm{Si})$\,$\approx$\,0.19 according to CHIANTI 8 \citep{DelZanna15b} for the Maxwellian distribution at log$(T_\mathrm{max}$\,[K])\,=\,4.9, and $\Delta s$\,=\,0.33$\arcsec f$, where $f$ is the path length filling factor in an \textit{IRIS} pixel, we obtain
\begin{equation}
	\tau_{0} \approx 0.26 f \frac{\left<N_\mathrm{e}\right>}{10^{10}\, \mathrm{cm}^{-3}}\,
	\label{Eq:tau_Si4_value}
\end{equation}
at the line center of \ion{Si}{4}. The corresponding numerical factor for the \ion{O}{4} and \ion{S}{4} lines is negligible \citep[see also][]{Doschek78}, of the order of 10$^{-6}$, because of the correspondingly lower $A_{ij}$ values for these intercombination lines.

Interestingly, for $\kappa$\,=\,2, the numerical factor changes to $\approx$\,1.5, mostly due to ($i$) decrease of $\Delta \lambda_\mathrm{D}$\,=\,$w_{\kappa=2}^\mathrm{(th)}$ with respect to Maxwellian (Table \ref{Table:3}), and ($ii$) increase in $N(\mathrm{Si}^{+3})/N(\mathrm{Si})$ to about \,$\approx$\,0.39 for such $\kappa$ at log$(T_\mathrm{max}$\,[K])\,=\,4.10 \citep[cf. Fig. 1 in][]{Dudik14a}.

We however note that the value of $\tau_{0}$ depends inversely on the choice of $\Delta\lambda_\mathrm{D}$. Taking the observed $\Delta\lambda_\mathrm{D}$\,=\,$w_\kappa$ instead of $w^\mathrm{(th)}_\kappa$ would lead to a substantial decrease,
\begin{equation}
	\tau_{0} \approx 0.02 f \frac{\left<N_\mathrm{e}\right>}{10^{10}\, \mathrm{cm}^{-3}}\,
	\label{Eq:tau_Si4_value2}
\end{equation}
for a Maxwellian distribution. The corresponding numerical factor for $\kappa$\,=\,2 is about 0.06.

If $\tau_0$\,$>$\,0, optical thickness effects should change the shape, and thus the width, of the line profile. Assuming that the source function $S_\lambda$\,=\,const., the emergent intensity $I^*$ is given by \citep[e.g.][]{Doschek04,Hubeny14}
\begin{equation}
	I^*(\lambda) = \int\limits_{0}^{\tau(\lambda)} S_\lambda \exp{(-t_\lambda)} \mathrm{d}t_\lambda \\
	\nonumber = S_\lambda \left[1 - \exp(-\tau(\lambda))\right]\,,
	\label{Eq:tau_profile}
\end{equation}
where $\tau(\lambda)$ is given by Eq. (\ref{Eq:tau_Si4}). If we now assume that $\Phi(\lambda)$\,=\,$I_\kappa(\lambda)/I_0$, i.e., a line profile given by a $\kappa$-distribution (Eq. \ref{Eq:kappa_fit}), the FWHM$^*_\kappa$ of such optically thick profile is given by
\begin{equation}
	{\mathrm{FWHM}^*_\kappa(\tau_0)}^2 = 8 (\kappa -3/2)  w_\kappa^2 \left[\left(\frac{\tau_0}{\mathrm{ln}(2)- \mathrm{ln}(\exp(-\tau_0)+1)}\right)^\frac{1}{\kappa} -1 \right]\,,
	\label{Eq:FWMH_kappa_tau}
\end{equation}
which reverts to the expression for FWHM$_\kappa^2$ (Eq. \ref{Eq:w_kappa}) for $\tau_0$\,$\to$\,0.
For a Gaussian profile $\Phi(\lambda)$\,=\,$I_\mathrm{G}(\lambda)/I_0$ (Eq. \ref{Eq:Gauss_profile}), we obtain
\begin{equation}
	{\mathrm{FWHM}^*_\mathrm{G}(\tau_0)}^2 = 8 w_\mathrm{G}^2 \left[\mathrm{ln}(\tau_0) - \mathrm{ln}\left(\mathrm{ln}(2) - \mathrm{ln}(1+ \mathrm{e}^{-\tau_0})\right)\right]\,.
	\label{Eq:FWMH_Gauss_tau}
\end{equation}
The FWHM$_\kappa^*(\tau_0)$\,/\,FWHM$_\kappa$ ratios are shown in the left panel of Fig. \ref{Fig:FWHM_kappa_tau}. The corresponding profiles given by Eq. (\ref{Eq:tau_profile}) are shown for the Gaussian and $\kappa$\,=\,2 cases in the middle and right panels, respectively. The profiles for $\tau_0$\,$\neq$\,0 have progressively increased width, which occurs dominantly in the peak of the line. We note that such profiles could still be fitted with a $\kappa$-distribution, but the goodness-of-fit decreases with increasing $\tau_0$, since a significant mis-match in the peak occurs if $\tau$\,$>$\,1. The resulting $\kappa$ obtained from the fit also increase with $\tau_0$. For example, a profile with $I_0$\,=\,10$^3$\,DN, $\kappa$\,=\,2 and $\tau_0$\,=\,1 could be fitted with $\kappa_\mathrm{fit}$\,=\,2.36\,$\pm$0.06, while a $\kappa$\,=\,5 profile with the same optical thickness would yield $\kappa_\mathrm{fit}$\,=\,11.1 $\pm$1.3. These resulting $\kappa_\mathrm{fit}$ are only weakly depdendent on the $I_0$.

So is the line optically thick? To hint at the answer of this question, we make use of the formulae derived above. Taking $\tau_0$\,=\,1.5 (see discussion following Eq. \ref{Eq:tau_Si4_value}), we obtain for $\kappa$\,=\,2 that the observed FWHM of the line should be a factor of $\approx$\,1.34 higher than if the line is optically thin. Considering now the example spectrum investigated in Sect. \ref{Sect:4}, which has nearly the same FWHM$_\kappa$ for all five TR lines (Table \ref{Table:2} and Fig. \ref{Fig:Fits}), and taking into account that the intercombination lines are optically thin, we are forced to conclude that the \ion{Si}{4} line is also optically thin, as suggested by Eq. (\ref{Eq:tau_Si4_value2}). It is the large non-thermal width of these TR lines that lead to suppression of $\tau_0$: Essentially, a line with large non-thermal width means that the potential absorbers are spread throughout the wavelength range of its profile, leaving correspondingly fewer absorbers at any given wavelength $\lambda$.

This result also mean that the large residuals for the $\kappa$-fit of the \ion{Si}{4} line in the example spectrum studied in Sect. \ref{Sect:4.2.2} cannot be explained by optically thick effects. The relative symmetry of the $\kappa$-fit residuals around 1402.8\,\AA~(Fig. \ref{Fig:Fits}) could then suggest presence of a weak additional component, with peak intensity of at most $\approx$\,4\% of the dominant $\kappa$-component.

%
%
\begin{figure*}
	\centering
	\includegraphics[width=17.6cm,bb= 0   0 980 374,clip]{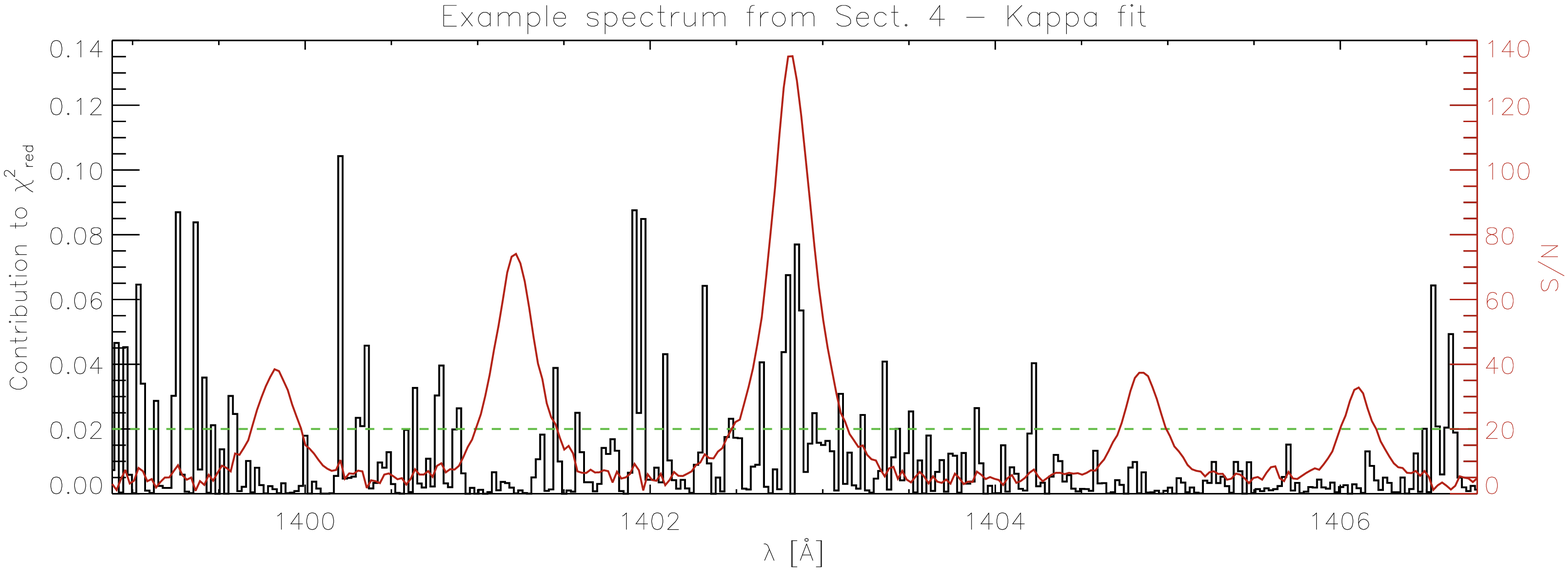}
	\includegraphics[width=6.13cm,bb= 10  0 400 350,clip]{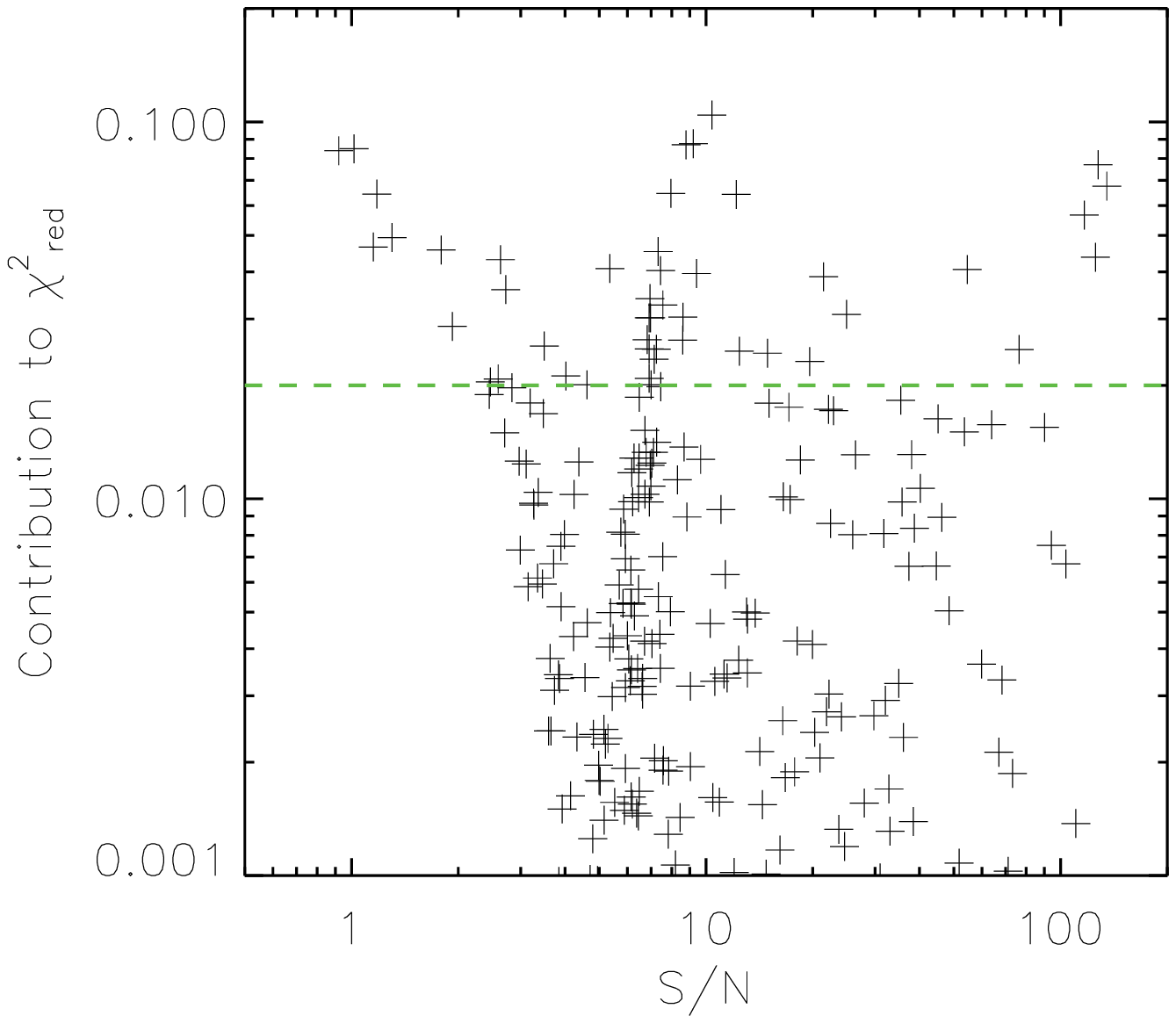}
	\includegraphics[width=4.87cm,bb= 90  0 400 350,clip]{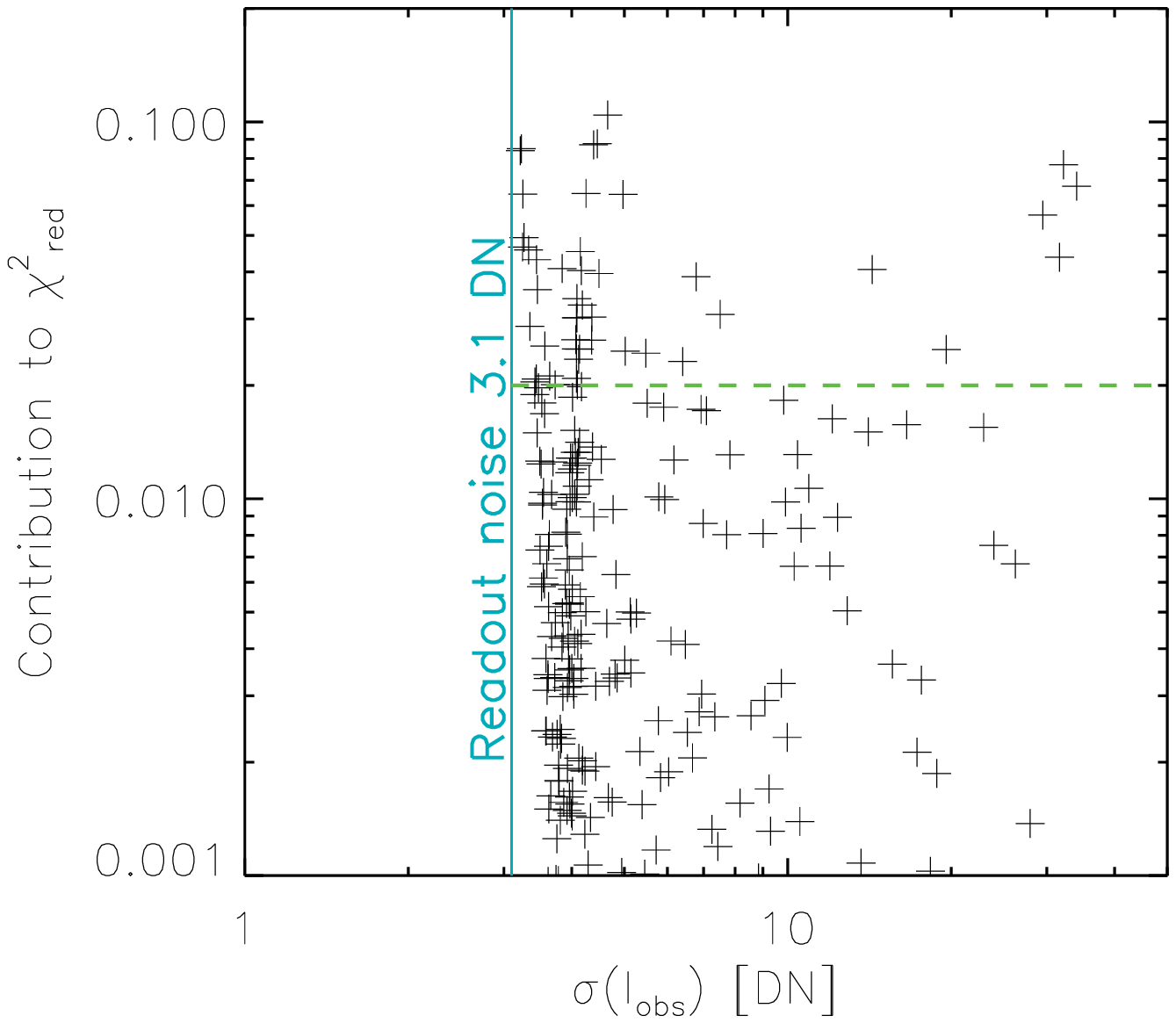}
	\includegraphics[width=5.19cm,bb= 90  0 420 350,clip]{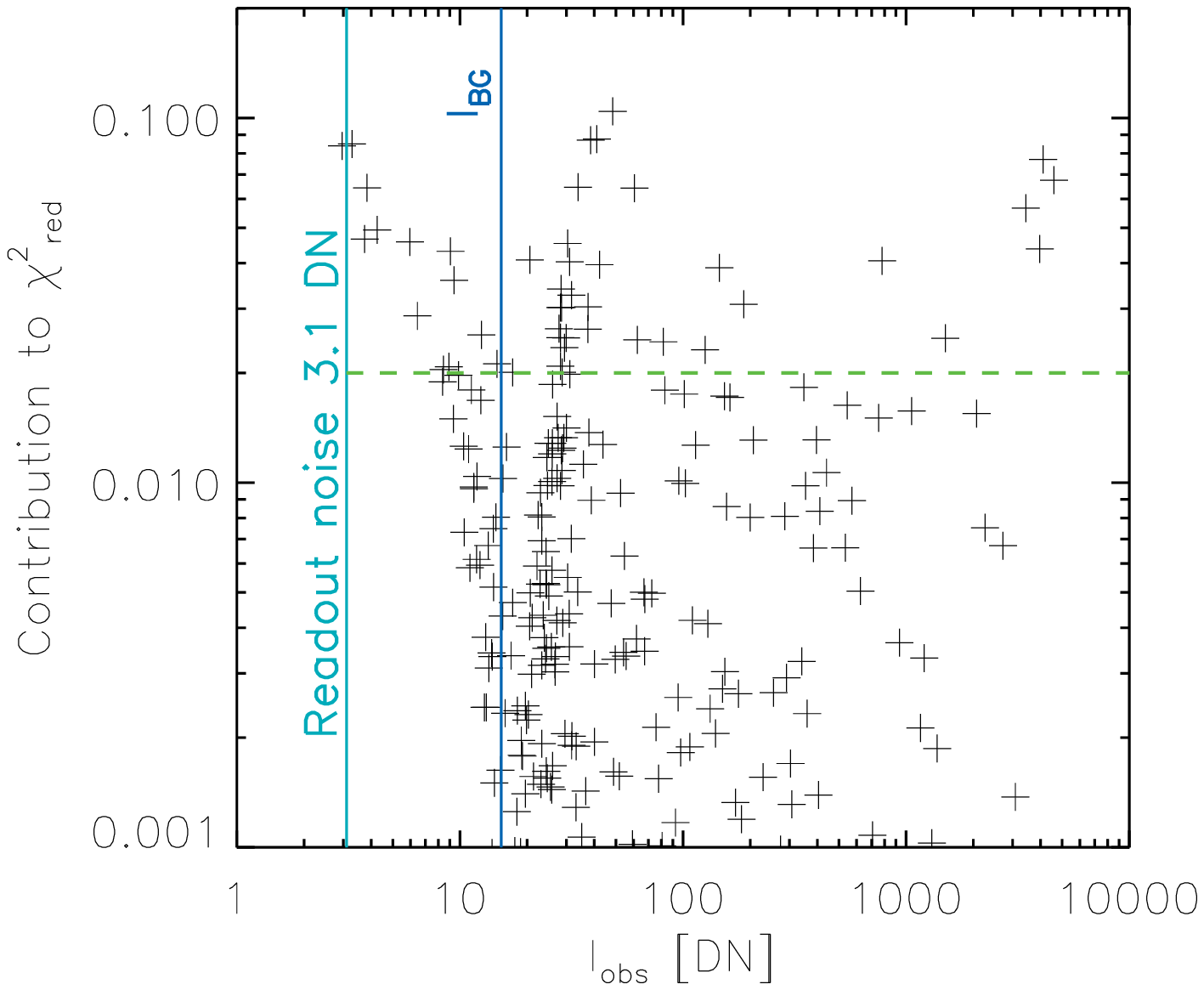}
\caption{Sources of the $\chi^2_\mathrm{red}$. Top: Contributions of individual spectral bins to the total $\chi^2_\mathrm{red}$ (thin black lines). The S/N = $I_\mathrm{obs}(\lambda)$\,/\,$\sigma\left(I_\mathrm{obs}(\lambda)\right)$ is overplotted in dark red. The S/N is larger where the TR lines are observed. Bottom: Scatterplots showing the dependence of contributions to $\chi^2_\mathrm{red}$ on S/N, the signal $I_\mathrm{obs}$, and the noise $\sigma(I_\mathrm{obs})$. The readout noise, equal to 3.1\,DN, is shown by azure vertical line, while the fitted pseudo-continuum $I_\mathrm{BG}$ (see Sect. \ref{Sect:4.2.2}) is shown by vertical blue line. The horizontal dashed green line shows the value of $\chi^2_\mathrm{red}$\,=\,0.02. See text for details.
\label{Fig:chi2r}}
\end{figure*}
%
%
\section{What dominates the $\chi^2_\mathrm{red} \approx$ 3?}
\label{Sect:B}

In Appendix \ref{Sect:A}, we found that the optical thickness effects in the \ion{Si}{4} 1402.8\,\AA~line were not a likely source that increases the $\chi^2_\mathrm{red}$. Why is then the $\chi^2_\mathrm{red}$ almost never lower than 3?

To answer this question, in Fig. \ref{Fig:chi2r} we plot the contribution of individual spectral bins $\lambda_i$ to the total $\chi^2_\mathrm{red}$. This is done for the $\kappa$-fit of the example spectrum reported in Sect. \ref{Sect:4.2.2}. The S/N ratio, equal to $I_\mathrm{obs}(\lambda_i)$\,/\,$\sigma\left(I_\mathrm{obs}(\lambda_i)\right)$, is overplotted in dark red color. The values of S/N follow approximately $I_\mathrm{obs}^{1/2}$. This is because for strong lines, the noise is dominated by the photon noise, while in the pseudo-continuum, the readout noise of 3.1\,DN can be an important contribution.

The top panel of Fig. \ref{Fig:chi2r} shows that there are numerous spectral bins that contribute more than 0.02, some up to 0.08--0.11 to the total $\chi^2_\mathrm{red}$. Here, the value of 0.02 was chosen as an indicative one, since larger values produce a net contribution of about $\approx$2 to the overall $\chi^2_\mathrm{red}$. Within the \ion{Si}{4} 1402.8\,\AA~line, there are 6 such spectral bins, contributing a net $\approx$0.31 to the overall $\chi^2_\mathrm{red}$. Very few such bins are located in other TR lines. Therefore, the majority of dominant contributors to $\chi^2_\mathrm{red}$ come from the pseudo-continuum.

Could these contributions from pseudo-continuum arise from weak, barely resolvable spectral lines that were not fitted? In Sect. \ref{Sect:4.2.2} we reported already that fitting small ``lumps and bumps'' above the pseudo-continuum does not help decrease the $\chi^2_\mathrm{red}$. To find out which pseudo-continuum pixels dominate the $\chi^2_\mathrm{red}$, in Fig. \ref{Fig:chi2r} we plot the dependence of the contribution to $\chi^2_\mathrm{red}$ on the S/N, as well as $I_\mathrm{obs}$ and $\sigma(I_\mathrm{obs})$. The results indicate that there are two dominant branches: one having very low S/N\,$\approx$\,1, and the other having S/N\,$\approx$\,7--10. Some contribution comes also from the high residuals in the \ion{Si}{4} line; these can be identified easily since their S/N is among the highest present, above 100.

The first branch originates in pixels where both the signal and noise is dominated by the readout noise. These pixels are close to the 3.1\,DN readout noise limit, shown in vertical azure line in the bottom panels of Fig. \ref{Fig:chi2r}. The other branch, with S/N\,$\approx$\,7--10, arise from pseudo-continuum pixels having intensities several times larger than $I_\mathrm{BG}$, which is shown by the vertical blue line. At least some of these could be due to weak lines from low ionization stages, such as \ion{Fe}{2} 1399.97\,\AA~(already mentioned in Sect. \ref{Sect:5.1.1}), 1401.7\,\AA, 1404.12\,\AA, etc., as well as \ion{He}{2} 13998.95\,\AA, 1403.98\,\AA, and a few unidentified lines at 1400.31\,\AA, 1401.96\,\AA, and so on. More details on these weak lines can be found in \citet{Sandlin86}, \citet{Keenan02}, \citet{Young15}, \citet[][Figure 2 therein]{Tian15}, and \citet[][Figure 14 therein]{Polito16a}. However, since the width of
such lines is small, about 2 wavelength bins in the present spectrum, these lines cannot be clearly recognized in the spectrum if they are weak, much less reliably fitted with a single Gaussian having 3 free parameters. Although including badly constrained fits to ``lumps and bumps'' would remove some of the contributions to the overall $\chi^2_\mathrm{red}$ as shown in Fig. \ref{Fig:chi2r}, doing so would also increase the number of fit parameters $N_\mathrm{fit}$, and thus reduce the $\nu$ factor (see Eq. \ref{Eq:chi2_reduced}), which in turn \textit{increases} the $\chi^2_\mathrm{red}$ if too many pseudo-features are forced to be fitted.

\end{document}